%% file: Goswami_et_al_2021.tex
\documentclass{aa}  

\usepackage{hyperref}
\usepackage{float}
\usepackage{tabularx}
\usepackage{multirow}
\usepackage{footnote}
\usepackage{subfigure}
\usepackage{ae,aecompl}
\usepackage{graphicx}   
\usepackage{amsmath}   
\usepackage{amssymb}   
\usepackage[switch]{lineno} 
\usepackage{txfonts}

\begin{document}

   \title{Spectroscopic study of CEMP-(s \& r/s) stars\thanks{Based [in part] on data collected at Subaru Telescope, which is operated by the National Astronomical Observatory of Japan.}\thanks{[Part of] the data are retrieved from the JVO portal (\url{http://jvo.nao.ac.jp/portal}) operated by the NAOJ.}\thanks{Tables 4 and 6 are only available in electronic form at the CDS via \url{https://cdsarc.u-strasbg.fr/ftp/vizier.submit//828/}}}

   \subtitle{Revisiting classification criteria and formation scenarios, highlighting i-process nucleosynthesis}

   \author{Partha Pratim Goswami
          \inst{1,2}
          \and
          Rajeev Singh Rathour
          \inst{1,3}
          \and
          Aruna Goswami\inst{1}
          }

   \institute{Indian Institute of Astrophysics, Koramangala, Bangalore
    560034, India  \\
              \email{partha.pg@iiap.res.in; aruna@iiap.res.in}
         \and
         Pondicherry University, R.V. Nagar, Kalapet, 605014, Puducherry, India
         \and
             Indian Institute of Science Education and Research, Pune, Maharashtra,
    411008, India\\
             \email{rajeev.rathour@students.iiserpune.ac.in}
             }

   \date{Received 24 April 2020; accepted 09 December 2020}

  \abstract
{ The origin of the enhanced abundances of both s- and r-process 
elements observed in a subclass of carbon-enhanced metal-poor (CEMP) stars, denoted CEMP-r/s stars, still remains poorly understood. The i-process
nucleosynthesis has been suggested as one of the most promising mechanisms for the origin  of these stars.}
 {Our aim is to  better understand  the chemical signatures and formation 
 mechanism(s) of five previously  claimed potential CH star candidates HE~0017+0055, HE~2144$-$1832, HE~2339$-$0837, HD~145777, and CD$-$27~14351
 through a detailed systematic follow-up spectroscopic study based on
 high-resolution spectra.}
   {The stellar atmospheric parameters, the effective temperature T$_{eff}$, the microturbulent  velocity $\zeta$, the surface gravity log g, and the metallicity [Fe/H] are derived from local thermodynamic equilibrium analyses using model atmospheres. Elemental abundances of C, N, $\alpha$-elements, iron-peak elements, and several neutron-capture elements are estimated using the equivalent width measurement technique as well as spectrum synthesis calculations in some cases. 
   In the context of the double enhancement observed in four of the programme stars, we have critically examined whether the literature i-process model yields ([X/Fe]) of heavy elements can explain the observed abundance distribution.}
   {The estimated metallicity [Fe/H] of the programme  stars ranges from $-$1.63 to $-$2.74. All  five stars show enhanced abundance for Ba, and four of them exhibit  enhanced abundance for Eu. Based on our analysis, HE~0017+0055, HE~2144$-$1832, and HE~2339$-$0837 are found to be CEMP-r/s stars, whereas HD~145777 and CD$-$27~14351 show characteristic properties of CEMP-s stars. 
   From a detailed analysis of different classifiers of CEMP  stars, we have identified the one which best describes the CEMP-s and CEMP-r/s stars. We found that for both CEMP-s and CEMP-r/s stars, [Ba/Eu] and [La/Eu] exhibit positive values and [Ba/Fe]~$\geq$~1.0. However, CEMP-r/s stars satisfy [Eu/Fe]~$\geq$~1.0, 0.0~$\leq$~[Ba/Eu]~$\leq$~1.0, and/or 0.0~$\leq$~[La/Eu]~$\leq$~0.7. CEMP-s stars normally show [Eu/Fe]~$<$~1.0 with [Ba/Eu]~$>$~0.0 and/or [La/Eu]~$>$~0.5. If [Eu/Fe]~$\geq$~1.0, then the condition on [Ba/Eu] and/or [La/Eu] for a star to be a CEMP-s star is [Ba/Eu]~$>$~1.0 and/or [La/Eu]~$>$~0.7. Using a large sample of similar stars from the  literature we have examined whether the ratio of heavy-s to light-s process elements [hs/ls] alone can be used as a classifier, and if there are any limiting values for [hs/ls]  that can be used to distinguish between CEMP-s and CEMP-r/s stars. Even though they  peak at different values of [hs/ls], CEMP-s and CEMP-r/s stars show an overlap in the range 0.0~$<$~[hs/ls]~$<$~1.5, and  hence this ratio cannot be used to distinguish between CEMP-s and CEMP-r/s stars. We have noticed a similar overlap in the case of [Sr/Ba]  as well, in the range $-$1.6~$<$~[Sr/Ba]~$<$~$-$0.5, and hence this ratio also cannot be used to separate the two subclasses.}
   {}


   \maketitle
%

\section{Introduction}
\label{sec:introduction}
The origin and evolution of neutron-capture elements in our
Galaxy are still unclear. That is why CH stars, with their metal-poor counterpart carbon-enhanced metal-poor (CEMP) stars have been studied for a very long 
time.  Both CH and CEMP stars are characterised by the presence of a strong G band of CH. While the broad category of objects defined to include stars with [C/Fe]\footnote[1]{\textbf{Notation:} [A/B] = log(N$_{A}$/N$_{B}$)$_{*}$ $-$ log(N$_{A}$/N$_{B}$)$_{\odot}$, where N$_{A}$ and N$_{B}$ are number densities.}$>$ 1.0 and [Fe/H] $<$ $-$1.0 are referred to as CEMP stars \citep{beers2005discovery}, CH stars normally exhibit a metallicity range of that of Ba stars.  The main differences between CH and Ba stars  lie in the carbon abundance and the value of  C/O. Unlike CH stars Ba stars do not  exhibit enhancement of carbon.  It has  been pointed out by several authors  that the s-process enhanced CEMP stars,  the so-called CEMP-s stars,   are low-metallicity analogues of CH stars and Ba stars (see \citealt{lucatello2005}, \citealt{starkenburg_et_al.2014}, and references therein). When a star ascends to the giant branch, the abundance of carbon at the surface decreases due to the mixing with the first dredge-up affected internal material. \citet{spite_et_al_2005} observed that in highly evolved metal-poor red giants the abundance of carbon decreases further due to the influence of extra mixing in the evolutionary path, which increases the abundance of nitrogen. Including these evolutionary effects, \citet{aoki2007carbon} put forward a slightly different classification scheme that considers  carbon abundance ([C/Fe]$\geq$ +0.7) along with the luminosity of the star. Some authors \citep{lee2013carbon, skuladottir2015first}  use [C/Fe]$\geq$ +0.7 to define  CEMP stars.
  The subclasses of CEMP stars help us to uncover the processes by which the neutron-capture elements are produced.
Depending upon the enhancement of elements produced by slow (\textquoteleft s') and rapid (\textquoteleft r') neutron-capture processes, CEMP stars are divided into different subclasses \citep{beers2005discovery, jonsell2006, masseron2010aholistic, abate2016cemp-rs, Frebel_review_2018,hansen2019abundances}. The early subclassification  of CEMP class was given by \citet{beers2005discovery}:
\begin{itemize}
        \item CEMP: [C/Fe] $>$ 1.0;
        \item CEMP-s:  [Ba/Fe]$>$+1.0, and [Ba/Eu]$>$0.5
        (characterised by enhancement of  barium, which is an s-process indicator);
        \item CEMP-r: [Eu/Fe]$>$+1.0 
                (characterised by enhancement of europium, which is an r-process indicator);
        \item CEMP-r/s: 0.0$<$[Ba/Eu]$<$0.5
        (enhanced in both barium and europium);
        \item CEMP-no: [Ba/Fe]$<$0
        (not enhanced in heavy elements).
 \end{itemize}
\citet{abate2016cemp-rs} adopted the following classification:   
 \begin{itemize}
        \item CEMP: [C/Fe] $>$ 1.0;
        \item CEMP-s:  [Ba/Fe]$>$+1.0 and [Ba/Eu]$>$0; 
        \item CEMP-r: [Eu/Fe]$>$+1.0 and [Ba/Eu]$\leq$0; 
        \item CEMP-r/s: [Eu/Fe]$>$+1.0, [Ba/Fe]$>$+1.0, and [Ba/Eu]$>$0.0; 
        \item CEMP-no: [Ba/Fe]$\leq$1.0 and [Eu/Fe]$\leq$1.0;
        
        These classification criteria for CEMP-s and CEMP-r/s stars are also adopted by \citet{jonsell2006} and \citet{masseron2010aholistic}.
 \end{itemize}
\citet{Frebel_review_2018} adopted the CEMP star definition of \citet{aoki2007carbon}, and put forward a classification scheme of CEMP stars as follows:
\begin{itemize}
    \item CEMP: [C/Fe] $>$ 0.7 for log(L/L$_{\odot}$) $\leq$ 2.3 \& [C/Fe] $\geq$ [3.0 $-$ log(L/L$_{\odot}$)] for log(L/L$_{\odot}$) $>$ 2.3;
    \item r I: 0.3$\leq$[Eu/Fe]$\leq$+1.0 and [Ba/Eu]$<$0.0;
    \item r II: [Eu/Fe]$>$+1.0 and [Ba/Eu]$<$0.0;
    \item r$_{lim}$: [Eu/Fe]$<$0.3, [Sr/Ba]$>$0.5, and [Sr/Eu]$>$0.0;
    \item CEMP-s: [Ba/Fe]$>$+1.0, [Ba/Eu]$>$+0.5, [Ba/Pb]$>$ $-$1.5;
    \item CEMP-r+s: 0.0$<$[Ba/Eu]$<$+0.5 and $-$1.0$<$[Ba/Pb]$<$ $-$0.5;
    \item CEMP-i: 0.0$<$[La/Eu]$<$0.6 and [Hf/Ir]$\sim$1.0 (some authors use the `CEMP-i' nomenclature to indicate CEMP-r/s stars).
    The r-process enriched stars may or may not be carbon enhanced.
\end{itemize}

\citet{hansen2019abundances} recently gave a new scheme of classification based on [Sr/Ba]:
\begin{itemize}
    \item CEMP: [C/Fe] $>$ 1.0;
    \item CEMP-no: [Sr/Ba]$>$0.75; 
    \item CEMP-s: $-$0.5$<$[Sr/Ba]$<$0.75; 
    \item CEMP-r/s: $-$1.5$<$[Sr/Ba]$<$ $-$0.5; 
    \item CEMP-r: [Sr/Ba]$<$ $-$1.5. 
\end{itemize}

A number of  different scenarios  describing  the origin of enhanced heavy elements on  the  surface chemical composition of CEMP-s and CEMP-r/s stars are available in the literature \citep{jonsell2006,lugaro2009}. For s-process enrichment a  binary AGB nucleosynthesis model is considered where the star we observe (secondary) is in a binary configuration with an evolved star (primary). This primary star completes the AGB phase and becomes a white dwarf, and in the process expels  s-process
enriched matter which is then accreted by  the secondary star via two major mass-transfer mechanisms:  Roche-lobe overflow (RLOF) and wind accretion \citep{Abate_et_al_WRLOF2013}.
Most of the CEMP-s stars are confirmed as  binary systems through  long-term radial velocity monitoring \citep{McClure1983, McClure1984, McClure1990, lucatello2005, jorissen2016rv, Hansen2016binaries}.  In addition,  nucleosynthesis
occurring  in the inter-shell region of the secondary star may also contribute to  both the light and heavy s-process element enrichment when the synthesised  material is brought to the surface via different processes like convective mixing (due to temperature gradient), non-convective processes like thermohaline mixing (due to density gradient) \citep{Stancliffe_et_al.2007}, rotation mechanisms, and third dredge-up (TDU).  
Several proposed scenarios are also being put forward for the origin of the CEMP-r/s stars by several authors:  the primary, after passing through the   AGB phase, explodes as  a type 1.5 supernova \citep{zijlstra2004low, wanajo2006enrichment};  the enrichment   of r-process elements is produced  via an accretion-induced collapse \citep{qian2003stellar, cohen2003abundance} and enriches the secondary;  a triple star system having a massive star is responsible 
for enriching the secondary star \citep{cohen2003abundance};   a primordial 
origin (i.e.  the environment, in which the CEMP-r/s star  was born, was  
already enriched  by r-process elements) \citep{bisterzo2011s}. Another 
nucleosynthesis process, generally termed  the intermediate (i) neutron-capture process, is a neutron-capture regime at neutron densities intermediate  
between those for s-process and r-process and has also been 
invoked \citep{cowan1977}. Multiple stellar sites such as very metal-poor AGB stars \citep{Campbell_&_Lattanzio_2008, Cristallo_et_al_2009, Campbell_et_al_2010, Stancliffe2011}, very late thermal pulse (VLTP) in post-AGB stars \citep{Herwig_et_al_2011}, super-AGB stars \citep{Doherty_et_al_2015, Jones_et_al_2016}, low-metallicity massive stars \citep{Bannerjee_et_al_2018,Clarkson_et_al_2018}, and rapidly accreting white dwarfs \citep{Denissenkov_et_al_2017, Cote_et_al_2018, Denissenkov_et_al_2019}  are expected to meet the conditions necessary for the i-process.

 From medium-resolution 
spectroscopic analysis of a sample of  carbon star candidates 
from the Hamburg/ESO survey \citep{christlieb2001}, several potential CH star candidates were identified by \citet{goswami2005ch} and \citet{Goswami2010CH}. In this work we present results from a follow-up high-resolution spectroscopic analysis of three such potential CH star candidates HE~0017+0055, HE~2144$-$1832,  and HE~2339$-$0837, along with two potential CH star candidates HD~145777 and CD$-$27~14351  listed in the CH star catalogue of \citet{bartkevicius1996new}. 
A detailed  systematic spectroscopic study of these  five previously  claimed potential CH stars is conducted  for a better understanding of  their chemical signatures and formation mechanism(s). 
 
The paper is organised as follows.  In Section~\ref{sec:previous_studies}, we present a brief summary of the  earlier studies or any reported measurements on our programme  stars  available in the literature.  The details of observations and data reduction are presented in Section~\ref{sec:data_reduction}.  Determination of photometric  temperatures is discussed in Section~\ref{sec:photometric_temp}. Estimation of radial velocity and derivation of stellar atmospheric parameters are presented in Section~\ref{sec:radial_velocity_stellar_atm_parameters}. Abundance analysis results along with a discussion about abundance uncertainties are presented  in Section~\ref{sec:abundance_analysis}. Section~\ref{sec:kinematic_analysis} discusses the results from the kinematic analysis of our programme stars.  A comprehensive discussion on several proposed formation scenarios of CEMP-r/s stars is presented in    Section~\ref{sec:discussion}.
 A detailed comparison of the observed abundances  with i-process model predictions is also presented in this section along with a discussion on  different classifiers of CEMP-s and CEMP-r/s stars. Details of estimations of [hs/Fe], [ls/Fe], and [hs/ls] from heavy element abundances  of a sample of CEMP-s and CEMP-r/s stars from the literature are also presented in this section. Conclusions are drawn in  Section~\ref{sec:conclusion}.

\section{Previous studies of the programme stars: A summary, and the novelty of this work}
\label{sec:previous_studies}

\paragraph*{}
\textbf{HD~145777} \\
We present the first-time abundance analysis for this object. 
\citet{bidelman1956carbon} classified HD~145777 as a CH star differing from the earlier classifications by \citet{mayall1940new} and \citet{sanford1944radial}, who assigned this object  to  spectral class R3 and R4 respectively.  \citet{bergeat2001effective} and \citet{McDonald2012}   performed studies on several stars including HD~145777 and derived effective temperature using the spectral energy distribution (SED) method of temperature calibration.  Our spectroscopic estimate differs by less than $\sim$ 100 K from these studies.  First-time abundance estimates for several elements including C through Zn, and neutron-capture elements are presented based on high-resolution spectroscopy.

\paragraph*{}
\textbf{CD$-$27~14351} \\
\citet{McDonald2012}   reported estimates of effective temperatures for
a large sample of \textit{Hipparcos} stars including CD$-$27~14351 using 
the spectral energy distribution (SED) method of temperature calibration. \citet{Drisya2017} performed   a detailed chemical composition study for
this object and reported it  to be  a CEMP-r/s star with a high value 
of [Eu/Fe] = 1.65.  These authors  also obtained a negative 
        value ($-$0.05) for [hs/ls] for this object, in contrast to 
        all  CEMP-r/s stars known so far. A  survey of the literature shows that,  
        in general,  CEMP-r/s stars exhibit
a positive value ($\sim$ 0.4 $<$ [hs/ls] $<$ 1.7) for [hs/ls].
This discrepancy prompted us to re-investigate the nature of this object. We have therefore re-examined its spectrum covering the wavelength range 
3500 to 9000 \AA\,. 
While the temperature estimate of our work differs from  \citet{McDonald2012} by $\sim$ 100 K, our estimates for  the atmospheric parameters are similar to those of  \citet{Drisya2017}. 
However, our estimates of elemental abundances for Ce and Eu, differ from those of  \citet{Drisya2017}  by 0.74 dex and 1.26 dex, respectively.

\paragraph*{}
\textbf{HE~0017$+$0055} \\
This object was discovered by \citet{Stephenson1989} as an R-type star, and was assigned  number 39 in the General Catalogue of Galactic Carbon Stars. The object is also listed in the list of faint high-latitude carbon stars  of \citet{christlieb2001}. 
\citet{kennedy2011} estimated the atmospheric parameters and the abundances of C, N, and O for this object.  \citet{jorissen2016HE0017} did a more detailed analysis of this object, deriving the abundances of some of the neutron-capture elements (Y, Zr, La, Ce, Nd, Sm, Eu, Dy, and Er) along with C and N. The effective temperature value for the object estimated by these authors differ from our estimate (by $\sim$120$-$180 K). While \citet{kennedy2011} estimated the log g and [Fe/H] to be 0.18 and $-$2.72, respectively, \citet{jorissen2016HE0017} adopted log g = 1.0 and estimated [Fe/H] = $-$2.40. Both these studies   adopt  a value of 2 km s$^{-1}$ for microturbulence.   \citet{jorissen2016HE0017} determined a low value ($\sim$ 4) for the carbon isotopic ratio $^{12}$C/$^{13}$C, which does not differ much from the value of  1.3 estimated by   \citet{goswami2005ch} based on  medium-resolution spectra. 
From  a long-term radial velocity monitoring programme
\citet{jorissen2016rv} found this object to exhibit  low-amplitude velocity variations  with a 
period of 384 days superimposed on a long-term trend. The 384-day period   was attributed either to a low-mass inner companion or to stellar pulsation. The differences in the estimates of the different groups  prompted us to re-investigate   this object based on a high-resolution spectrum.

\paragraph*{}

\textbf{HE~2144$-$1832}\\
This  object was  studied by \citet{Stephenson1989} and found to be an R-type  star. \citet{hansen2016abundances} 
reported  estimates of stellar atmospheric parameters and abundance estimates for four elements for this object: C, N, Ba, and Sr. This study was based on spectra obtained using X-shooter spectrograph \citep{Vernet_2011} covering wavelength regions 3000 - 5000 \AA, 5500 -10000 \AA, and 10000 - 25000 \AA\ at  spectral resolutions of 4350, 7450, and 5300 respectively. We  conducted a detailed chemical composition study for this object using a higher resolution (R ${\sim}$ 60,000) spectrum  with high S/N.  New estimates for C, N, Ba, and Sr, and first-time abundance estimates for several other  elements  including neutron-capture elements  are presented in this work. 
We also estimated  $^{12}$C/$^{13}$C  ${\sim}$ 2.5 for this object,  which is not too different from the estimated value of 2.1  reported by  \citep{goswami2005ch} based on a low-resolution spectroscopic study.

\paragraph*{}
\textbf{HE~2339$-$0837} \\
\citet{kennedy2011} reported the atmospheric parameters and the abundances of C and O for this object based on medium-resolution spectra. Detailed chemical abundance studies have not been reported in the literature for this object. We present a first-time detailed abundance analysis for this object based on a high-resolution spectrum.

\vskip 0.4cm
Regarding the binary nature of the programme stars,  HD~145777, HE~2144$-$1832, and HE~0017$+$0055 are established as radial velocity variables based on long-term radial velocity monitoring programmes, and hence \citet{jorissen2016rv} suggested these objects to be in long-period  binaries. Information on binarity is not available in the literature  for HE~2339$-$0837 and CD$-$27~14351.

\section{Observations and data reduction }
\label{sec:data_reduction}
\begin{figure}
        \centering
        \includegraphics[height=9cm,width=9cm]{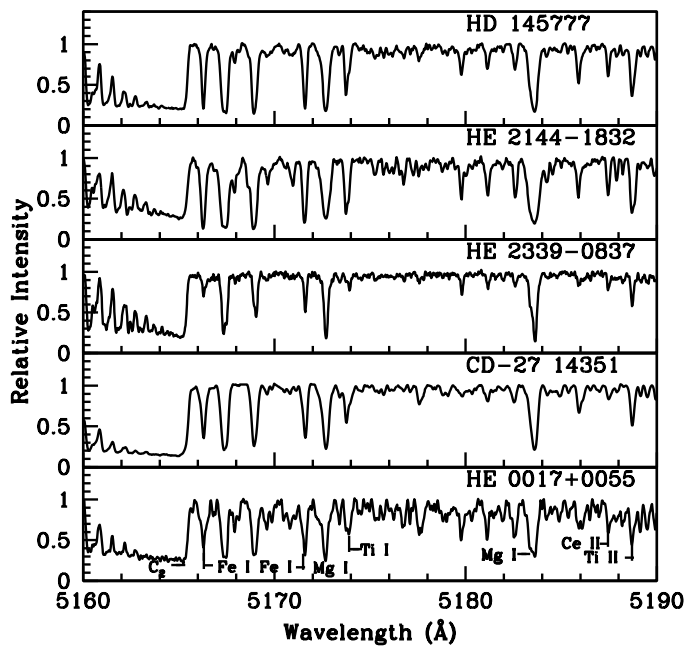}
        \caption{ Sample spectra of the programme stars in the  wavelength 
                region 5160 to 5190 {\bf  {\rm \AA}}.}
\label{fig:sample_spec}
\end{figure}

High-quality high-resolution spectra of HD~145777, HE~0017$+$0055, and HE~2144$-$1832 were obtained using  the Hanle Echelle SPectrograph (HESP) attached to the 2m  Himalayan Chandra Telescope (HCT) at the  Indian Astronomical Observatory (IAO), Hanle. The detector is a 4K x 4K CCD  with a pixel size of 15 $\mu$. The wavelength  coverage spans    3,500$-$10,000 {\rm \AA} at a spectral resolution  ({$\lambda$/$\delta\lambda$}) of 60,000.   Data is reduced following a standard procedure  using  Image Reduction and Analysis Facility (IRAF) software packages. Spectroscopic reduction procedures such as trimming, bias subtraction, flat normalisation, and extraction are applied to the raw data. Wavelength calibration is done using a high-resolution Th-Ar arc spectrum. A high-resolution spectrum ($R \sim$ 48,000)  from the Fiber-fed Extended Range Optical Spectrograph (FEROS), attached to the  1.52m  ESO telescope at La Silla, Chile,  is used for 
CD$-$27~14351. The spectrum covers  3520 $-$ 9200 {\rm \AA} in   the wavelength region. For HE~2339$-$0837, a high-resolution spectrum ($R \sim$ 50,000)   is taken from the SUBARU archive (\url{http://jvo.nao.ac.jp/portal}) acquired with the High Dispersion Spectrograph (HDS) of the 8.2m Subaru Telescope. The wavelength coverage of the observed spectra spans from about 4020 {\rm \AA} to 6775 {\rm \AA}, with a gap of about 100 {\rm \AA} (from 5340 {\rm \AA} to 5440 {\rm \AA}) due to the physical spacing of the CCD detectors.
The spectra are continuum fitted using the task continuum in  IRAF and dispersion corrected. A few  sample spectra are shown in Figure~\ref{fig:sample_spec}. Table~\ref{tab:basicdata} gives the basic data for the programme stars.

{\footnotesize
\begin{table*}
\centering
\caption{ \bf{Basic data for the programme stars}}
\label{tab:basicdata}
\scalebox{0.92}{
\begin{tabular}{lccccccccccc}
\hline
Star Name         & RA$(2000)$   & Dec.$(2000)$   & B      & V      & J       & H      & K     & Exposure  & Date of Obs.  & Source      \\
                &              &                &        &        &         &        &       & (seconds) &               & of spectrum \\
\hline
HD~145777       & 16 13 13.87  &$-$15 12 01.25  & 11.55  & 10.31  & 7.73    & 7.07   & 6.84  &   2400    & 01-06-2017    & HESP        \\
CD$-$27~14351   & 19 53 08.00  &$-$27 28 14.97  & 11.82  & 9.70   &  7.02   & 6.30   & 6.14  &   1200    & 14-07-2000    & FEROS       \\
HE~0017$+$0055  & 00 20 21.60  &$+$01 12 06.81  & 12.99  & 11.66  & 9.31    & 8.70   & 8.50  &   2700    & 21-09-2017    & HESP        \\
HE~2144$-$1832  & 21 46 54.66  &$-$18 18 15.59  & 12.65  & 10.97  & 8.77    & 8.18   & 7.96  &   2700    & 08-11-2017    & HESP        \\
HE~2339$-$0837  & 23 41 59.93  &$-$08 21 18.61  & 15.32  & 14.00  & 12.63   & 12.11  & 12.03 &    900    & 27-06-2004    & SUBARU/HDS         \\

\hline

\end{tabular}}
\end{table*}
}

\section{Photometric temperatures}
\label{sec:photometric_temp}
The photometric temperatures of the programme stars were determined using 
broad-band colours, optical and IR, with colour--temperature calibrations 
available for main-sequence stars \citep{Alonso1996} and giants 
\citep{alonso1999effective}, and are based on the infrared flux method (IRFM). The  procedure followed is as described in 
\citet{goswami2006, goswami2016}. As \citet{Alonso1996, alonso1999effective} reported,  the uncertainty on temperature calculations using the IRFM method is about
 ${\sim}$ 90 K.  Precise photometric data, reliable reddening estimates, and metallicity  information  are required  when using this method.
 We   estimated the photometric temperatures  of the stars at  several assumed metallicity values. The estimated temperatures,
along with the adopted metallicities, are listed in 
Table~\ref{tab:photometric_temp}.
Temperature estimates obtained using calibration relations involving (J-K), (J-H), and (V-K) colours give values that  differ  by  about  ${\pm}$ 200 K. We do not consider the empirical T$_{eff}$ 
scale for the B-V colour indices as   this calibration relation may not
give reliable estimates, due to the effect of CH molecular absorption in the B band. The severe blending of the spectra by molecular lines affects the photometric results to a significant extent \citep{yoon2020}. The  photometric temperature estimates obtained using the J-K calibration relation 
are used as an initial guess for  selecting model atmospheres to estimate spectroscopic temperature of the objects in an   iterative process as this empirical calibration is independent of metallicity \citep{Alonso1996, alonso1999effective}.

{\footnotesize
\begin{table*}
\centering
\caption{\bf{Temperatures from  photometry }}
        \label{tab:photometric_temp}
\begin{tabular}{llllllllllll}
\hline
Star Name         & $T_{eff}$& $T_{eff}$& $T_{eff}$& $T_{eff}$& $T_{eff}$& $T_{eff}$& T$_{eff}$ & Spectroscopic \\
                  &          & $(-1.5)$ & $(-2.0)$ &  $(-2.5)$& $(-1.5)$ & $(-2.0)$ & $(-2.5)$  & estimates   \\
                  &  (J-K)   &   (J-H)  &   (J-H)  &  (J-H)   &   (V-K)  &  (V-K)   &   (V-K)  & \\
\hline
HD~145777         & 4072     & 4295     & 4273     & 4234     & -        &    -     &    -    &  4160  \\
CD$-$27~14351     & 4097     & 4119     & 4099     & -        & -        &    -     &    -    & 4320\\
HE~0017$+$0055    & 4261     & 4478     & 4456     & 4414     & 4084     & 4080     &    -    & 4370\\
HE~2144$-$1832    & 4264     & 4558     & 4536     & 4493     & 4171     & 4168     &    -    & 4190 \\
HE~2339$-$0837    & 4812     & 4797     & 4773     & 4727     & 5086     & 5093     &  5107   &  4940 \\
\hline
\end{tabular}

The numbers in  parentheses below $T_{eff}$ indicate the metallicity values at which the temperatures are calculated. \\
Temperatures are given in Kelvin.\\
\end{table*}
}

\section{Radial velocities and stellar atmospheric parameters}
\label{sec:radial_velocity_stellar_atm_parameters}
The radial velocities of the programme stars are determined by measuring the
shift in the wavelengths with respect to the laboratory wavelengths, for a large number of unblended and clean lines in their spectra. For the rest frame laboratory wavelength we   use the Arcturus spectrum \citep{hinkle_et_al_2000} as a template. The object Arcturus was chosen so as to have homogeneity in the analysis as it belongs to the giant class and has a comparable temperature as the objects under study. Except for HD 145777, the rest are found to be 
high-velocity objects. Estimated mean radial velocities along with the standard deviation from the mean values, after correcting for heliocentric motion
are presented in  Table~\ref{tab:radial_velocity_table}. The literature values are also presented for  comparison.
We   also  use the FXCOR package in IRAF  over the whole spectrum to cross-check these calculations and we find  them to be consistent with those 
obtained from  line-to-line measurement of clean lines of different elements. 

The stellar atmospheric parameters, the effective temperature T$_{eff}$, the surface gravity log g, the micro-turbulent velocity ${\xi}$, and the metallicity
[Fe/H], are determined using a set of clean unblended Fe I and Fe II lines with excitation potential in the range 0.0 $-$ 6.0 eV and equivalent width 20 $-$
180 m{\rm \AA}.  Due to the presence of molecular lines and bands of carbon all over the spectra, the lines of Fe and other elements are severely blended. After strong filtration for clean, unblended, and symmetrical lines in the spectra of the objects, the number of (Fe~I, Fe~II) lines used for  determination of the atmospheric parameters are (33, 4), (33, 3), (21, 3), (27, 2), and (19, 4)  for HD~145777, CD$-$27~14351, HE~0017+0055, HE~2144$-$1832, and HE~2339$-$0837, respectively. The lines used are presented in  Table~\ref{tab:Fe_linelist} along with the measured equivalent widths and atomic line information. References to the log gf values are given in the last column. A few  Fe lines that are not severely blended are also included in the list for which we have used the method of de-blending (with the SPLOT task in IRAF) for measuring equivalent widths. 

\begin{figure}
        \centering
        \includegraphics[height=7cm,width=9cm]{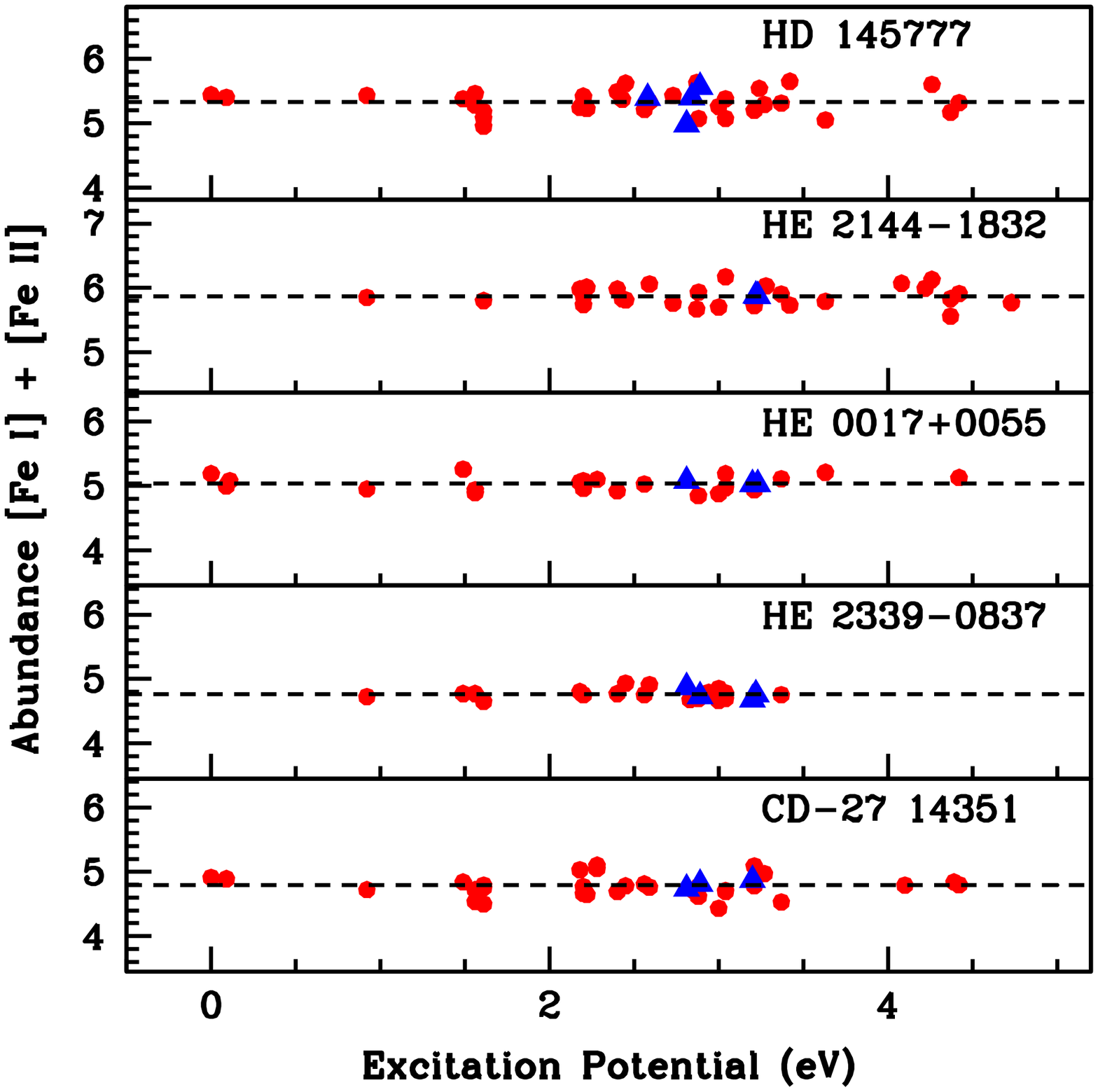}
                \includegraphics[height=7cm,width=9cm]{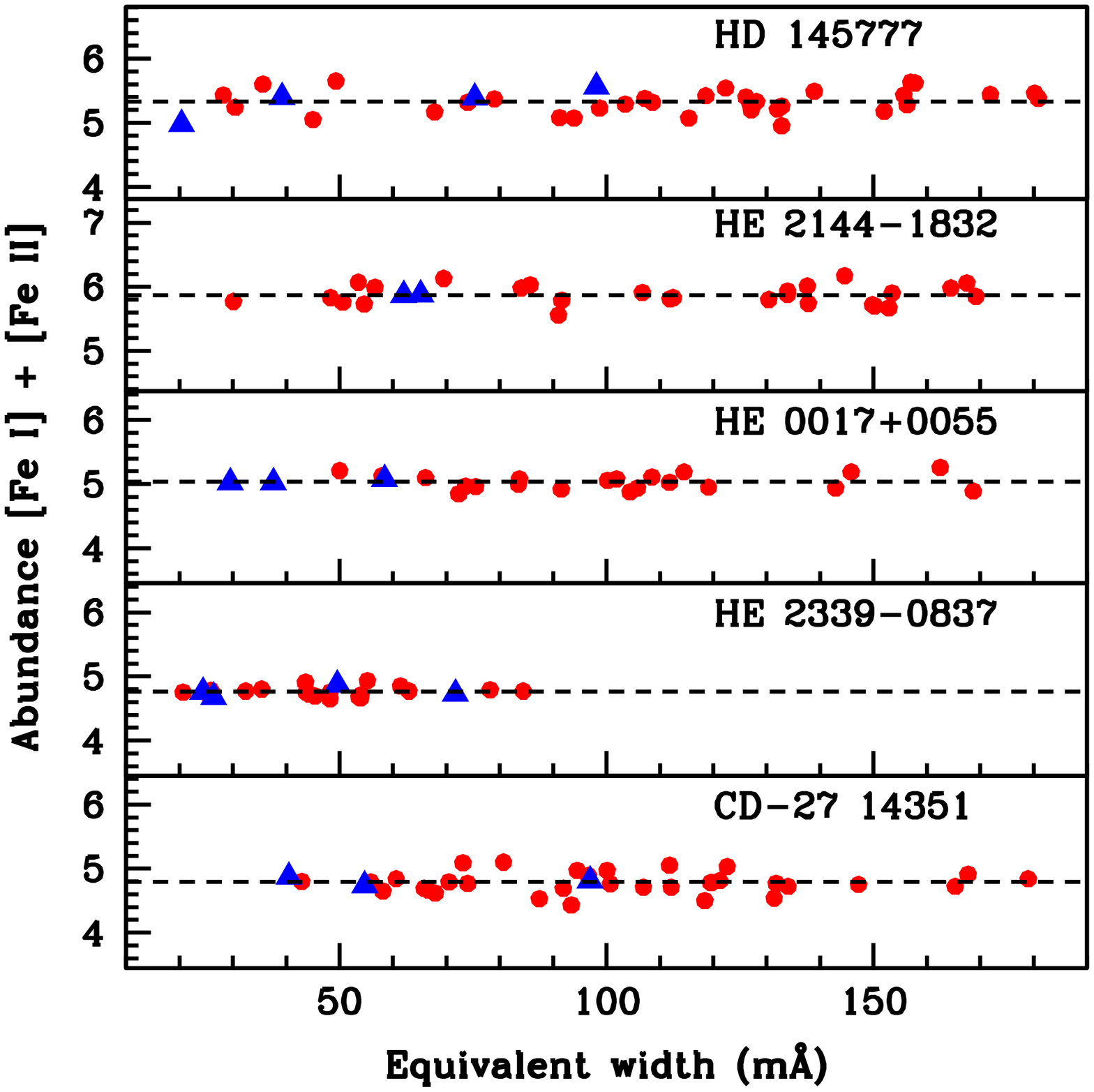}
        \caption{Iron abundances of stars  for individual Fe I and Fe II lines as a function of excitation potential (top panel) and equivalent width (bottom panel). The solid red circles indicate Fe I lines and solid blue triangles indicate Fe II lines.}
\label{fig:balance}
\end{figure}

An initial model atmosphere is selected from the Kurucz grid of model atmospheres with no convective overshooting (\url{http://kurucz.hardvard.edu/grids.html}) 
corresponding to  the photometric temperature estimate and the initial guess of log g value for giants and/or dwarfs. 
The effective temperature is determined by forcing the slope of Fe abundances versus the excitation potential of the measured
Fe I lines to zero (Figure~\ref{fig:balance}, top panel). At this particular temperature,  the micro-turbulent velocity  is fixed to be that value for which there is no dependence of the abundances derived from
the Fe  lines on the reduced equivalent width (Figure~\ref{fig:balance}, bottom panel). At these fixed values of temperature and micro-turbulent velocity, the  surface gravity is obtained by demanding the abundances derived from both Fe I and Fe II lines to be nearly the same.
The abundances obtained from the Fe I and Fe II lines give the metallicity. Thus, starting with the initially selected model atmosphere, the final model atmosphere is obtained following  the  iterative method, which is then  adopted to carry out  further abundance analysis.
  The  analysis is facilitated by using an updated version of 
  MOOG software \citep{Sneden_1973_MOOG}  in its updated 2013 version, which 
  assumes local thermodynamic equilibrium (LTE) conditions. Solar abundances are adopted from \citet{asplund2009}. The absorption lines due to Fe I are affected by 3D non-LTE
(NLTE) effects \citep{Amarsi_et_al_2016}. The NLTE effect is
negligible in the case of Fe II lines for [Fe/H] $\geq$ $-$2.50, and
increases with decreasing metallicity \citep{Amarsi_et_al_2016}.
However, we do not consider  the NLTE corrections in our analysis.
\citet{Ezzeddine_et_al_2017} showed that the departure of
[Fe/H]$_{NLTE}$ from [Fe/H]$_{LTE}$ anti-correlate with [Fe/H]$_{LTE}$
growing from $\sim$ 0.0 dex at [Fe/H] $\sim$ $-$1.0 to $\sim$ 1.0 dex at [Fe/H] $\sim$ $-$8.0. From  
Figure 2 and equation (1) of \citet{Ezzeddine_et_al_2017}, we  
see that the NLTE corrections on the metallicity of our programme
stars range from 0.08 to 0.23, which are well within the uncertainty 
limit of  our [Fe/H] estimates. The estimated  stellar  parameters are listed in Table~\ref{tab:atm_paracomp}. 
 {\footnotesize
\begin{table}
\centering
\caption{\bf{ Radial velocities of the programme stars. }}
\label{tab:radial_velocity_table} 
\begin{tabular}{ccc}
\hline
Star Name         & V$_{r}$            & V$_{r}$           \\
                & ($km s^{-1}$)      & ($km s^{-1}$)       \\
                & (our estimates)    &  \textit{a}  \\
\hline
HD~145777       & 16.23$\pm 0.40$    & 20.14 $\pm 0.47$    \\
CD$-$27~14351   & 61.1$\pm 0.50$ $^{\textit{b}}$    & 59.70$\pm 0.90$ \\
HE~0017$+$0055  & $-$80.56$\pm 0.52$ & $-$80.73$\pm 0.23$  \\
HE~2144$-$1832  & 137.19$\pm 0.68$   & 141.24 $\pm 0.64$   \\
HE~2339$-$0837  & 169.83$\pm 0.68$   &  -                  \\
\hline
\end{tabular}

\textit{a} $-$ \citet{gaia2018}; \textit{b} $-$ \citet{Drisya2017}

\end{table}
}

{\footnotesize

        \begin{table*}
\centering
\caption{\bf{Equivalent widths (in m\r{A}) of Fe lines used to derive atmospheric parameters.}}
\label{tab:Fe_linelist} 
\scalebox{0.87}{
\begin{tabular}{cccccccccc}
\hline
Wavelength   &Element    &E$_{low}$ &   log gf  &  HD~145777     & CD$-$27~14351 &HE~0017$+$0055 & HE~2144$-$1832 & HE~2339$-$0837   & References  \\
(\r{A})      &           & (eV)     &           &                &               &               &               &                &        \\
\hline 
4466.573     &  Fe I     & 0.11     & $-$4.464  &      -         &       -       &  101.9 (5.08) &       -       &        -        &     1    \\
4476.019     &           & 2.85     & $-$0.570  &      -         &  112.1 (4.71) &        -      &       -       &        -        &     2   \\
4871.318     &           & 2.87     & $-$0.410  &  157.0 (5.63)  &       -       &        -      &  152.9 (5.67) &        -        &     1   \\
\hline
\end{tabular}}

The numbers in   parentheses in Cols. 5--9 give the derived abundances from the respective line.\\
References: 1. \citet{FMW1988}, 2. \citet{BK1974} \\
\textbf{Note:} This table is available in its entirety  online.
A portion is shown here for guidance regarding its form and content.\\

\end{table*}}

{\footnotesize
\begin{table*}
\centering
\caption{\bf{Derived atmospheric parameters of our programme stars and literature values. }}
\label{tab:atm_paracomp} 
\begin{tabular}{lccccccccc}
\hline     
Star Name       & T$_{eff}$  &log g  & $\zeta    $   & [Fe I/H]          &  [Fe II/H]        & [Fe/H] & Ref \\
                &    (K)     & (cgs) & (km s$^{-1}$) &                   &                   &        &     \\
\hline
HD~145777       & 4160      & 0.90 & 2.02            & $-$2.17$\pm 0.18$ & $-$2.17$\pm 0.25$ & $-$2.17  & 1   \\
                & 4245      & -    &  -            &  -                  &   -               &  -       & 2   \\
                & 4216      & -    &  -            &  -                  &  -                &  -       & 7   \\
\hline
CD$-$271~4351   & 4320      & 0.50 & 2.58          & $-$2.72$\pm 0.16$   &$-$2.69$\pm 0.07$  & $-$2.71  & 1   \\
                & 4335      & 0.50 & 2.42          &-                    & -                 & $-$2.62  & 5   \\
                & 4223      & -    &  -            &  -                  &  -                &  -       & 7   \\
\hline
HE~0017$+$0055  & 4370      & 0.80 & 1.94         &$-$2.47$\pm 0.12$     &$-$2.45$\pm 0.03$  &$-$2.46   & 1   \\
                & 4250      & 1.00 & 2.00          & -                   & -                 & $-$2.40  & 4   \\
                & 4185      & 0.18 & 2.00          & -                   & -                 & $-$2.72  & 6   \\
\hline
HE~2144$-$1832  & 4190      & 0.60 & 1.87        &$-$1.63$\pm 0.15$     &$-$1.63$\pm 0.00$  &  $-$1.63  & 1   \\
                & 4200      & 0.60 & 2.20        &  -                   & -                   &$-$1.70  & 3   \\
\hline
HE~2339$-$0837  & 4940      & 1.40 & 1.55        &$-$2.74$\pm 0.08$     &$-$2.74$\pm 0.08$   &$-$2.74   & 1   \\
                & 4939      & 1.60 & 2.00        & -                    & -                  &$-$2.71   & 6   \\
\hline
\end{tabular}

References: 1. Our work, 2. \citet{bergeat2001effective}, 3. \citet{hansen2016abundances}, 4. \citet{jorissen2016HE0017}, 5. \citet{Drisya2017}, 6. \citet{kennedy2011}, 7. \citet{McDonald2012}.  \\          
\end{table*}
}

{\footnotesize
\begin{table*}
\centering
\caption{\bf{Equivalent widths (in m\r{A}) of lines used to calculate elemental abundances.}}
\label{tab:Elem_linelist1} 
\scalebox{0.87}{
\begin{tabular}{cccccccccc}
\hline
Wavelength   &Element    &E$_{low}$ &   log gf  &  HD~145777     & CD$-$27~14351 &HE~0017$+$0055 &HE~2144$-$1832 & HE~2339$-$0837 & References  \\
(\r{A})      &           & (eV)     &           &                &               &               &               &                &             \\
\hline 
5682.633     &  Na I     & 2.10     & $-$0.700  &  40.2 (4.61)   &  52.9 (5.05)  &      -        &  69.8 (5.05)  &  36.3 (5.06)   &     1     \\
5688.205     &           & 2.10     & $-$0.450  &  45.1 (4.44)   &  51.0 (4.77)  &      -        &  93.7 (5.16)  &  47.2 (5.02)   &     1     \\
6160.747     &           & 2.10     & $-$1.260  &     -          &       -       &      -        &  45.1 (5.20)  &       -        &     1     \\
4571.096     &  Mg I     & 0.00     & $-$5.691  &     -          &       -       &      -        &      -        &       -        &     2     \\
5528.405     &           & 4.35     & $-$0.620  & 149.7 (6.42)   & 170.9 (6.70)  &      -        & 176.5 (6.71)  & 142.2 (7.03)   &     3     \\
\hline
\end{tabular}}

The numbers in  parentheses in Cols. 5--9 give the derived abundances from the respective line.\\ 
References: 1. \citet{KP1975}, 2. \citet{LV1974}, 3. \citet{LZ1971}\\
\textbf{Note:} This table is available in its entirety  online.
A portion is shown here for guidance regarding its form and content.\\

\end{table*}}

\section{Results}
\label{sec:abundance_analysis}
\subsection{ Abundance analysis }
The abundances of various elements are determined from the measured equivalent widths of absorption lines due to neutral and ionised elements
using the local thermodynamic equilibrium (LTE) analysis. 
Only the symmetric and clean lines are used  for our analysis.
As equivalent width measurement results depend on personal bias, to avoid it we  also used the  \texttt{Tool for Automatic Measurement of Equivalent width (TAME)} \citep{Tame_kang_2012} to verify our measurements. \texttt{TAME} measures equivalent widths, determining the local continuum and using de-blending wherever necessary. A master line list (Table~\ref{tab:Elem_linelist1}) 
is generated including all the elements, taking the excitation potential and log gf values from the Kurucz database of atomic line list and the measured
equivalent widths.
 For our analysis we made use of the LTE line analysis and spectrum synthesis 
code MOOG 2013\footnote[2]{\url{http://www.as.utexas.edu/˜chris/moog.html}} 
 \citep{Sneden_1973_MOOG}. The adopted 
model atmospheres are selected from the Kurucz grid of model atmospheres with no convective overshooting (\url{http://kurucz.harvard.edu/grids.html}).
Elemental abundances of C, N, $\alpha$-elements, iron-peak elements, and several neutron-capture elements  are estimated.
We  used spectrum synthesis calculations for the elements  showing hyperfine splitting (e.g. Sc, V, Mn, Ba, La, and Eu). Several studies \citep{ Andrievsky_et_al.2009,
Andrievsky_et_al.2011, Hansen_et_al.2013, Tremblay_et_al.2013,
Gallagher_et_al.2016, Sitnova_et_al.2016, Nordlander_et_al.2017} in the past 
have focused  on the 3D or NLTE or 3D-NLTE effects  on the
abundances of both heavy and light elements. However, we  did not apply any NLTE corrections to our LTE estimates as the NLTE correction factors are negligible. The abundance results are presented in Tables~\ref{tab:abundances}~and~\ref{tab:abundances2}. A comparison of our estimated abundance ratios with the literature values is presented in Table~\ref{tab:abuncomp}.
 We   also estimated the carbon isotopic ratio $^{12}$C/$^{13}$C and calculated [ls/Fe], [hs/Fe], and [hs/ls] for the stars, where ls  represents Sr, Y, and Zr and hs   represents Ba, La, Ce, and Nd. These results are presented in Table~\ref{tab:abundanceratios}.

\subsubsection{Carbon, nitrogen, oxygen}
\label{sec:cno}
\begin{figure}
        \centering
        \includegraphics[height=7.5cm,width=9cm]{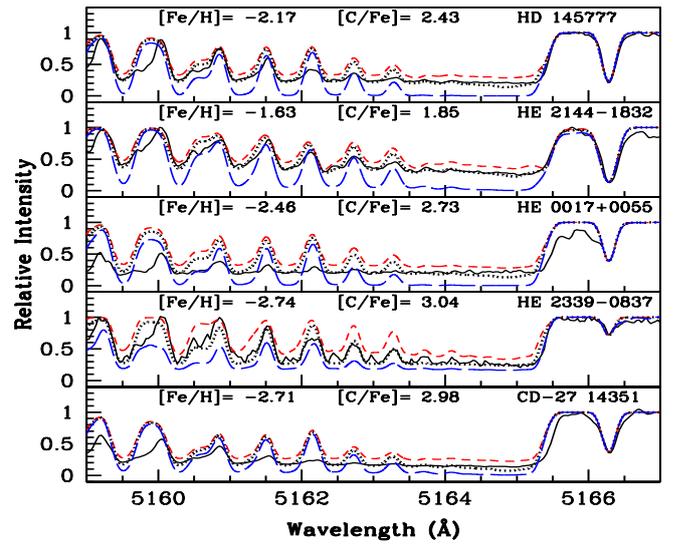}
        \caption{Spectral synthesis plot of C$_{2}$ band around 5165 {\rm \AA}. The dotted lines indicate the synthesised spectra and the solid lines indicate the observed spectra. Two alternative synthetic spectra are shown corresponding to $\Delta$[C/Fe] = +0.01 (long-dashed line) and $\Delta$[C/Fe] = $-$0.01 (short-dashed line) in panel 1 (for HD~145777); $\Delta$[C/Fe] = +0.05 (long-dashed line) and $\Delta$[C/Fe] = $-$0.05 (short-dashed line) in panel 2 (for HE~2144$-$1832); $\Delta$[C/Fe] = +0.10 (long-dashed line) and $\Delta$[C/Fe] = $-$0.10 (short-dashed line) in  panel 3 (for HE~0017+0055); $\Delta$[C/Fe] = +0.3 (long-dashed line) and $\Delta$[C/Fe] = $-$0.3 (short-dashed line) in panel 4 (for HE~2339$-$0837);  and $\Delta$[C/Fe] = +0.01 (long-dashed line) and $\Delta$[C/Fe] = $-$0.01 (short-dashed line) in panel 5 (for CD$-$27~14351).}
\label{fig:C2}
 \end{figure}

\begin{figure}
        \centering
        \includegraphics[height=7.5cm,width=9cm]{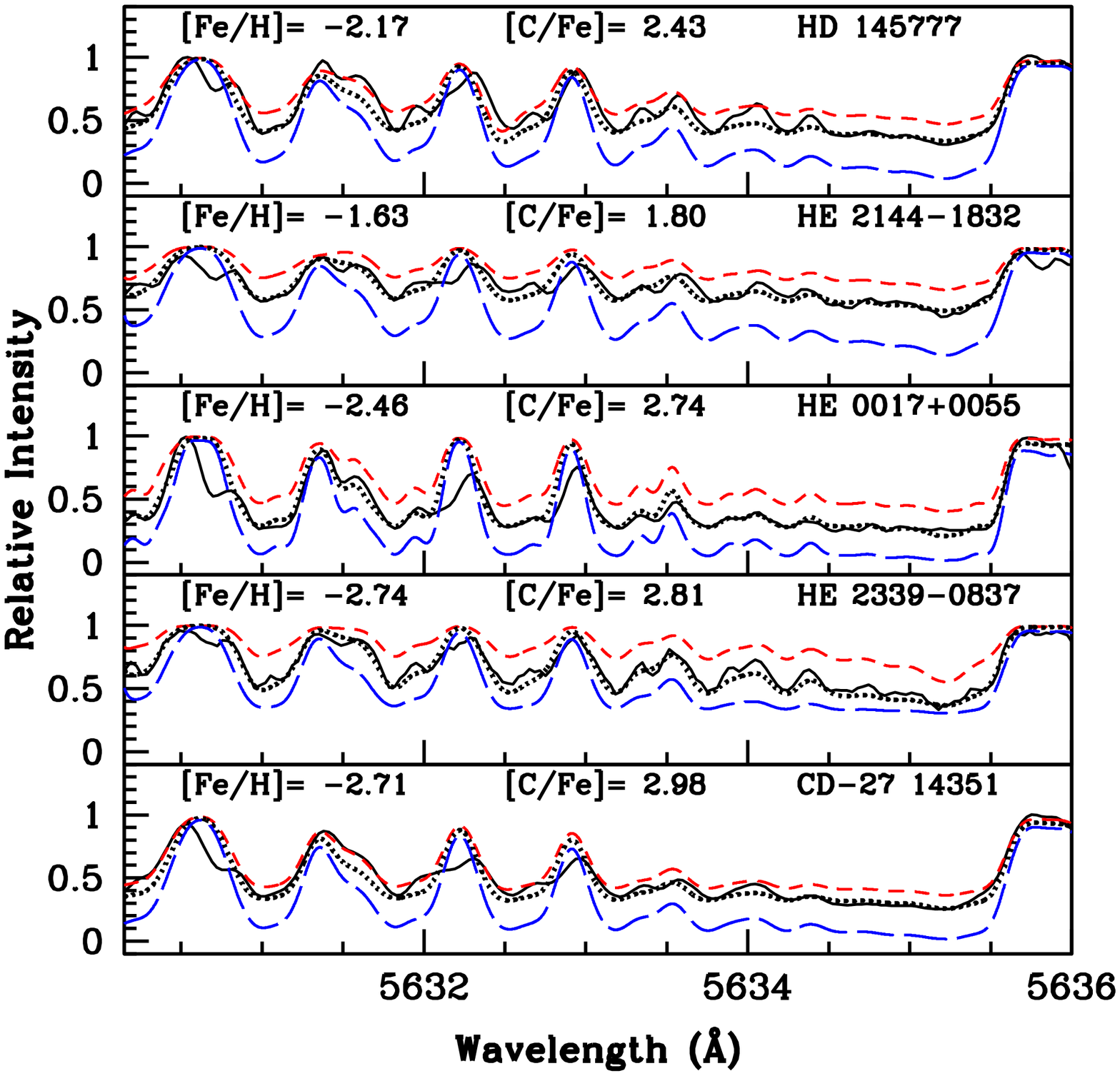}
        \caption{Spectral synthesis plot of C$_{2}$ band around 5635 {\rm \AA}. The dotted lines indicate the synthesised spectra and the solid lines indicate the observed spectra. Two alternative synthetic spectra are shown corresponding to $\Delta$[C/Fe] = +0.02 (long-dashed line) and $\Delta$[C/Fe] = $-$0.02 (short-dashed line) in panel 1 (for HD~145777); $\Delta$[C/Fe] = +0.15 (long-dashed line) and $\Delta$[C/Fe] = $-$0.15 (short-dashed line) in panel 2 (for HE~2144$-$1832); $\Delta$[C/Fe] = +0.05 (long-dashed line) and $\Delta$[C/Fe] = $-$0.05 (short-dashed line) in  panel 3 (for HE~0017+0055); $\Delta$[C/Fe] = +0.3 (long-dashed line) and $\Delta$[C/Fe] = $-$0.3 (short-dashed line) in panel 4 (for HE~2339$-$0837);  and $\Delta$[C/Fe] = +0.03 (long-dashed line) and $\Delta$[C/Fe] = $-$0.03 (short-dashed line) in panel 5 (for CD$-$27~14351).}
\label{fig:C2_5635}
 \end{figure}

\begin{figure}
        \centering
        \includegraphics[height=7.5cm,width=9cm]{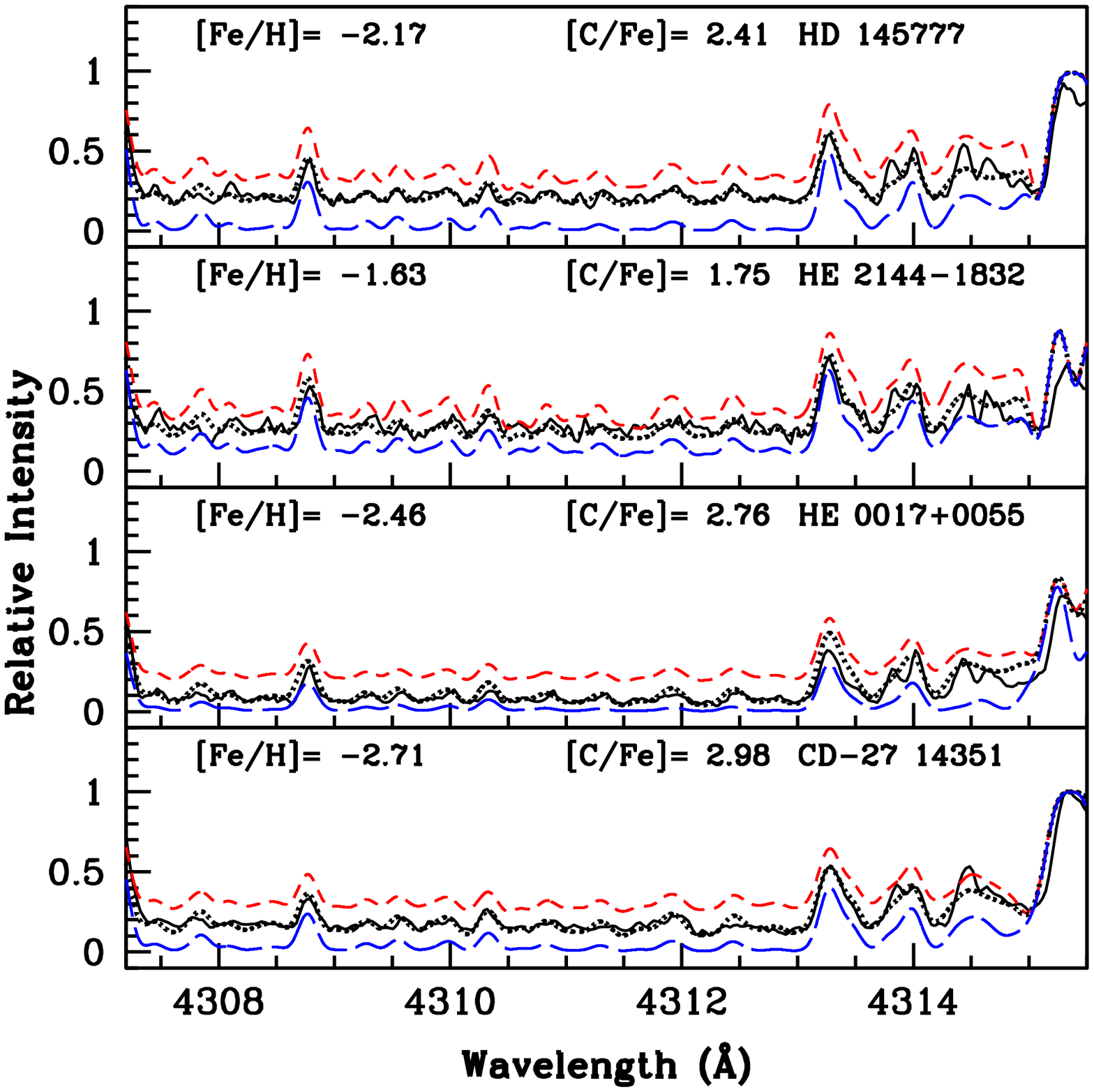}
        \caption{Spectral synthesis plot of CH band around 4310 {\rm \AA}. The dotted lines indicate the synthesised spectra and the solid lines indicate the observed spectra. Two alternative synthetic spectra are shown corresponding to $\Delta$[C/Fe] = +0.05 (long-dashed line) and $\Delta$[C/Fe] = $-$0.05 (short-dashed line) in panel 1 (for HD~145777); $\Delta$[C/Fe] = +0.15 (long-dashed line) and $\Delta$[C/Fe] = $-$0.15 (short-dashed line) in panel 2 (for HE~2144$-$1832); $\Delta$[C/Fe] = +0.30 (long-dashed line) and $\Delta$[C/Fe] = $-$0.30 (short-dashed line) in  panel 3 (for HE~0017+0055); and $\Delta$[C/Fe] = +0.10 (long-dashed line) and $\Delta$[C/Fe] = $-$0.10 (short-dashed line) in panel 4 (for CD$-$27~14351).}
\label{fig:CH}
 \end{figure}

The abundance of oxygen could not be estimated as the oxygen lines 
are found to be blended and not usable for abundance estimates.
The abundance of carbon is estimated  using the spectrum synthesis 
calculation of the  C$_{2}$ molecular bands near 5165 {\rm \AA} and 
5635 {\rm \AA}, and the CH molecular band near 4310 {\rm \AA}. All  three 
bands  yield almost the same abundance of  carbon for  the respective stars 
(Tables~\ref{tab:abundances}~and~\ref{tab:abundances2}). However, 
the CH molecular band in the spectrum of HE~2339$-$0837 could not be used 
as the band is 
saturated. Carbon is found to be enhanced ([C/Fe] $>$ 1) in all the stars.
Our estimates of carbon abundance are higher than the carbon 
abundance reported for HE~2144$-$1832 in \citet{hansen2016abundances}. This discrepancy may  be attributed to the lower resolution (R $\sim$ 7450) of the  spectra used in  \citet{hansen2016abundances}, whereas the  resolution of our spectrum is R $\sim$ 60,000. One more reason for the  discrepancy could be a difference in the adopted oxygen abundance. Estimates of carbon abundance depend on the adopted initial value of oxygen for the spectrum synthesis calculations. As the abundance of oxygen could not be estimated on our spectrum, we  consider a solar value for oxygen (log$\epsilon$ = 8.69). The spectrum 
synthesis fits of C$_{2}$ and CH molecular bands for the stars are shown in 
Figures~\ref{fig:C2},~\ref{fig:C2_5635},~and~\ref{fig:CH}.

\begin{figure}
        \centering
        \includegraphics[height=7.5cm,width=9cm]{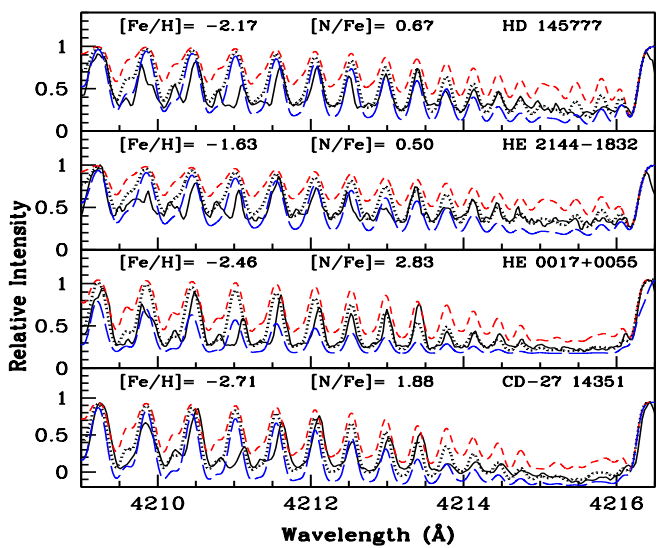}
        \caption{Synthesis of CN band around 4215 {\rm \AA}. The dotted line represents synthesised spectra and the solid line indicates the observed spectra. Two alternative fits, the short-dashed line corresponding to $\Delta$[N/Fe] = $-$0.3 and the long-dashed line  corresponding to $\Delta$[N/Fe] = $+$0.3 are shown to illustrate the sensitivity of the line strengths to N abundance.} 
 \label{fig:CN}
 \end{figure}

The abundance of nitrogen is derived using the spectrum synthesis 
calculations of the only useful CN band at 4215 {\rm \AA} (Figure~\ref{fig:CN})
using the estimated carbon abundance. Nitrogen is found to be enhanced 
in HE~0017$+$0055 and CD$-$27~14351 with [N/Fe] = 2.83 and 1.88, respectively. HD~145777 and HE~2144$-$1832 show moderate enhancement with [N/Fe] = 0.67 and 0.50, respectively. The CN band at 4215 {\rm \AA} is saturated in the spectrum of  HE~2339$-$0837, and hence the abundance of N could not be estimated for this object. 

\begin{figure}
        \centering
        \includegraphics[height=7.5cm,width=9cm]{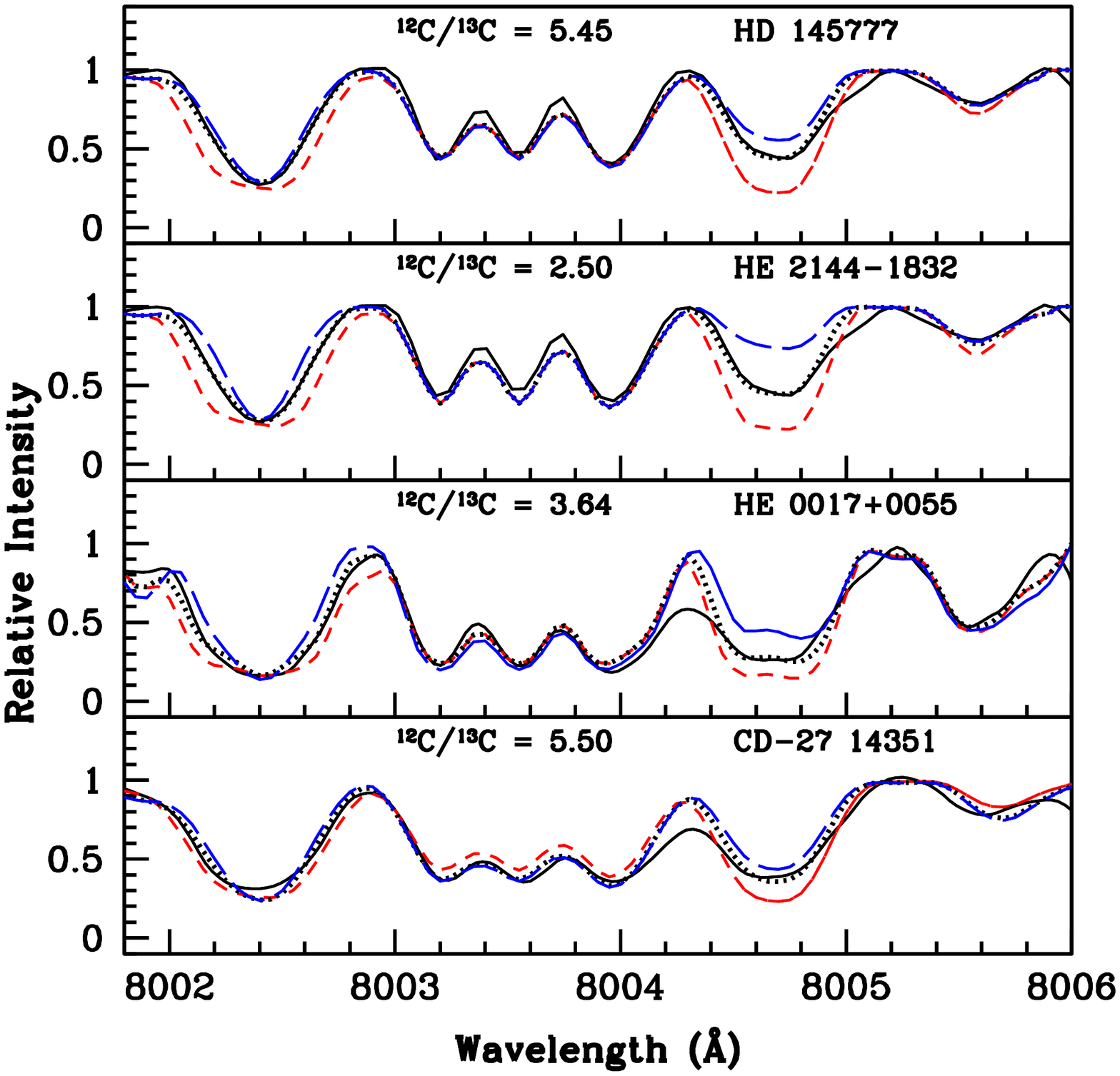}
        \caption{Spectral synthesis fits (dotted curves) of the CN features around 8005 {\rm \AA} obtained with the adopted C and N abundances and $^{12}$C/$^{13}$C values (dotted curve). The observed spectrum is shown by a solid curve. Two alternative fits with $^{12}$C/$^{13}$C $\sim$ 1 (short-dashed line) and 12 (long-dashed line) are shown to illustrate the sensitivity of the line strengths to the isotopic carbon abundance ratios.}
\label{fig:C12C13}
\end{figure}

\begin{figure}
        \centering
        \includegraphics[height=7.5cm,width=9cm]{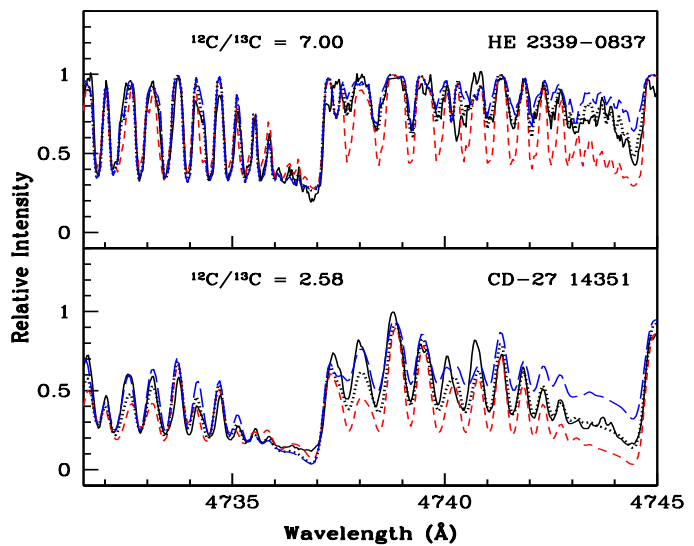}
        \caption{Spectral synthesis fits (dotted curves) of the C$_{2}$ features around 4740 {\rm \AA} obtained with the adopted C abundances and $^{12}$C/$^{13}$C values (dotted curve). The observed spectrum is shown by a solid curve. Two alternative fits with $^{12}$C/$^{13}$C $\sim$ 1 (short-dashed line) and 15 (long-dashed line) are shown to illustrate the sensitivity of the line strengths to the isotopic carbon abundance ratios.}
\label{fig:C12C13_4740}
\end{figure}

We   estimated the carbon isotopic ratio $^{12}$C/$^{13}$C using 
the spectrum synthesis calculation of the $^{12}$CN and $^{13}$CN features near 
8005 {\rm \AA} (Figure~\ref{fig:C12C13}) and the $^{12}$C$^{13}$C and 
$^{13}$C$^{13}$C features near 4740 {\rm \AA} (Figure~\ref{fig:C12C13_4740}). 
The values of $^{12}$C/$^{13}$C derived using these two  features are presented in Table~\ref{tab:abundanceratios}.
 We   used the wavelengths, lower excitation potentials, and log gf 
values of different molecular transitions for the C$_{2}$ band at 
5165 {\rm \AA} and CN bands from \citet{brooke2013}, \citet{ram2014}, 
and \citet{sneden2014}. The line lists of the C$_{2}$ bands at 5635 {\rm \AA} 
and 4740 {\rm \AA}, and the CH band at 4310 {\rm \AA} are taken from the {\textit{linemake}}\footnote[3]{{\textit{linemake}} contains laboratory atomic data (transition probabilities, hyperfine and isotopic substructures) published by the Wisconsin Atomic Physics and the Old Dominion Molecular Physics
groups. These lists and accompanying line list assembly software
have been developed by C. Sneden and are curated by V. Placco
at https://github.com/vmplacco/linemake.} atomic and molecular line database.

\subsubsection{Na, Mg, Ca, Sc, Ti, V}
While HD~145777 and HE~2144$-$1832 show moderate enhancement in Na with [Na/Fe] = 0.45 and 0.53, respectively, HE~2339$-$0837 and CD$-$27~14351 exhibit enhanced abundance of Na with [Na/Fe]~$>$~1. Na abundance could not be estimated for HE~0017$+$0055 as the Na lines are severely blended and not usable for abundance estimates. 

Magnesium and calcium are also found to be moderately enhanced in the stars HD~145777 and HE~2144$-$1832. While HE~2339$-$0837 and CD$-$27~14351 show enhanced abundance of Mg with [Mg/Fe]~$>$~1,  Ca is found to be moderately enhanced
with [Ca/Fe] in the range 0.4 to 0.9. We could not determine the abundance of Mg and Ca for HE~0017$+$0055 as no good lines were found.

We could estimate Sc abundance only for HD~145777 using spectrum synthesis calculations of the line Sc II 6245.637 {\rm \AA} considering the hyperfine splitting contributions from \citet{prochaska&mcwilliam2000}. This object shows mild enhancement of Sc with [Sc/Fe] = 0.77.

Titanium is estimated using lines from both neutral and ionised species for all the programme stars except HE~2339$-$0837, where Ti abundance is estimated only using Ti II lines. 
While Ti is found to be mildly enhanced in HD~145777, HE~2144$-$1832, HE~0017$+$0055, and HE~2339$-$0837, the  object CD$-$27~14351 shows an overabundance with [Ti/Fe] $\sim$ 0.96. Although it is seen that due to NLTE effects, the abundances derived from Ti I and Ti II lines may differ
\citep{johnson2002},  we have derived similar  abundances from  Ti I and Ti II lines for our programme  stars. The equal abundances found
from    Ti I and Ti II lines  ensure the log g values estimated for our programme stars.

The abundances  of V are found to be near solar with  [V/Fe] $\sim$ $-$0.11 and 0.06  for HD~145777 and  HE~2144$-$1832,  respectively. V is highly abundant in  CD$-$27~14351 with [V/Fe] = 1.11. For HE~0017+0055 and HE~2339$-$0837 abundance of V could not be estimated.

\subsubsection{Cr, Mn, Co, Ni, Zn}
The abundance of Cr is derived using seven Cr I lines (Table~\ref{tab:Elem_linelist1}).
Cr is underabundant with [Cr/Fe] = $-$0.21, $-$0.12, $-$0.50, and $-$0.34 for HD~145777, HE~2144$-$1832, HE~0017$+$0055, and HE~2339$-$0837, respectively. CD$-$27~14351 shows near solar abundance of Cr.

The abundance of Mn is estimated using  spectrum synthesis calculations of the Mn I lines at  4765.846 {\rm \AA} and 4766.418 {\rm \AA} for HE~2144$-$1832, and the   Mn II 5432.543 {\rm \AA} line is used for
HD~145777.  Mn is  found to be underabundant in   the stars HD~145777 and HE~2144$-$1832. 
The abundance of Mn could not be estimated for the other three objects as no suitable lines were detected for abundance determination.

The abundance of Co could be derived only for HD~145777 and HE~2144$-$1832,
for which we   obtained near solar values  with [Co/Fe] = 0.03 and 0.06, respectively. The abundance of Co could not be estimated for the other three objects as no good lines were found  for abundance estimation.

The abundance of Ni could be  estimated only for HE~2144$-$1832 using equivalent width measurement of five Ni I lines (Table~\ref{tab:Elem_linelist1}). Ni shows moderate enhancement in the star with [Ni/Fe] = 0.41. The abundance of Ni could not be estimated for the other objects as no good lines were found. 

The abundance of Zn determined using  Zn I 4810.528 {\rm \AA} line  gives [Zn/Fe] = 0.63, $-$0.08, and 0.52 for the stars HD~145777, HE~2144$-$1832, and HE~0017$+$0055, respectively. The abundance of Zn could not be estimated for the other two objects as no suitable lines were detected for abundance determination.
 
\subsubsection{ Sr, Y, Zr}

The abundance of Sr is estimated using spectrum synthesis calculations of the  
Sr I 4607.327 {\rm \AA} line for all the stars except HE~0017+0055 as 
this line is severely blended. While HD~145777 and HE~2144$-$1832 show 
moderate enhancement of Sr, it is found to be overabundant with 
[Sr/Fe] = 1.32 and  1.74 in  HE~2339$-$0837 and CD$-$27~14351, 
respectively. The estimated Sr abundance is found to be  $\sim$ 0.8 dex
lower than that obtained by  \citet{hansen2016abundances} for the star 
HE~2144$-$1832 (Table~\ref{tab:abuncomp}), which they had obtained using the 
Sr II 4077.709 {\rm \AA} line on a spectrum of spectral resolution 
R ${\sim}$ 7450, and S/N = 6 (at 4000 {\rm \AA}). This line is severely 
blended in our spectrum, which was obtained at a higher resolution 
(R ${\sim}$ 60,000), and it could not be used for abundance analysis. The 
abundance  estimated from the blended Sr II line in 
\citet{hansen2016abundances} may be the reason for the observed discrepancy.

We  estimated the abundance of Y using six lines (Table~\ref{tab:Elem_linelist1}). HE~0017$+$0055 and HE~2339$-$0837 show mild overabundance with [Y/Fe] ${\sim}$ 0.58 and  0.67, respectively. The other three objects show overabundance of Y with [Y/Fe]~$>$~1. 

Four Zr I lines and four Zr II lines (Table~\ref{tab:Elem_linelist1}) are used to derive Zr abundance in the objects. Based on the availability, Zr I lines are used for  HD~145777 and HE~2144$-$1832, and Zr II lines are used for  HE~0017+0055 and HE~2339$-$0837. For the object CD$-$27~14351, both Zr I and Zr II lines could be used.   Zr is found to be enhanced in all  five stars.

We could not detect lines due to niobium (Nb) and technetium (Tc)  in any of the spectra.

\subsubsection{ Ba, La, Ce,  Pr, Nd}
\begin{figure*}
     \begin{center}
\centering
  {%
            \label{fig:Ba_31}
            \includegraphics[height=7.0cm,width=8.5cm]{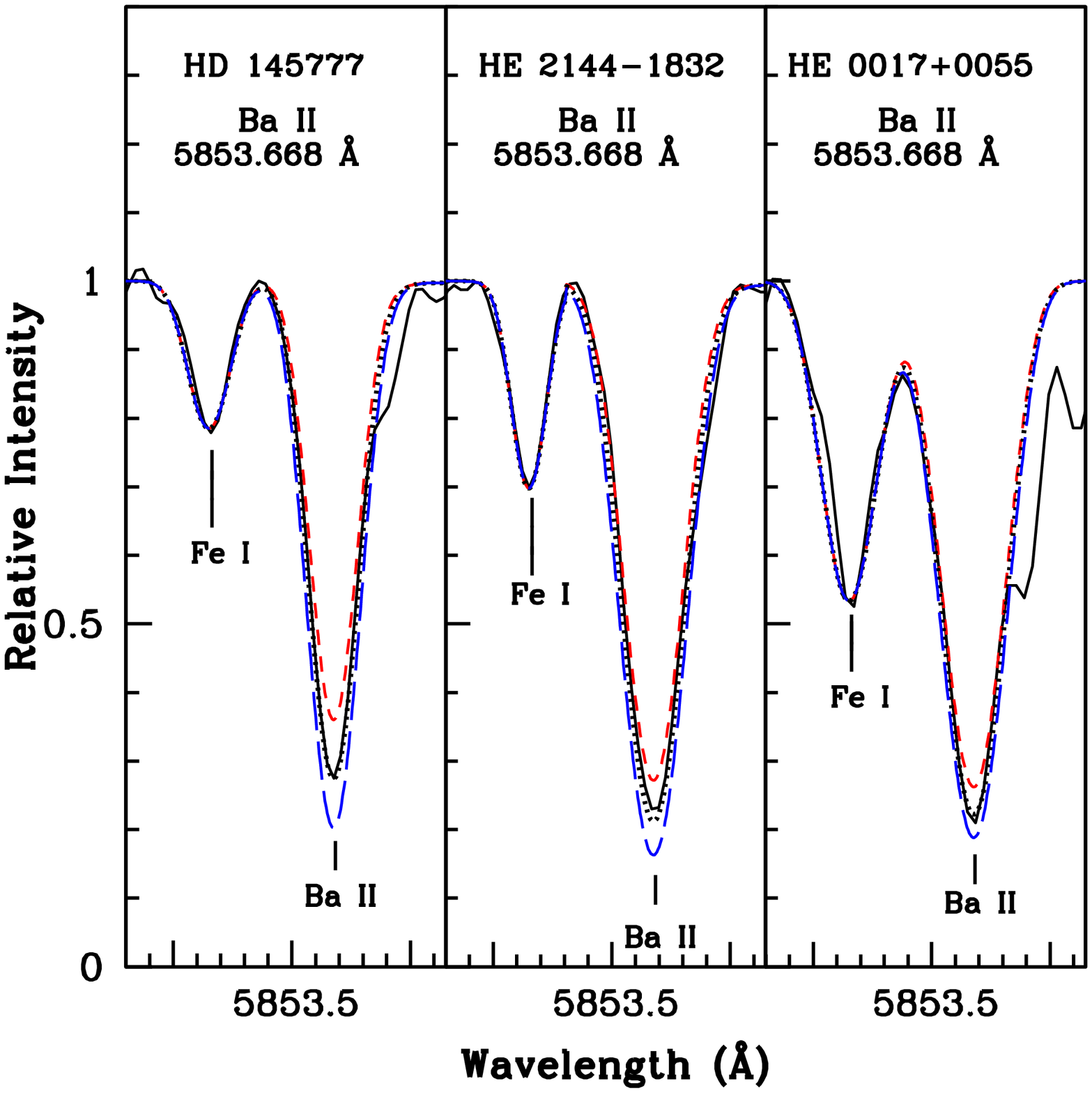}
        }%
 {%
            \label{fig:Ba_32}
            \includegraphics[height=7.0cm,width=8.5cm]{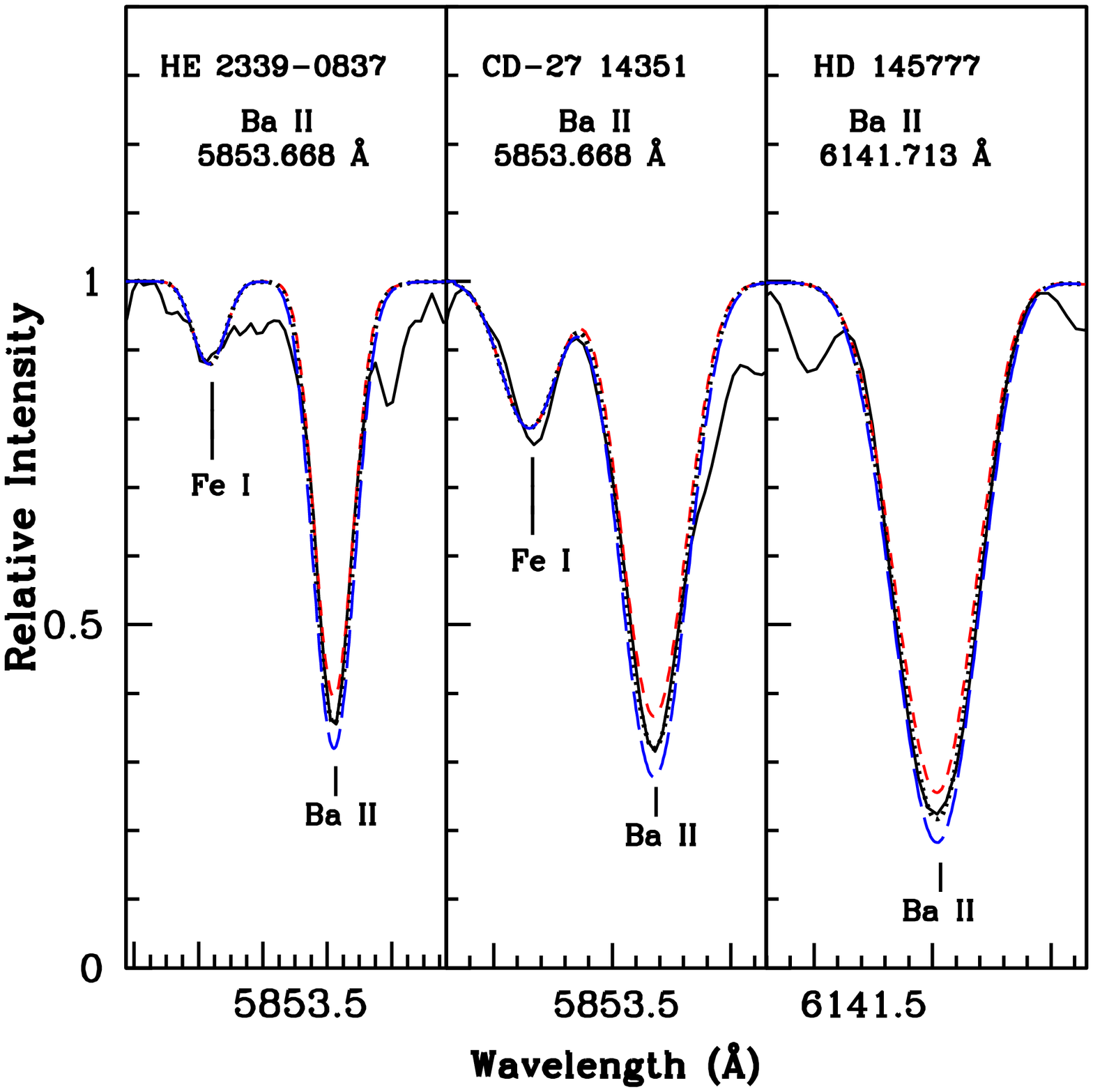}

        }
    \caption{Spectral synthesis of Ba II 5853 {\rm \AA}  and 6141 {\rm \AA} lines. Dotted line represents synthesised spectra and the solid line indicates the observed spectra. Two alternative fits, short dashed line corresponding to $\Delta$[Ba II/Fe] = $-$0.3 and long dashed line  corresponding to $\Delta$[Ba II/Fe] = $+$0.3 are shown to illustrate the sensitivity of line strength to Ba abundance.} %
   \label{fig:subfigures_Ba}
       \end{center}

\end{figure*}

The abundance of Ba is derived using spectrum synthesis calculations of 
Ba II 5853.668 {\rm \AA} line for all  five stars. Spectrum synthesis 
calculation of Ba II 6141.713 {\rm \AA} is also used to find the Ba abundance 
for HD~145777, HE~2339$-$0837, and CD$-$27~14351 
(Figure~\ref{fig:subfigures_Ba}). Both the lines gave similar abundance 
for the stars with the highest standard deviation of 0.15 for 
HE~2339$-$0837. The  Ba II 6141.713 {\rm \AA} line  could not be used 
to determine the abundance of Ba for the other two stars as the line is 
heavily blended in these two stars. The other two frequently used lines 
Ba II 4554.029 {\rm \AA} and Ba II 4934.076 {\rm \AA} are found to be severely blended in the spectra of the programme stars, 
and hence  could not be used for abundance analysis. Hyperfine splitting 
contributions of the two lines used in the analysis are taken from \citet{mcwilliam1998}. Ba is enhanced in all the stars with  1.27 $\leq$ [Ba/Fe] $\leq$ 2.30.

The abundance of La is estimated using the spectrum synthesis 
calculation of La II 4921.776 {\rm \AA} and found to be overabundant 
in   all  five stars with 1.37 $\leq$ [La/Fe] $\leq$ 2.46. La II 4808.996 {\rm \AA} line is also used  for HE~0017$+$0055. Hyperfine splitting contributions of La II 4921.776 {\rm \AA} line are taken from \citet{jonsell2006}.

The abundance of Ce, Pr, and Nd  are estimated using the equivalent width measurement technique. Twelve Ce II lines, 3 Pr II lines, and 13 Nd II lines  (Table~\ref{tab:Elem_linelist1}) are examined to derive the abundance of Ce, Pr, and Nd. Ce is found to be overabundant in all  five stars. For the object CD$-$27~14351 we find a much lower value ([Ce/Fe] = 1.89) than the $\sim$ 2.63 found by \citet{Drisya2017}. \citet{Drisya2017} used two lines Ce II 4460.207 {\rm \AA} and 4527.348 {\rm \AA} for Ce abundance estimate. We  used three lines Ce II 4460.207 {\rm \AA}, 4483.893 {\rm \AA,} and 5187.458 {\rm \AA}. For the line Ce II 4460.207 {\rm \AA}, which  is common in both the studies, our  measured equivalent width is $\sim$ 160 m{\rm \AA}, about 14 m{\rm \AA} higher than  in \citet{Drisya2017}. This line, however, returned an  abundance value that is $\sim$ 0.7 dex lower. In order to check whether the difference in abundance is due to the adopted model atmosphere, we   calculated the abundance of Ce using the measured equivalent width ($\sim$ 146 m{\rm \AA}) of this line by \citet{Drisya2017} and also the model atmosphere adopted by them. The derived abundance is found to be [Ce/Fe] = 1.72, which is closer to our estimated value. The other line, Ce II 4527.348 {\rm \AA}, used in their study  is found to be severely blended in our  spectrum.

Praseodymium (Pr) is found to be overabundant in all the stars with 1.67 $\leq$ [Pr/Fe] $\leq$ 2.42. Neodymium (Nd) is also found to be  enhanced in all the stars with 1.37 $\leq$ [Nd/Fe] $\leq$ 2.55. 

\subsubsection{Sm, Eu}
The abundance of Sm is derived using six Sm II lines (Table~\ref{tab:Elem_linelist1}). No Sm lines could be found for CD$-$27~14351, hence Sm abundance could not be estimated for this object. The other four objects show enhanced abundance of Sm.

\begin{figure}
        \centering\
        \includegraphics[height=9cm,width=9cm]{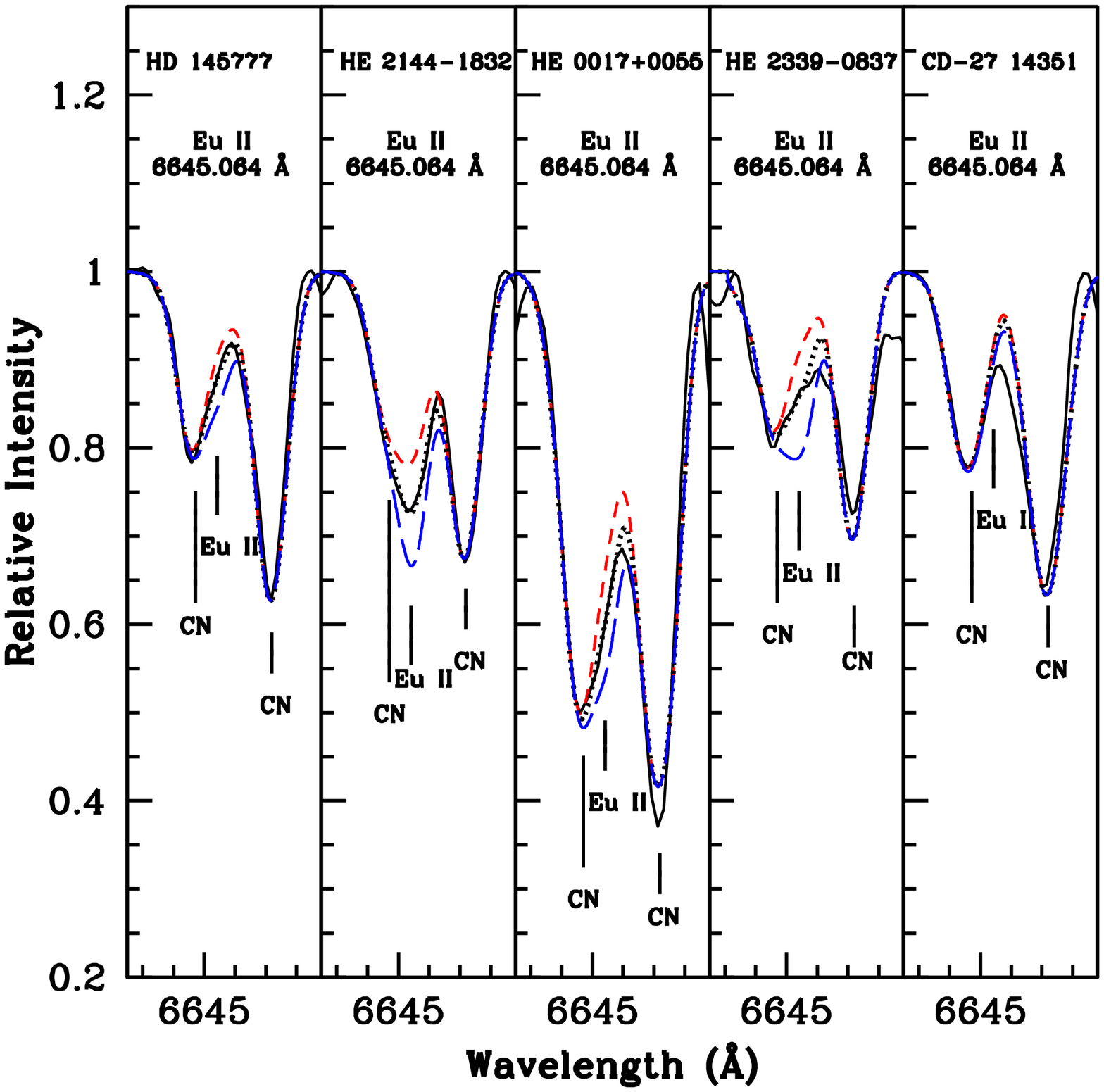}
        \caption{ Synthesis of Eu II  6645 {\rm \AA} line. Dotted line represents synthesised spectra and the solid line indicates the observed spectra. Two alternative fits, short dashed line  corresponding to $\Delta$[Eu II/Fe] = $-$0.3 and long dashed line  corresponding to $\Delta$[Eu II/Fe] = $+$0.3 are shown to illustrate the sensitivity of line strength to Eu abundance.}
\label{fig:eu}
\end{figure}

The abundance of Eu is derived using the spectrum synthesis calculation of the
Eu II 6645.064 {\rm \AA} line (Figure~\ref{fig:eu}) considering hyperfine 
splitting contributions from \citet{worley2013}.
The other frequently used lines, Eu II 4129.725 {\rm \AA}, 
4205.042 {\rm \AA}, and  6437.640 {\rm \AA}, are severely blended, and hence  
could not be used for abundance analysis. Eu shows overabundance 
with 0.80 $\leq$ [Eu/Fe] $\leq$ 2.14,  except in  CD$-$27~14351, for which  
we could  estimate  an upper limit  with   [Eu/Fe] ${\sim}$ 0.39. As  
shown in Figure~\ref{fig:eu}, the 
Eu II 6645.064 {\rm \AA} line is blended with two CN lines. It can be clearly seen that the Eu II 6645.064 {\rm \AA} line is absent in the spectrum of the object CD$-$27~14351;  a misidentification  of  the CN line at 6644.750 {\rm \AA} as the Eu II 6645.064 {\rm \AA} line could  result in a much higher abundance of Eu; this may explain the very high abundance recorded for Eu by \citet{Drisya2017}.

{\footnotesize
\begin{table*}
\centering
\caption{\bf{Elemental abundances in HD~145777, CD$-$27~14351, and HE~0017$+$0055}}
\label{tab:abundances}
\scalebox{0.84}{
\begin{tabular}{l r c|l r r|l r r|l r r}
\hline
& \multicolumn{6}{c}{HD~145777} & \multicolumn{2}{c}{CD$-$27~14351} & \multicolumn{3}{c}{HE~0017$+$0055}\\
\hline
Element & Z  & solar $log{\epsilon}^a$ & $log{\epsilon}$ &[X/H]& [X/Fe] & $log{\epsilon}$ & [X/H] & [X/Fe] & $log{\epsilon}$ & [X/H] & [X/Fe]\\
            &    &                         &    (dex)       &       &        &   (dex)         &       &        &   (dex)         &       &\\
\hline
C (C$_{2}$, 5165 {\rm \AA}) & 6  & 8.43 & 8.69 (syn)    & 0.26  & 2.43   & 8.70 (syn)      & 0.27  & 2.98   & 8.70 (syn)      &  0.27 & 2.73   \\
C (C$_{2}$, 5635 {\rm \AA}) & 6  & 8.43 & 8.69 (syn)    & 0.26  & 2.43   & 8.70 (syn)      & 0.27  & 2.98   & 8.71 (syn)      &  0.28 & 2.74   \\ 
C (CH, 4310 {\rm \AA})      & 6  & 8.43 & 8.67 (syn)    & 0.24  & 2.41   & 8.70 (syn)      & 0.27  & 2.98   & 8.73 (syn)      &  0.30 & 2.76   \\
N (CN, 4215 {\rm \AA})      & 7  & 7.83 & 6.33 (syn)    & $-$1.50 & 0.67 & 7.00 (syn)      & $-$0.83 & 1.88 & 8.20 (syn)      &   0.37 & 2.83   \\
Na {\sc i}  & 11 & 6.24 & 4.52$\pm$0.12 (2)  & $-$1.72 & 0.45   & 4.91$\pm$0.20 (2)  & $-$1.33 & 1.38 & -                 &   -    & -      \\
Mg {\sc i}  & 12 & 7.60 & 6.25$\pm$0.23 (2)  & $-$1.35 & 0.82   & 6.70 (1)           & $-$0.90 & 1.81 & -                 &   -    & -      \\
Ca {\sc i}  & 20 & 6.34 & 4.66$\pm$0.21 (9)  & $-$1.68 & 0.49   & 4.54$\pm$0.20 (5)  & $-$1.80 & 0.91 & -                 &   -    & -      \\
Sc {\sc ii} & 21 & 3.15 & 1.75 (syn)         & $-$1.40 & 0.77   & -                  &  -      &  -   & -                 &   -    & -      \\
Ti {\sc i}  & 22 & 4.95 & 3.44$\pm$0.17 (2)  & $-$1.51 & 0.66   & 3.19$\pm$0.20 (3)  & $-$1.76 & 0.95 & 3.13 (1)          &$-$1.82 & 0.64   \\
Ti {\sc ii} & 22 & 4.95 & 3.43$\pm$0.08 (5)  & $-$1.52 & 0.65   & 3.20$\pm$0.13 (5)  & $-$1.75 & 0.96 & 3.14$\pm$0.05 (3) &$-$1.81 & 0.65   \\
V {\sc i}   & 23 & 3.93 & 1.65 (syn)         & $-$2.28 &$-$0.11 & 2.33 (syn)         & $-$1.60 & 1.11 & -                 &   -    & -      \\
Cr {\sc i}  & 24 & 5.64 & 3.26$\pm$0.13 (6)  & $-$2.38 &$-$0.21 & 3.11$\pm$0.18 (3)  & $-$2.53 & 0.18 & 2.68 (1)          &$-$2.96 &$-$0.50 \\
Mn {\sc i}  & 25 & 5.43 &       -            &    -    & -      & -                  &  -      &  -   & -                 &   -    & -      \\
Mn {\sc ii} & 25 & 5.43 & 2.45 (syn)         & $-$2.98 &$-$0.81 & -                  &  -      &  -   & -                 &   -    & -      \\
Fe {\sc i}  & 26 & 7.50 & 5.33$\pm$0.18 (33) & $-$2.17 & -      & 4.78$\pm$0.16 (33) & $-$2.72 & -    & 5.03$\pm$0.12 (21)&$-$2.47 & -      \\
Fe {\sc ii} & 26 & 7.50 & 5.33$\pm$0.25 (4)  & $-$2.17 & -      & 4.81$\pm$0.07 (3)  & $-$2.69 & -    & 5.05$\pm$0.03 (3) &$-$2.45 & -      \\
Co {\sc i}  & 27 & 4.99 & 2.85 (1)           & $-$2.14 & 0.03   & -                  &  -      & -    & -                 &   -    & -      \\
Ni {\sc i}  & 28 & 6.22 & -                  & -       & -      & -                  &  -      & -    & -                 &   -    & -      \\
Zn {\sc i}  & 30 & 4.56 & 3.02 (1)           & $-$1.54 & 0.63   & -                  &  -      & -    & 2.62 (1)          &$-$1.94 & 0.52   \\
Sr {\sc i}  & 38 & 2.87 & 1.37 (syn)         & $-$1.50 & 0.67   & 1.90 (syn)         & $-$0.97 & 1.74 & -                 &   -    & -      \\
Y {\sc ii}  & 39 & 2.21 & 1.26$\pm$0.02 (2)  & $-$0.95 & 1.22   & 1.47$\pm$0.10 (3)  & $-$0.74 & 1.97 & 0.33$\pm$0.08 (3) &$-$1.88 & 0.58   \\
Zr {\sc i}  & 40 & 2.58 & 1.46$\pm$0.14 (2)  & $-$1.12 & 1.05   & 2.07 (1)           & $-$0.51 & 2.20 & -                 &   -    & -      \\
Zr {\sc ii} & 40 & 2.58 & -                  & -       & -      & 2.09 (1)           & $-$0.49 & 2.22 & 1.67$\pm$0.20 (3) &$-$0.91 & 1.55   \\
Ba {\sc ii} & 56 & 2.18 &1.28 (syn)$\pm$0.03 & $-$0.90 & 1.27   & 1.29 (syn)$\pm$0.09& $-$0.89 & 1.82 & 2.02 (syn)        &$-$0.16 & 2.30   \\
La {\sc ii} & 57 & 1.10 & 0.30 (syn)         & $-$0.80 & 1.37   & $-$0.05 (syn)      & $-$1.15 & 1.56 & 1.10 (syn)$\pm$0.20&0.0     & 2.46  \\
Ce {\sc ii} & 58 & 1.58 & 1.20$\pm$0.09 (5)  & $-$0.38 & 1.79   & 0.76$\pm$0.16 (3)  & $-$0.82 & 1.89 & 1.23$\pm$0.06 (2) &$-$0.35 & 2.11   \\
Pr {\sc ii} & 59 & 0.72 & 0.22$\pm$0.00 (2)  & $-$0.50 & 1.67   & $-$0.03 (1)        & $-$0.75 & 1.96 & 0.68 (1)          &$-$0.04 & 2.42   \\
Nd {\sc ii} & 60 & 1.42 & 0.73$\pm$0.18 (5)  & $-$0.69 & 1.48   & 0.08$\pm$0.16 (6)  & $-$1.34 & 1.37 & 1.21$\pm$0.11 (4) &$-$0.21 & 2.25   \\
Sm {\sc ii} & 62 & 0.96 & 0.42$\pm$0.03 (2)  & $-$0.54 & 1.63   & -                  & -       & -    & 0.48$\pm$0.04 (2) &$-$0.48 & 1.98   \\
Eu {\sc ii} & 63 & 0.52 &$-$0.85 (syn)       & $-$1.37 & 0.80   &$< -$1.80 (syn)     &$< -$2.32&$<$ 0.39& 0.20 (syn)        &$-$0.32 & 2.14 \\
\hline 
\end{tabular}}
 
 $^{a}$ \citet{asplund2009}. The numbers in  parentheses in Cols. 4, 7, and 10 show the number of lines used for the abundance determination.\\
\end{table*}
}

{\footnotesize
\begin{table*}
\centering
\caption{\bf{Elemental abundances in HE~2144$-$1832 and HE~2339$-$0837}}
\label{tab:abundances2}
\begin{tabular}{l r c|l r r|l r r}
\hline
& \multicolumn{6}{c}{HE~2144$-$1832} & \multicolumn{0}{c}{HE~2339$-$0837}\\
\hline
    Element    & Z  & solar $log{\epsilon}^a$ & $log{\epsilon}$ &[X/H]& [X/Fe] & $log{\epsilon}$ & [X/H] & [X/Fe]\\
            &    &                         &    (dex)       &       &        &   (dex)         &       &    \\
\hline
C (C$_{2}$, 5165 {\rm \AA}) & 6  & 8.43 & 8.65 (syn)        & 0.22    & 1.85    & 8.73 (syn)         & 0.30    & 3.04      \\
C (C$_{2}$, 5635 {\rm \AA}) & 6  & 8.43 & 8.60 (syn)        & 0.17    & 1.80    & 8.50 (syn)         & 0.07    & 2.81      \\
C (CH, 4310 {\rm \AA})      & 6  & 8.43 & 8.55 (syn)        & 0.12    & 1.75    &   -                &   -     &   -       \\
N (CN, 4215 {\rm \AA})      & 7  & 7.83 & 6.70 (syn)        & $-$1.13 & 0.50    & -                  & -       & -         \\
Na {\sc i}  & 11 & 6.24 & 5.14$\pm$0.08 (3) & $-$1.10 & 0.53    & 5.04$\pm$0.03 (2)  & $-$1.20 & 1.54      \\
Mg {\sc i}  & 12 & 7.60 & 6.64$\pm$0.11 (2) & $-$0.96 & 0.67    & 7.03 (1)           & $-$0.57 & 2.17      \\
Ca {\sc i}  & 20 & 6.34 & 5.11$\pm$0.17 (8) & $-$1.23 & 0.40    & 4.09$\pm$0.14 (4)  & $-$2.25 & 0.49      \\
Ti {\sc i}  & 22 & 4.95 & 3.53$\pm$0.13 (6) & $-$1.42 & 0.21    & -                  & -       & -         \\
Ti {\sc ii} & 22 & 4.95 & 3.51$\pm$0.05 (3) & $-$1.44 & 0.19    & 2.62$\pm$0.15 (4)  & $-$2.33 & 0.41      \\
V {\sc i}   & 23 & 3.93 & 2.36 (syn)        & $-$1.57 & 0.06    & -                  & -       & -         \\
Cr {\sc i}  & 24 & 5.64 & 3.89$\pm$0.18 (4) & $-$1.75 & $-$0.12 & 2.56 (1)           & $-$3.08 &$-$0.34    \\
Mn {\sc i}  & 25 & 5.43 & 3.20 (syn)$\pm$0.0& $-$2.23 & $-$0.60 & -                  &   -     & -         \\
Fe {\sc i}  & 26 & 7.50 & 5.87$\pm$0.15 (27)& $-$1.63 & -       & 4.76$\pm$0.08 (19) & $-$2.74 & -         \\
Fe {\sc ii} & 26 & 7.50 & 5.87$\pm$0.00 (2) & $-$1.63 & -       & 4.76$\pm$0.08 (4)  & $-$2.74 & -         \\
Co {\sc i}  & 27 & 4.99 & 3.42$\pm$0.03 (3) & $-$1.57 & 0.06    & -                  &   -     & -         \\
Ni {\sc i}  & 28 & 6.22 & 5.00$\pm$0.16 (5) & $-$1.22 & 0.41    & -                  &   -     & -         \\
Zn {\sc i}  & 30 & 4.56 & 2.85 (1)          & $-$1.71 &$-$0.08  & -                  &   -     & -         \\
Sr {\sc i}  & 38 & 2.87 & 1.90 (syn)        & $-$0.97 & 0.66    & 1.45 (syn)         & $-$1.42 & 1.32      \\
Y {\sc ii}  & 39 & 2.21 & 1.74$\pm$0.08 (3) & $-$0.47 & 1.16    & 0.14$\pm$0.10 (3)  & $-$2.07 & 0.67      \\
Zr {\sc i}  & 40 & 2.58 & 1.92$\pm$0.18 (3) & $-$0.66 & 0.97    & -                  & -       & -         \\
Zr {\sc ii} & 40 & 2.58 & -                 & -       & -       & 1.48 (1)           & $-$1.10 & 1.64      \\
Ba {\sc ii} & 56 & 2.18 & 2.04 (syn)        & $-$0.14 & 1.49    &1.65 (syn)$\pm$0.15 & $-$0.53 & 2.21      \\
La {\sc ii} & 57 & 1.10 & 1.00 (syn)        & $-$0.10 & 1.53    & 0.60 (syn)         & $-$0.50 & 2.24      \\
Ce {\sc ii} & 58 & 1.58 & 1.65$\pm$0.17 (6) & 0.07    & 1.70    & 1.21$\pm$0.13 (3)  & $-$0.37 & 2.37      \\
Pr {\sc ii} & 59 & 0.72 & 0.81$\pm$0.17 (3) & 0.09    & 1.72    & 0.24$\pm$0.22 (2)  & $-$0.48 & 2.26      \\
Nd {\sc ii} & 60 & 1.42 & 1.40$\pm$0.17 (7) & $-$0.02 & 1.61    & 1.23$\pm$0.10 (5)  & $-$0.19 & 2.55      \\
Sm {\sc ii} & 62 & 0.96 & 1.11$\pm$0.21 (3) & 0.15    & 1.78    & 0.51$\pm$0.14 (3)  & $-$0.45 & 2.29      \\
Eu {\sc ii} & 63 & 0.52 &$-$0.10 (syn)      & $-$0.62 & 1.01    &$-$0.38 (syn)       & $-$0.90 & 1.84      \\
\hline 
\end{tabular}
 
 $^{a}$ \citet{asplund2009}. The numbers in parentheses in Cols. 4 and 7 show the number of lines used for the abundance determination.\\
\end{table*}
}

{\footnotesize
\begin{table*}
\centering
\caption{\bf{Comparison of the abundances of our programme stars with the literature values}}
\label{tab:abuncomp}    
\begin{tabular}{lcccccccc}
\hline                       
Star Name       & [Fe/H]     & [C/Fe]$^{*}$      &  [N/Fe]    & [Sr/Fe]   & [Y/Fe]     & [Zr/Fe]    & [Ba II/Fe] &  Ref \\
\hline
HD~145777       &$-$2.17     &  2.42      &  0.67      & 0.67      &  1.22      &  1.05      &  1.27      &   1  \\
CD$-$27~14351   &$-$2.71     &  2.98      &  1.88      & 1.74      &  1.97      &  2.21      &  1.82      &   1  \\
                &$-$2.62     &  2.89      &  1.89      & 1.73      &  1.99      &  -         &  1.77      &   4  \\
HE~0017$+$0055  &$-$2.46     &  2.74      &  2.83      &  -        &  0.58      &  1.55      &  2.30      &   1  \\
                &$-$2.40     &  2.17      &  2.47      &  -        &  0.50      &  1.60      &$>$ 1.90    &   3  \\
                &$-$2.72     &  2.31      &  0.52      &  -        &  -         &  -         &  -         &   5  \\
HE~2144$-$1832  &$-$1.63     &  1.80      &  0.50      & 0.66      &  1.16      &  0.97      &  1.49      &   1  \\
                &$-$1.70     &  0.80      &  0.60      & 1.50      &  -         &  -         &  1.30      &   2  \\
HE~2339$-$0837  &$-$2.74     &  2.93      &  -         & 1.32      &  0.67      &  1.64      &  2.21      &   1  \\
                &$-$2.71     &  2.71      &  -         &  -        &  -         &  -         &  -         &   5 \\  
\hline                       
Star Name       & [La II/Fe] & [Ce II/Fe] & [Pr II/Fe] &[Nd II/Fe] & [Sm II/Fe] & [Eu II/Fe] &            &  Ref \\
\hline
HD~145777       &  1.37      &  1.79      &  1.67      &  1.48     &  1.63      &  0.80      &            &   1  \\
CD$-$27~14351   &  1.56      &  1.89      &  1.96      &  1.37     &  -         &$<$ 0.39    &            &   1  \\
                &  1.57      &  2.63      &   -        &  1.26     &  -         &  1.65      &            &   4  \\
HE~0017$+$0055  &  2.46      &  2.11      &  2.42      &  2.25     &  1.98      &  2.14      &            &   1  \\
                &  2.40      &  2.00      &   -        &  2.20     &  1.90      &  2.30      &            &   3  \\
                &  -         &  -         &   -        &  -        &  -         &  -         &            &   5  \\
HE~2144$-$1832  &  1.53      &  1.70      &  1.72      &  1.61     &  1.78      &  1.01      &            &   1  \\
                &  -         &  -         &   -        &  -        &  -         &  -         &            &   2  \\
HE~2339$-$0837  &  2.24      &  2.37      &  2.26      &  2.55     &  2.29      &  1.84      &            &   1  \\
                &  -         &  -         &   -        &  -        &  -         &  -         &            &   5  \\
\hline   
\end{tabular}

* $-$ Abundance of carbon is the average abundance derived from different molecular bands.\\

References: 1. Our work, 2. \citet{hansen2016abundances}, 3. \citet{jorissen2016HE0017}, 4. \citet{Drisya2017}, 5. \citet{kennedy2011}. \\          
\end{table*}
}

{\footnotesize
\begin{table*}
\centering
\caption{\bf{Observed abundance ratios}}
        \label{tab:abundanceratios}
\scalebox{0.95}{
\begin{tabular}{lcccccccc}
\hline
Star Name       & [Fe/H]  & [ls/Fe]  & [hs/Fe]  & [hs/ls] &  $^{12}$C/$^{13}$C              & $^{12}$C/$^{13}$C & [Ba/Eu]          &   Ref  \\
                &         &          &          &         & (From CN band at 8005 {\rm \AA})& (From C$_{2}$ band at 4740 {\rm \AA})&  &        \\                
\hline
HD~145777       &$-$2.17  & 0.98     & 1.48     & 0.50    &    5.45        &   5.50  & 0.47    &  1   \\
CD$-$27~14351   &$-$2.71  & 1.97     & 1.66     &$-$0.31  &    5.50        &   2.58  &  -      &  1   \\
                &$-$2.62  & 1.82     & 1.77     &$-$0.05  &    10.1        &     -   & 0.14    &  4   \\
HE~0017$+$0055  &$-$2.46  & 1.07     & 2.28     & 1.21    &    3.64        &   4.00  & 0.16    &  1   \\
                &$-$2.40  & 1.05     & 2.20     & 1.15    &     -          &   4.00  &  -      &  3   \\
                & -       & -        & -        & -       &     -          &   1.30  &  -      &  2   \\
HE~2144$-$1832  &$-$1.63  & 0.93     & 1.58     & 0.65    &    2.50        &   2.50  & 0.48    &  1   \\
                & -       & -        & -        & -       &     -          &   2.10  &  -      &  2   \\
HE~2339$-$0837  &$-$2.74  & 1.21     & 2.34     & 1.13    &     -          &   7.00  & 0.37    &  1   \\
\hline
\end{tabular}}

References: 1. Our work, 2. \citet{goswami2005ch}, 3. \citet{jorissen2016HE0017}, 4. \citet{Drisya2017}.  \\          
\end{table*}
}

\subsection{Abundance uncertainties}
\label{sec:Abundance Uncertainties}
Two components contribute to the total
uncertainties on elemental abundances: random error and systematic error. In order to derive the uncertainties on the elemental abundances, we  followed the procedure mentioned in \citet{shejeela_agb_2020}.
We  estimated the total uncertainties on log$\epsilon$ using the equation
\begin{equation}
\label{eqn:uncertainty}
\begin{split}
\sigma ^{2}_{log\epsilon} = & \sigma ^{2}_{ran} + \left( \dfrac{\delta log\epsilon}{\delta T}\right)^{2}\sigma ^{2}_{T_{eff}} + \left( \dfrac{\delta log\epsilon}{\delta logg}\right)^{2}\sigma ^{2}_{logg}\\
                            & + \left( \dfrac{\delta log\epsilon}{\delta \zeta}\right)^{2}\sigma ^{2}_{\zeta} + \left( \dfrac{\delta log\epsilon}{\delta [Fe/H]}\right)^{2}\sigma ^{2}_{[Fe/H]} 
\end{split}
,\end{equation}
where $\sigma_{ran}$ is the random error that arises due to the uncertainties on the factors like equivalent width measurement, oscillator strength, and line blending. We   adopted $\sigma_{ran}$ = $\dfrac{\sigma_{s}}{\sqrt N}$, where $\sigma_{s}$ is the standard deviation of the abundance of a particular species estimated using N number of lines of that species.

The typical uncertainties on the stellar atmospheric parameters T$_{eff}$, logg, $\zeta$, and [Fe/H] are denoted $\sigma_{T_{eff}}$, $\sigma_{logg}$, $\sigma_{\zeta}$, and $\sigma_{[Fe/H]}$, respectively. We  evaluated the partial derivatives appearing in Equation~\ref{eqn:uncertainty} for the star HE~2144$-$1832, varying the stellar parameters T$_{eff}$, logg, $\zeta,$ and [Fe/H] by $\pm$ 100 K, $\pm$ 0.2 dex, $\pm$ 0.2 km/s$^{-1}$, and $\pm$ 0.2 dex, respectively. Finally, the uncertainties on [X/Fe] are derived as
\begin{equation}
\sigma ^{2}_{[X/Fe]} =  \sigma ^{2}_{X} + \sigma ^{2}_{[Fe/H]}
.\end{equation}

The resulting differential abundances and the derived uncertainties  on [X/Fe] are presented in Table~\ref{tab:error}. WE note that the calculated uncertainties on [X/Fe] are overestimated because of the assumption of the uncorrelated nature of the uncertainties arising from the different stellar parameters in Equation~\ref{eqn:uncertainty}.

\section{Kinematic analysis}
\label{sec:kinematic_analysis}

Along with the physical parameters and composition it is also important to know the group of Galactic populations to which the programme stars  belong. 
To determine this we  calculated the space  velocity for the 
stars using the method of \citet{Johnson1987space_velocity}. 
Using the method given by \citet{bensby2003} and information such as parallax ($\pi$) and proper motion ($\mu_{\alpha}, \mu_{\delta}$) from the Gaia and SIMBAD databases \citep{gaia2018} and radial velocity (V$_{r}$) from our estimates, we   calculated the components of space velocity with respect to Local Standard of Rest (LSR) using the relation 

\begin{equation}
(U,V,W)_{LSR}=(U,V,W)+ (U,V,W)_{\odot}  ~ {km/s}
.\end{equation}
Here U, V, and W are the velocity vectors pointing towards the Galactic centre, the direction of Galactic rotation,
and the Galactic north pole, respectively. 
The solar U, V, W component velocities (11.1, 12.2, 7.3) km/s are taken from \citet{schonrich2010}. The total spatial velocity (V$_{spa}$) is given by

\begin{equation}
V_{spa}=\sqrt{U_{LSR}^{2}+V_{LSR}^{2}+W_{LSR}^{2}}
.\end{equation}

The detailed description of the calculations can be found in \citet{meenakshi2019chIII}. 
The estimated components of spatial velocity and the total spatial
velocity are presented in   Table~\ref{tab:kinematicresults}.
Following the procedures of \citet{reddy2006}, \citet{bensby2003, bensby2004}, and \citet{mishenina2004}, we    calculated the probability that the stars are 
member of the thin disc, the thick disc, or the halo population. 
The estimated metallicity and  spatial velocities indicate that three
stars are   members of the  thick disc population. The probability estimates 
for them being members of the thick disc population are 0.83, 0.84, and 0.65 for HD 145777, HE~2144$-$1832, and HE~0017$-$0055, respectively. The probability estimates are  0.9 and 1.0 that the objects HE~2339$-$0837 and CD$-$27~14351, respectively, are members of the  halo population (Table~\ref{tab:kinematicresults}).
 \begin{table*}
\centering
\caption{\bf {Spatial velocity and probability estimates}}
\begin{tabular}{lccccccc}
\hline
Star Name       & $U_{LSR} (km/s) $  & $V_{LSR} (km/s)$     &$ W_{LSR} (km/s)$    & $V_{spa}$ (km/s)   & $P_{thin} $ & $P_{thick}$& $ P_{halo}$\\
\hline
HD~145777       &$-$4.51 $\pm  3.22$ &$-$155.01 $\pm 17.49$  &   67.18 $\pm 5.62$   & 169.00 $\pm 13.86$ & 0           & 0.83       &  0.17\\
CD$-$27~14351   & 101.95 $\pm 4.04$  &$-$213.31 $\pm 24.66$  &$-$57.21 $\pm 4.17$   & 243.25 $\pm 20.88$ & 0           & 0.10       & 0.90 \\
HE~0017$+$0055  &  16.21 $\pm 2.02$  &$-$190.74 $\pm 29.87$  &$-$10.33 $\pm 15.90$  & 191.71 $\pm 30.31$ &  0          & 0.65       &  0.35\\
HE~2144$-$1832  & 234.33 $\pm 16.01$ &     8.57 $\pm 6.32$   & $-$9.90 $\pm 9.09$   & 234.69 $\pm 15.82$ & 0.01        & 0.84       &  0.15\\
HE~2339$-$0837  & 218.78 $\pm 34.88$ &$-$126.66 $\pm 37.83$  &$-$224.20$\pm 14.05$  & 337.90 $\pm 4.18$  & 0           &  0         &  1.0 \\
\hline
\label{tab:kinematicresults}    
\end{tabular}
\end{table*}

\section{Discussion}
\label{sec:discussion}
We begin with a brief discussion on the formation scenarios of CEMP-s  and CEMP-r/s stars from the literature, in the context of the stars in this study.

\subsection{Formation scenarios of CEMP-(s \& r/s) stars and their likelihoods}
\label{sec:formation_scenarios_rs}
\begin{figure*}
     \begin{center}
        \subfigure[AGB mass-transfer ]{%
            \label{fig:first}
                \def\svgwidth{\columnwidth}
                \scalebox{1.05}{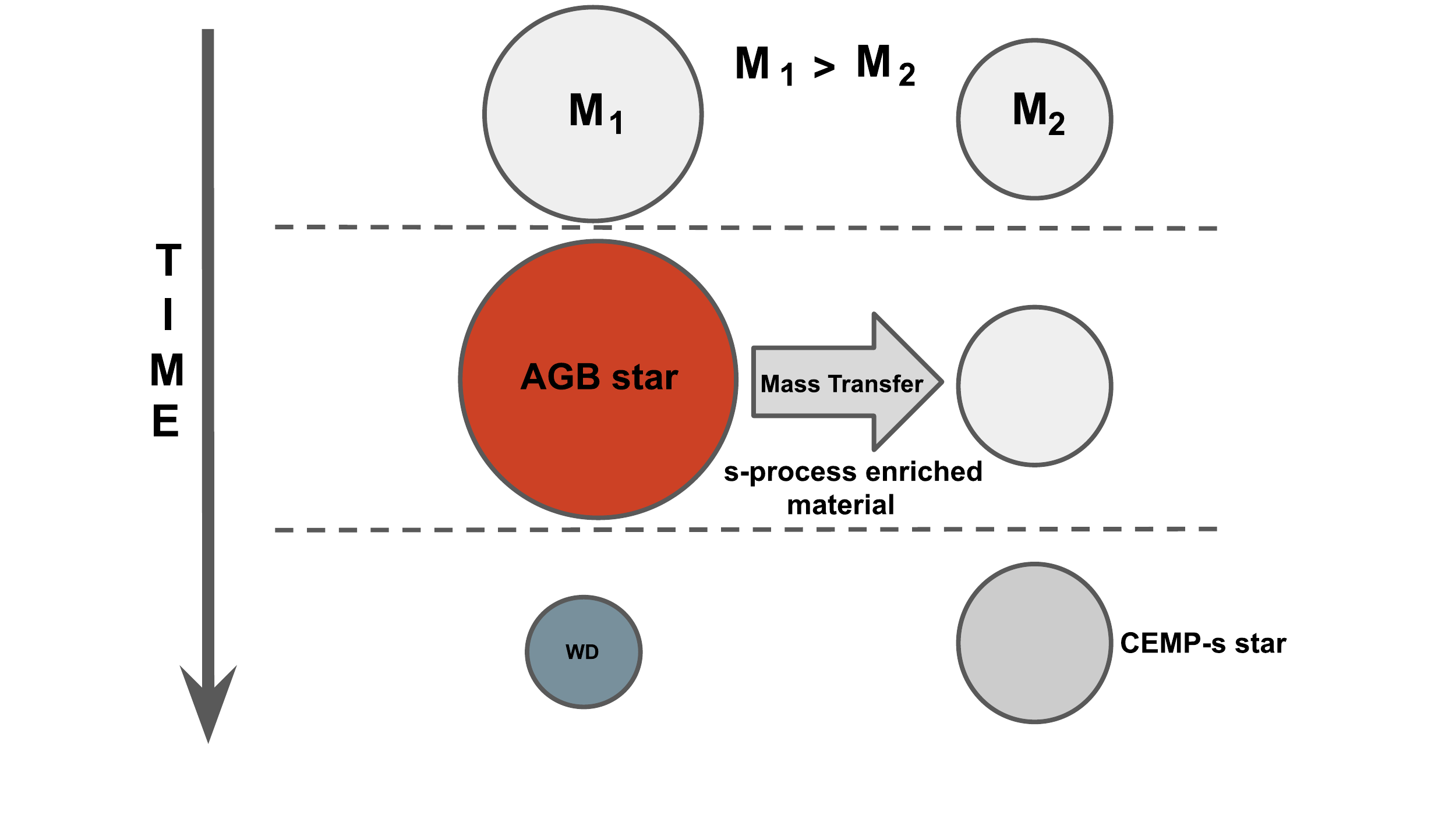}
        }%
        \subfigure[Self-pollution of a star formed from r-rich ISM]{%
            \label{fig:second}
                \def\svgwidth{\columnwidth}
                \scalebox{1.05}{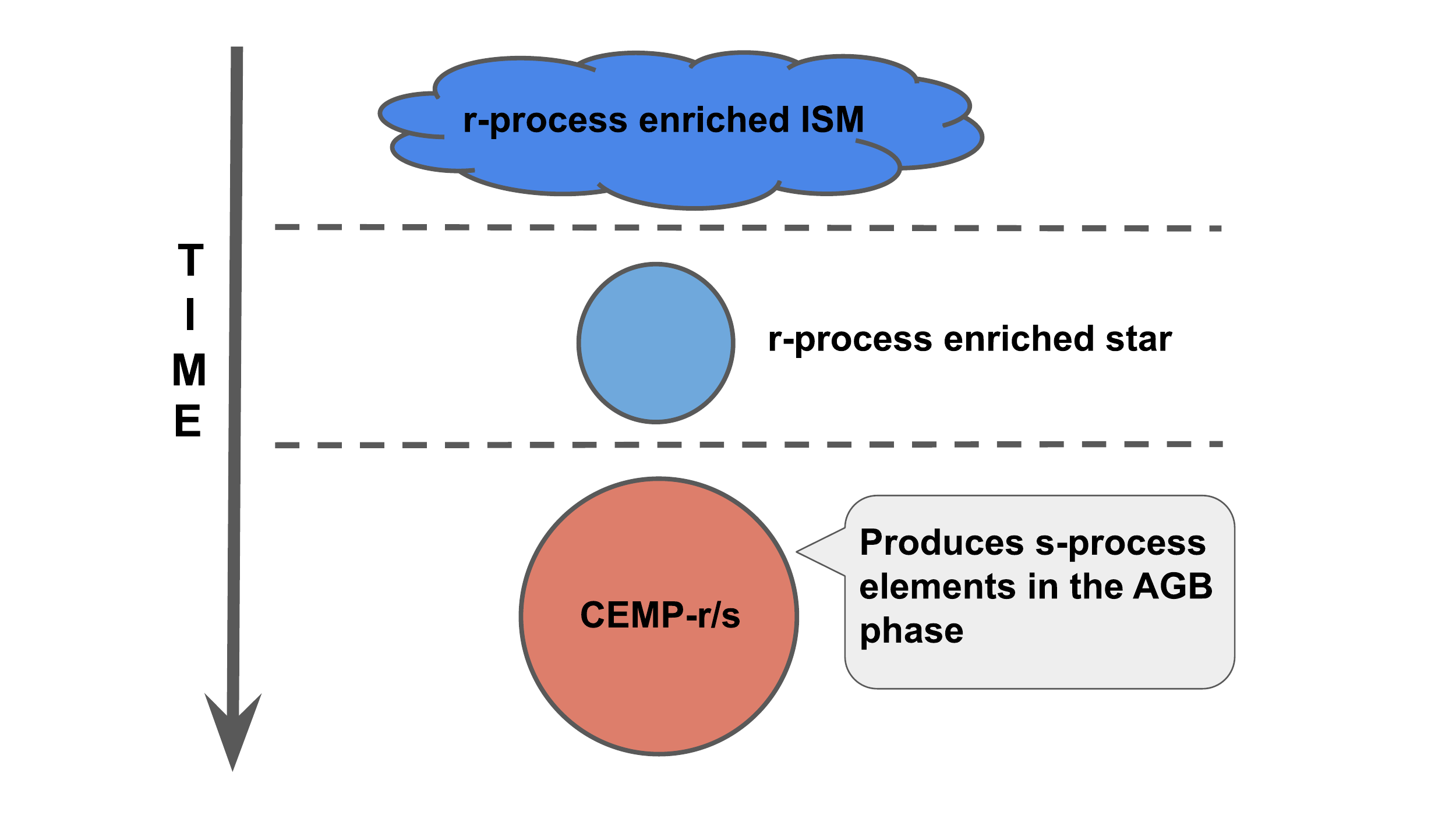}

        }\\ 
        \subfigure[AGB-pollution of a star in binary system formed from r-rich ISM]{%
           \label{fig:third}
                \def\svgwidth{\columnwidth}
                \scalebox{1.05}{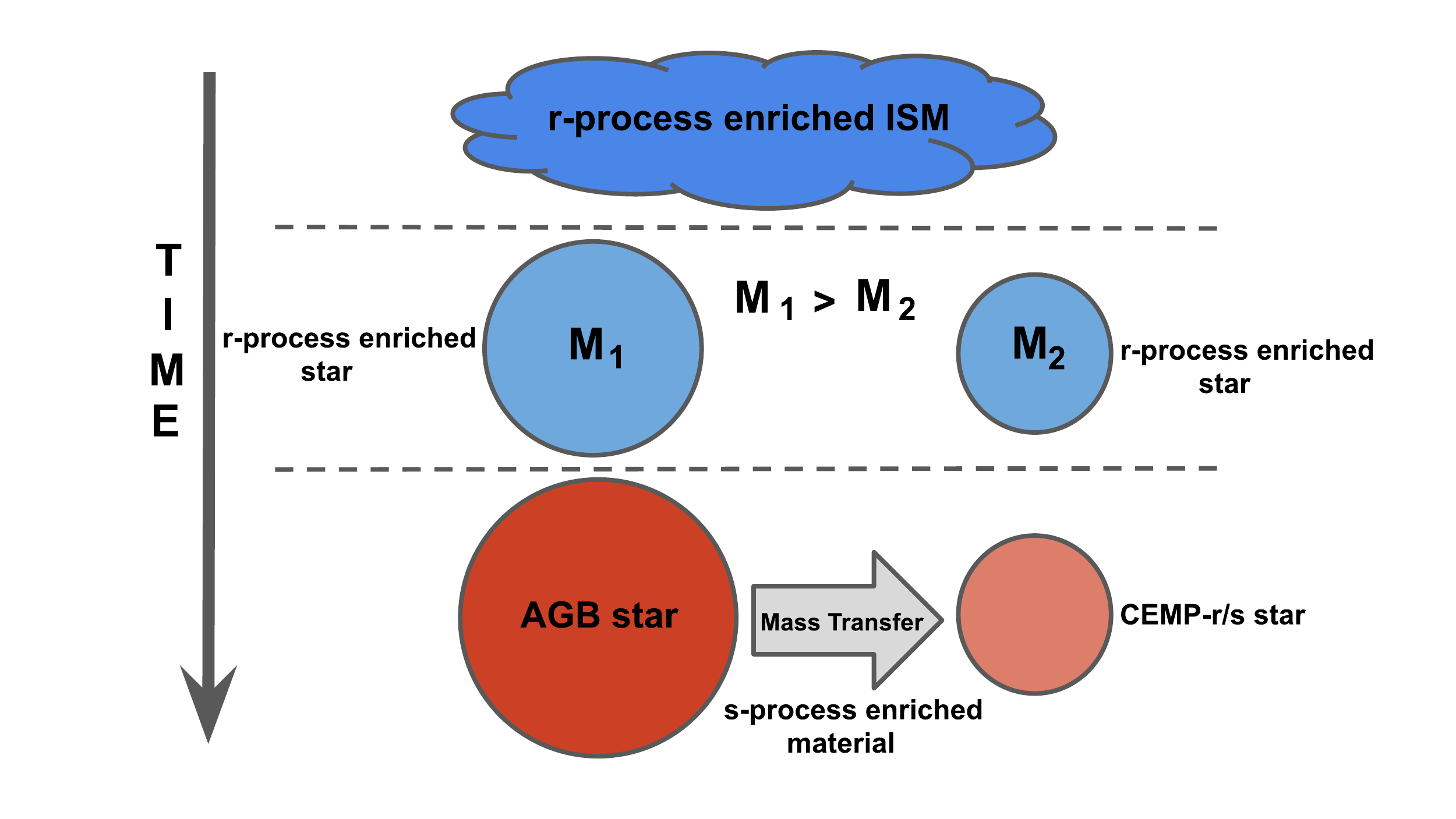}

        }%
        \subfigure[SN and AGB pollution of a star in triple system]{%
            \label{fig:fourth}
                \def\svgwidth{\columnwidth}
                \scalebox{1.05}{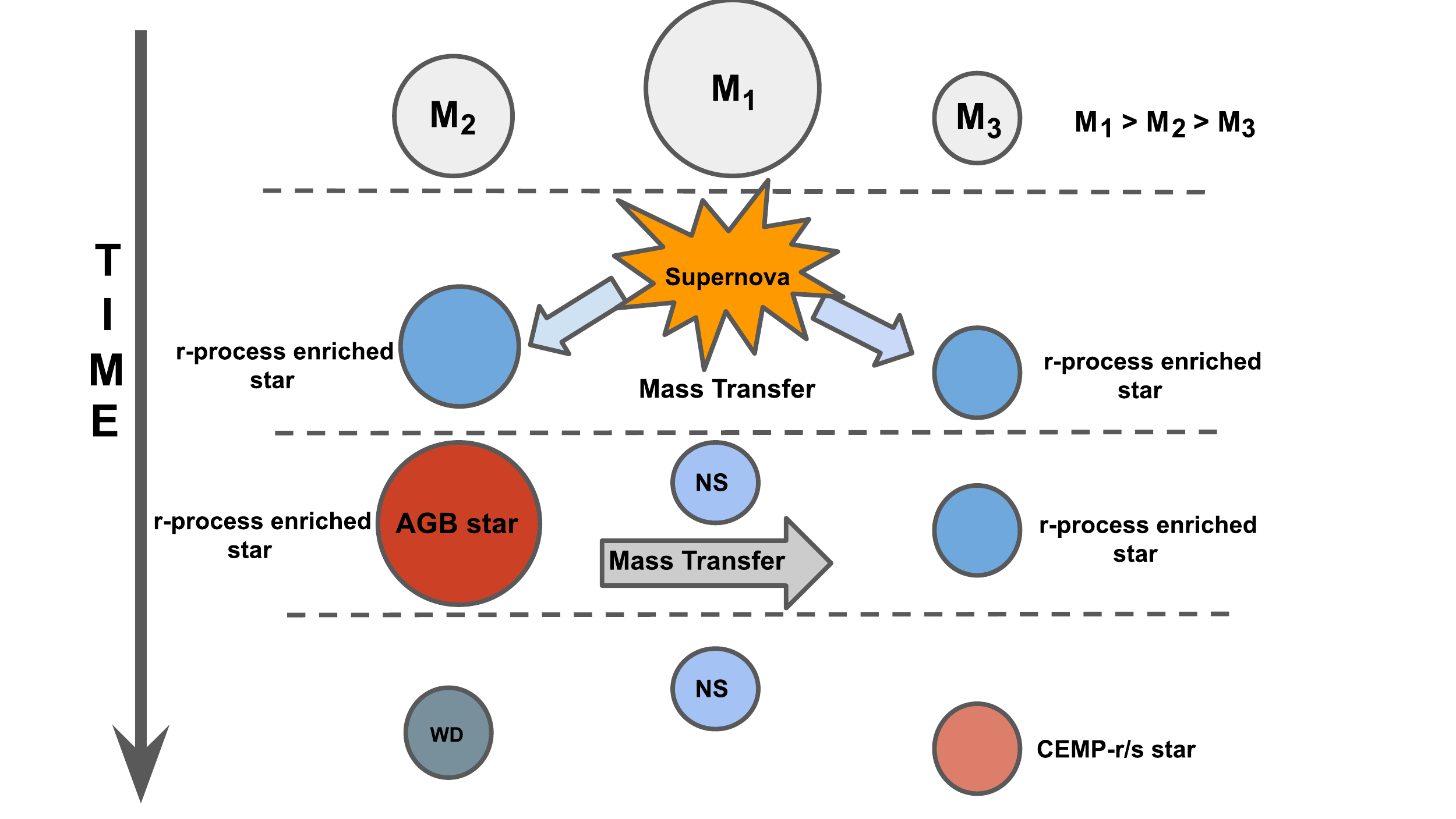}

        }\\
        \subfigure[AGB and 1.5 SN pollution of a star in binary system]{%
            \label{fig:fifth}
                \def\svgwidth{\columnwidth}
                \scalebox{1.05}{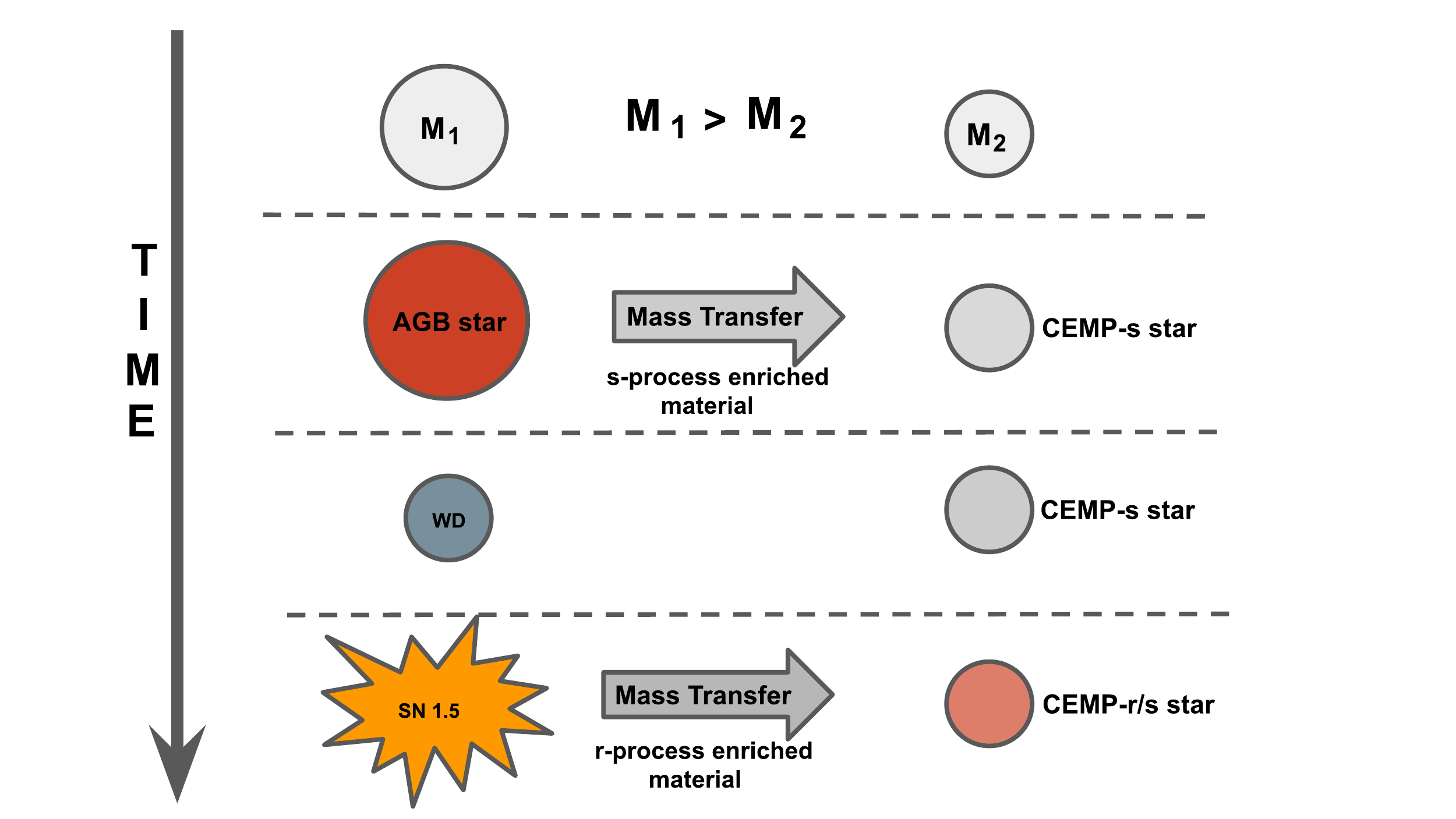}

        }%
        \subfigure[AGB and AIC pollution of a star in binary system]{%
            \label{fig:sixth}
                \def\svgwidth{\columnwidth}
                \scalebox{1.05}{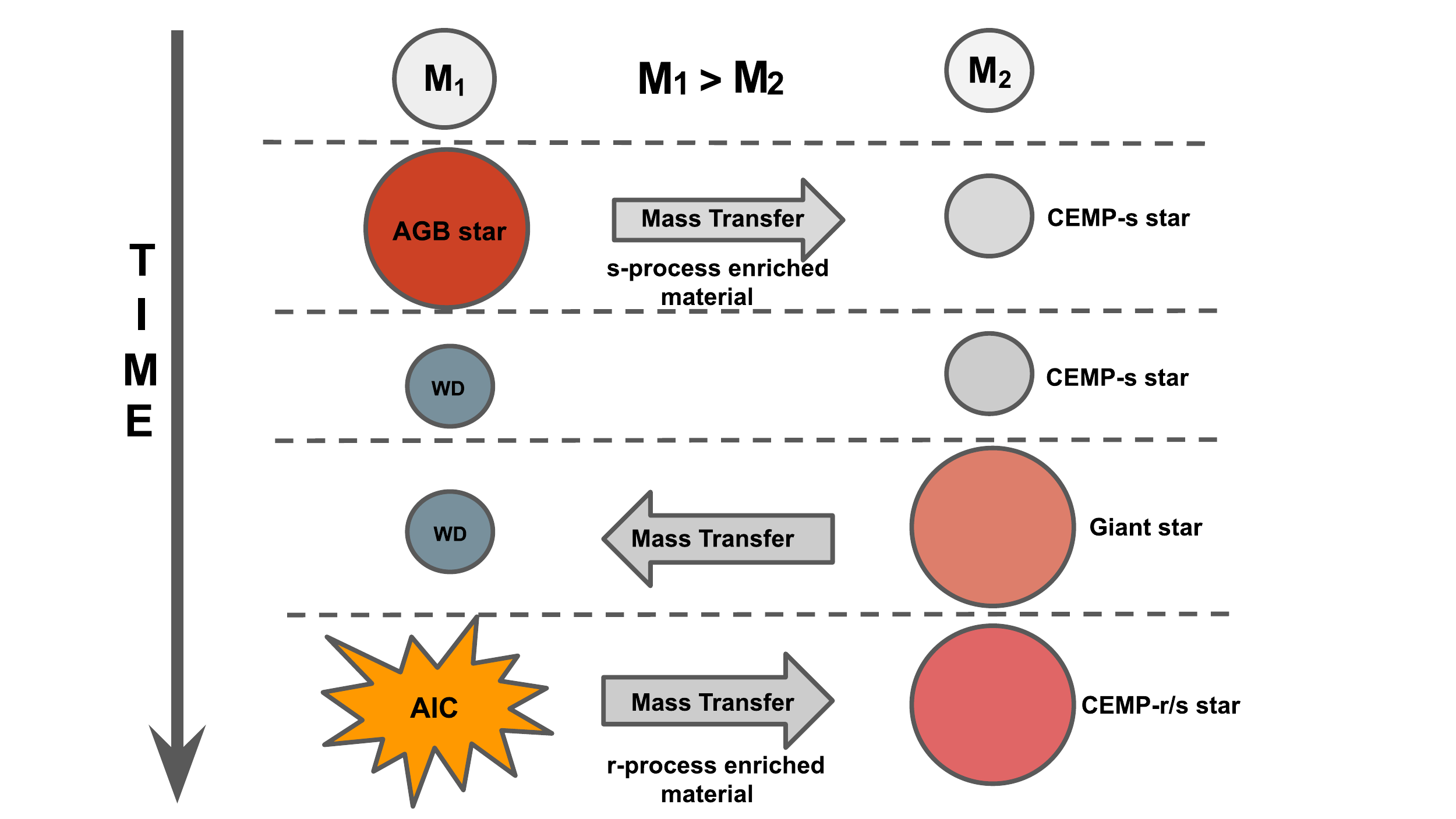}

        }\\

    \caption{Representation of different proposed formation scenarios of CEMP-s and CEMP-r/s stars. M$_{1}$, M$_{2}$, and M$_{3}$ represent the masses of the stars, where M$_{1}$~$>$~M$_{2}$~$>$~M$_{3}$. Here evolutions of single, binary, and triple star systems with different conditions are shown. The figures are discussed individually in Sect.~\ref{sec:formation_scenarios_rs}.
     }%
   \label{fig:subfigures}

       \end{center}
\end{figure*}

As discussed in Section~\ref{sec:introduction}, the widely accepted scenario to explain the enhancement of heavy elements exhibited by CEMP-s stars is that these objects are in  binary systems with a now invisible white dwarf companions (as shown in Figure~\ref{fig:first}). The observed enhancement of heavy elements in CD$-$27 14351 may  be attributed to a binary companion in such a binary system.
Various proposed formation scenarios that attempt to explain the unusual 
elemental abundance pattern of the CEMP-r/s stars are available in 
the literature. \citet{jonsell2006} discussed nine scenarios explaining their 
origin, and concluded that none of the explanations was satisfactory. 
\citet{lugaro2009} also discussed a few formation scenarios of this class 
of stars. In search of the most plausible formation mechanism, 
\citet{abate2016cemp-rs} calculated the frequency of CEMP-r/s stars among 
CEMP-s stars, considering different formation scenarios discussed by 
\citet{jonsell2006} and \citet{lugaro2009}, and compared that with the 
frequency of an observed sample of CEMP-r/s stars taken from the literature. 
A pictoral representation of some of these scenarios  relevant to the 
present study is presented in  Figure~\ref{fig:subfigures}.

{\bf{(i) Radiative levitation}}\\ 
This scenario is based on the fact that due to the large photon absorption cross sections of partially ionised heavy elements, they can be pushed outwards by radiative pressure that  causes the abundances of heavy elements in the atmospheres of hot stars (e.g. Przybylski's star) to appear much higher than the solar abundances. 
The observed abundance peculiarities of CEMP-r/s stars could also be thought of as a consequence of radiative levitation. However, as discussed in  \citet{cohen2003abundance}, \citet{jonsell2006}, and \citet{abate2016cemp-rs}, this scenario is   rejected as a  formation mechanism of CEMP-r/s stars. \citet{richard2002} and \citet{matrozis2016} have found from their simulations that radiative levitation is most effective in stars on the main sequence, and especially when they approach the turn-off as they have very thin convective envelopes. CEMP (-s \& -r/s) stars are generally observed to be  subgiants or giants and the effect of radiative levitation is found to be negligible in giants and 
low-temperature objects. All  four objects that we have found to be enhanced in both s- and r-process elements are  low-temperature (4160$-$4940 K) objects with log g values in the range  0.6 to 1.40 cgs units. Radiative levitation thus cannot be a  process responsible  for the observed overabundance of heavy elements in these stars. \\

{\bf{(ii) Self-pollution of a star formed from r-rich ISM}}\\
This scenario \citep{hill2000heavy, cohen2003abundance, jonsell2006} is shown in Figure~\ref{fig:second}. As discussed in  \citet{jonsell2006} and \citet{abate2016cemp-rs},  this hypothesis may be rejected as none of the CEMP-r/s stars observed to date  have been found to be in the evolutionary stage of AGB phase. The stars  we  studied are giants, and the  estimated low $^{12}$C/$^{13}$C values and the absence of Tc lines imply the extrinsic nature of the overabundance of carbon and heavy elements. This scenario thus cannot explain the abundance pattern observed in our programme  stars. \\

{\bf{(iii) SN and AGB pollution of a star in triple system}}\\
Figure~\ref{fig:fourth} shows this scenario \citep{cohen2003abundance,jonsell2006}. Both \cite{cohen2003abundance} and  \citet{jonsell2006} dismissed this scenario because it seems very unrealistic that the triple system survives such a nearby SN explosion for further mass transfer. \citet{abate2016cemp-rs} also rejected this hypothesis, failing to reproduce the observed frequency of CEMP-r/s stars. \\

{\bf{(iv) AGB and 1.5 SN pollution of a star in a binary system}}\\
This scenario is shown in Figure~\ref{fig:fifth} \citep{jonsell2006}. It was shown by \citet{zijlstra2004low} that at low metallicity, due to low mass-loss efficiency, the degenerate core of high-mass AGB star remains massive enough to reach the Chandrasekhar mass limit and explodes as a type 1.5 supernova \citep{iben1983}. However, type 1.5 supernova can disrupt the binary system by destroying the primary star \citep{nomoto1976, iben1983, lau2008}. As discussed in  \citet{abate2016cemp-rs},  this hypothesis is also dismissed  as most CEMP-r/s stars are found to be in binary systems \citep{lucatello2005}. \\
 \\
{\bf{(v) AGB and accretion-induced collapse (AIC) pollution of a star in a binary system}}\\
This scenario \citep{qian2003stellar, cohen2003abundance} is shown in Figure~\ref{fig:sixth}. The three phases of mass transfer seem problematic since the observed CEMP-r/s stars, in many cases, are found to be at the main-sequence turn-off making the accretion difficult \citep{lugaro2009}. As discussed in \citet{abate2016cemp-rs}, a narrow orbital separation is required for this scenario, so that even  after the first phase of mass transfer the stars stay close enough to fill the Roche-lobe for the next phase of mass transfer. However, taking this situation into account,  the observed frequency of CEMP-r/s stars could not be reproduced \citep{abate2016cemp-rs}. In addition, it is quite uncertain if this kind of collapse can produce r-process elements so as to match the observed abundance pattern of the CEMP-r/s stars \citep{qian1996, qian2003stellar}. \\

{\bf{(vi) Intermediate neutron-capture process (i-process)}}\\
In their simulations \citet{cowan1977} found that a significantly high 
neutron flux (higher than that of s-process) can be  produced by  mixing  
different amounts of hydrogen-rich material into the intershell region of 
AGB stars (also known as proton ingestion episodes or PIEs), leading to the 
occurrence of i-process nucleosynthesis in AGB stars.
This nucleosynthesis process operating at a neutron density 
(n $\sim$ 10$^{15}$ cm$^{-3}$), which is intermediate to that of s- and  
r-process neutron densities can produce both s- and r-process elements 
in a single stellar site \citep{Dardelet_2014, Hampel2016, Roederer_2016, 
Hampel_2019}. \citet{Hampel2016} successfully reproduced the abundance 
distribution of 20 CEMP-r/s stars with the help of an i-process model. 
Although it is     evident that the i-process can explain the 
abundance pattern seen in CEMP-r/s stars, the  astrophysical sites of the  
i-process are not clearly understood. A variety of sites have been proposed 
where PIEs can take place, favouring the conditions for i-process 
nucleosynthesis.

Figure~\ref{fig:first} shows the formation scenario of CEMP-r/s stars, which 
is similar to that suggested for CH, Ba, and CEMP-s stars. In this scenario 
these stars are secondary in binary systems, where the primary, which is slightly 
more massive, evolves to the AGB phase and transfers s-process rich elements to 
the secondary star. The only difference in the scenario for CEMP-r/s stars is 
that the primary produces both s- and r-process elements in the AGB phase 
with the help of i-process nucleosynthesis, and later pollutes the secondary 
with both s- and r-process elements. Recent simulations have shown that 
higher neutron densities of the order of 10$^{12-15}$ cm$^{-3}$ necessary 
for i-process are attained 
in very metal-poor AGB stars \citep{Campbell_&_Lattanzio_2008, 
Cristallo_et_al_2009, Campbell_et_al_2010, Stancliffe2011}. The 
reactions $^{13}$C($\alpha$, n)$^{16}$O and $^{22}$Ne($\alpha$, n)$^{25}$Mg 
are the two neutron sources proposed to operate in AGB stars.  The 
reaction $^{22}$Ne($\alpha$, n)$^{25}$Mg that  can produce a neutron density 
of $\sim$ 10$^{11-14}$ cm$^{-3}$  needs a temperature 
of $\sim$ 3$\times$10$^{8}$ K for activation, which is achieved only in 
intermediate-mass stars (M $\geq$ 3 M$\odot$) at the thermal pulse (TP) 
phase \citep{Lugaro2012}.  As this phase is very short (a few days), in spite 
of high neutron density, neutron exposure remains inefficient to produce the heavy-s process peak \citep{Fishlock_et_al_2014}. Again, the $^{13}$C($\alpha$, n)$^{16}$O neutron source gets activated at a   lower temperature ($\sim$ 1$\times$10$^8$ K), and hence can operate in the interpulse phase of both low- and intermediate-mass AGB stars. Although this reaction can produce a neutron density of $\sim$ 10$^{7}$ cm$^{-3}$, the longer timescale ($\sim$10$^5$ years) of the interpulse phase enables sufficient neutron exposure to produce the heavy-s peak. However, some authors \citep{masseron2010aholistic, jorissen2016HE0017} argue in favour of the $^{22}$Ne($\alpha$, n)$^{25}$Mg reaction as the primary  neutron source responsible for the production of the abundance peculiarity observed in CEMP-r/s stars. The isotopic ratio of Mg could provide sufficient clues about the neutron source. The operation of $^{22}$Ne($\alpha$, n)$^{25}$Mg reaction produces a large amount of $^{25}$Mg, which in turn produces $^{26}$Mg by neutron-capture. So, we can expect the $^{24}$Mg~:~$^{25}$Mg~:~$^{26}$Mg ratio to change from the terrestrial (79 : 10 : 11) value to a ratio with higher abundances of $^{25}$Mg and $^{26}$Mg \citep{Scalo1978a}. However, \citet{Zamora2004} attempted to determine Mg isotopic ratios from MgH bands in cool carbon stars and  concluded that at optical wavelengths it is not possible to derive the Mg isotopic ratios as the synthetic spectra are insensitive to variations of Mg isotopic ratios due to the presence of strong C$_{2}$ and CN molecular bands. Thus, the scope of finding a spectroscopic test to identify the neutron source responsible for the production of heavy elements in CEMP-r/s stars  still remains open.

Another site proposed for the i-process nucleosynthesis, supporting binary mass-transfer scenario for the formation of CEMP-r/s stars, is  the very late thermal pulse (VLTP) in post-AGB stars \citep{Herwig_et_al_2011}.  The abundance pattern of Sakurai's object (V4334 Sagittarii), a born-again giant, could not be reproduced by s-process yields. This object shows an overabundance of Rb, Sr, and  Y that is  two orders of magnitude higher than that of the second s-process peak \citep{Asplund_et_al_1999}. Assuming proton ingestion into the He-shell convection zone, \citet{Herwig_et_al_2011} carried out a 3D hydrodynamic simulation of a VLTP in a pre-white dwarf. They achieved a significantly high neutron density (~10$^{15}$ cm$^{-3}$) and could successfully reproduce the abundance distribution of Sakurai's object. The post-AGB stars in the Large and Small Magellanic Clouds (LMC and SMC, respectively) are found to exhibit an unusual abundance pattern \citep{van_Winckel_2003}. \citet{Lugaro_et_al_2015} illustrated that the s-process cannot explain the abundance pattern of these stars and proposed that i-process might explain better the abundances of the heavy elements along with the measured low abundance of Pb. Later, \citet{Hampel_2019} could satisfactorily fit the abundance patterns observed in seven Pb-poor post-AGB stars (including the post-AGB stars of LMC and SMC) with i-process models.

Simulations of super-AGB TP stars showed that proton ingestion into the 
convective He-burning shell is favoured following the production of 
i-process neutron density \citep{Doherty_et_al_2015, Jones_et_al_2016}. 
Metal-poor massive stars (20 $-$ 30 M$_{\odot}$) are also considered 
as i-process sites \citep{Bannerjee_et_al_2018,Clarkson_et_al_2018}. A 
single CEMP-r/s star can form from the ISM contaminated by the i-process 
material ejected by metal-poor massive stars \citep{Bannerjee_et_al_2018}.

One more proposed site for i-process nucleosynthesis is rapidly accreting 
white dwarf (RAWDs) \citep{Denissenkov_et_al_2017, Cote_et_al_2018, 
Denissenkov_et_al_2019}. \citet{Denissenkov_et_al_2019} proposed that 
single CEMP-r/s stars could be the tertiary (with a wider orbit) in a 
triple star system, where it orbits a close binary with a RAWD. Later the tertiary escapes from the triple star system being polluted by i-process material from the RAWD when the RAWD explodes as a SNIa. Due to the requirement of a particular sequence of events, population synthesis calculations can only decide the probability of the formation of CEMP-r/s stars from RAWDs \citep{Hampel_2019}.

\citet{Roederer_2016} reported a metal-poor star HD~94028 with underabundance of carbon ([C/Fe] = -0.06), and low Ba and Eu abundances. With the help of high-quality NUV spectra, they could estimate the abundances or upper limits of 64 species of 56 elements including the species whose features are seen mostly in the NUV. \citet{Roederer_2016} found that the star exhibits a supersolar [As/Ge] ratio, a solar [Se/As] ratio, and enhanced abundances of Mo and Ru. They could not reproduce this elemental pattern with any combination of s and r-process, but an additional contribution from the i-process could fit the peculiar pattern. The contribution of i-process has also been observed in pre-solar grains in pristine meteorites \citep{Fujiya_et_al_2013, Liu_et_al_2014}. These observations indicate more than one astrophysical sites for the i-process.

{\bf{(vii) AGB pollution of a star in binary system formed from r-rich ISM}}\\ 
This scenario \citep{hill2000heavy,cohen2003abundance,jonsell2006,ivans2005near,bisterzo2011s} is illustrated  in Figure~\ref{fig:third}. Although
\citet{bisterzo2011s,bisterzo2012} claim to reproduce, within the error bars,  the observed [hs/ls] (which is higher in CEMP-r/s stars than CEMP-s stars) considering the binary system formed from the r-rich molecular cloud, there are several arguments that stand against this scenario.
It was noted that in the case of independent s- and r-process enrichment, the correlation of the abundances of Ba and Eu cannot be reproduced by the AGB models \citep{abate2016cemp-rs}, and  also that this scenario cannot explain the large fraction of CEMP-r/s stars among the CEMP-s stars \citep{jonsell2006, lugaro2009}.

As the abundance patterns of most of the CEMP-r/s stars found in the 
literature can be explained with i-process models, some authors prefer the nomenclature `CEMP-i' to  `CEMP-r/s' \citep{Hampel2016, Frebel_review_2018, 
Hampel_2019} and calling the stars formed by AGB pollution in the binary system formed 
from r-rich ISM `CEMP-r+s stars' \citep{Gull_et_al_2018, 
Frebel_review_2018}.

\citet{Gull_et_al_2018} reported a red giant CEMP star 
RAVE J094921.8$-$161722 ([Fe/H] = $-$2.2, [C/Fe] = 1.35) with a surprising 
elemental abundance pattern. The star exhibits an enhanced abundance of Pb, 
indicating s-process contribution and Th, which is produced in r-process 
nucleosynthesis. \citet{Gull_et_al_2018} claimed that this object was the 
first bona fide CEMP-r+s star as its abundance pattern could be satisfactorily 
fitted only with AGB mass-transfer model taking into account initial r-process 
enhancement. \citet{Sbordone_et_al.2020} reported the abundance analysis of 
 object GIU J190734.24-315102.1 located in the Sagittarius (Sgr) dwarf 
spheroidal (dSph) galaxy, and claimed that the abundance pattern could be 
best fit only with a model considering AGB-pollution in a binary 
system pre-enriched with a neutron star--neutron star merger event. 
 \citet{Sbordone_et_al.2020}  classify this object  as the first 
 CEMP-r/s  star found in the Sgr dSph, but due to its formation scenario 
 this object may be referred to as a bona fide  CEMP-r+s star.

 The expected rate 
 of occurrence of CEMP-r+s stars among metal-poor stars is 
 2\% $-$ 3\% \citep{Gull_et_al_2018}. With  an occurrence rate 
 of 3\%, about two dozen r-II stars have been found to date \citep{Gull_et_al_2018}, although why  no other bona fide CEMP-r+s 
 stars have yet been detected  remains a puzzle \citep{Frebel_review_2018}.

\subsection{Comparison of the observed abundances of the programme stars with i-process model predictions}
\label{sec:i_process}
\citet{Hampel2016} calculated yields of heavy elements considering different constant neutron densities ranging from $10^{7}-10^{15}$ cm$^{-3}$ using single-zone nuclear network calculations. The nucleosynthesis is assumed to occur in the intershell region of AGB stars. The physical input parameters, such as temperature and density for the intershell region of a low-mass (1 M$_{\odot}$) low-metallicity (z = 10$^{-4}$) AGB star, are adapted from \citet{Stancliffe2011}. The constituents of the intershell region have been adapted from that of \citet{Abate2015_carbon-enhanced}. The temperature and density of the intershell region are considered to be $1.5 \times 10^{8}$ K and $\rho$ = 1600 gcm$^{-3}$, respectively. However, no significant changes in the results were   found when tested with a range of temperatures ($1 \times 10^{8}$  to $2.2 \times 10^{8}$ K) and densities ($\rho$ = 800 gcm$^{-3}$ to 3200 gcm$^{-3}$).  The run times of the models are adjusted in such a way that a high neutron exposure ($\tau$ $\sim$ 495 mb$^{-1}$) is ensured. Equilibrium abundance pattern between  heavy elements and the seed nuclei is established based on  such a high neutron exposure, and the element-to-element ratio becomes a function of constant neutron density. After the exposure the neutron flux is kept switched off for t = 10 Myr.

 During the time when neutron flux is switched on, it should be  noted that for lower neutron densities the elemental abundance pattern of typical s-process is produced with stable ls (Sr, Y, Zr) and hs (Ba, La, Ce) peaks. However,  with i-process neutron densities (n = $10^{12}-10^{15}$ cm$^{-3}$), the neutron-capture path goes further away from the valley of stability, making both ls and hs peak shift to lighter elements; in particular,  they form a peak at $^{135}$I. Then, after the neutron exposure is switched off, it is found that the decay of unstable isotopes produce stable ls and hs peak elements, for example  $^{135}$I decays to produce $^{135}$Ba. Abundances of Ba and Eu are found to increase with neutron density. This is how the i-process can modify the abundance pattern of neutron-capture elements. 
 
 We  used  the  model yields ([X/Fe]) of \citet{Hampel2016}, with neutron densities ranging from $10^{9}-10^{15}$ cm$^{-3}$, and  compared them with the observed abundances of our programme stars.   In order to find the neutron density responsible for the observed abundance distribution of the  programme stars, we  followed  the  procedure given in  \citet{Hampel2016}. We  used the equation 
\begin{equation}
X=X_{i}(1-d)+X_{\odot}d
,\end{equation}
where $X_{i}$ is the model yield, $X_{\odot}$ is the solar-scaled abundance, and $d$ is a dilution factor. 

\begin{figure}
        \centering
        \includegraphics[height=9cm,width=9cm]{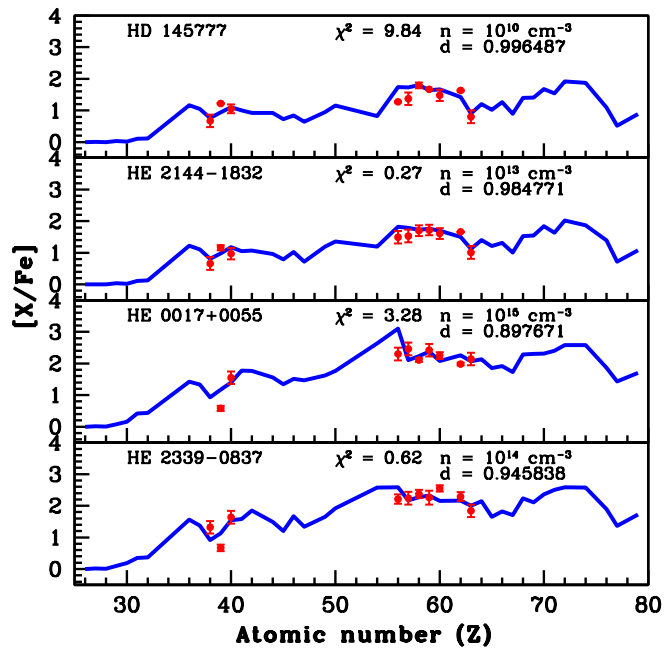}
        \caption{ Best-fitting i-process model (solid blue curve) for the stars. The points with error bars indicate the observed abundances.}
                \label{fig:iprocess}
\end{figure}

Figure~\ref{fig:iprocess} shows the best-fit models with appropriate neutron densities and corresponding dilution factors. It is seen that the i-process model with neutron densities of n $\sim$ $10^{13}$ cm$^{-3}$, $10^{15}$ cm$^{-3}$, and $10^{14}$ cm$^{-3}$ closely fit the observed abundances of HE~2144$-$1832, HE~0017$+$0055, and HE~2339$-$0837, respectively. The best model fit  for HD~145777 is found at a neutron density of n $\sim$ $10^{10}$ cm$^{-3}$.

\subsection{Classification of the programme stars}
\label{sec:classifiers}

As discussed in  Section~\ref{sec:introduction}, different authors \citep{beers2005discovery, jonsell2006, masseron2010aholistic, abate2016cemp-rs, Frebel_review_2018,hansen2019abundances} have used different criteria to 
 classify the CEMP stars into various subclasses.  
Four objects in our sample, HD~145777, HE~0017+0055, HE~2144$-$1832, and HE~2339$-$0837,   show overabundances of Ba and Eu along with other light-s and heavy-s elements. With 0~$<$~[Ba/Eu]~$<$~0.5,  all  four objects fall in the category of CEMP-r/s stars if the classification criteria of  \citet{beers2005discovery} for CEMP-r/s stars is followed.  But, according to the classification scheme of \citet{abate2016cemp-rs}, CEMP-r/s stars are those that have   [Ba/Fe] and [Eu/Fe] values that are  greater than unity. This criterion classifies HE~2144$-$1832 ([Ba/Fe] = 1.49 and [Eu/Fe] = 1.01), HE~0017$+$0055 ( [Ba/Fe] = 2.30 and  [Eu/Fe] = 2.14), and HE~2339$-$0837 ([Ba/Fe] = 2.21 and [Eu/Fe] = 1.84) as CEMP-r/s stars, and HD~145777 (with [Ba/Fe] = 1.27 and [Eu/Fe] = 0.80) as a CEMP-s star. All  four of these objects fall in the category of CEMP-r/s stars if we use the criteria 0.0~$<$~[La/Eu]~$<$~0.6 given by \citet{Frebel_review_2018} for the CEMP-i subclass (we use the `CEMP-r/s' nomenclature). We could not estimate the abundances of Hf, Ir, and Pb, hence the other criteria, namely [Hf/Ir] $\sim$ 1.0 for the CEMP-i subclass and [Ba/Pb] $>$ $-$1.5 for the CEMP-s subclass put forward by \citet{Frebel_review_2018}, could not be used. However, the lower limit given on [Ba/Eu] ($>$ 0.5) by \citet{Frebel_review_2018} for the CEMP-s subclass clearly indicates that these four stars cannot be classified as CEMP-s stars.  If we use  [Sr/Ba] as a classifier,  as discussed in \citet{hansen2019abundances}  with values  [Sr/Ba] $\sim$ $-$0.60, $-$0.83, and $-$0.89 respectively for   HD~145777, HE~2144$-$1832, and HE~2339$-$0837, they fall  in the category of CEMP-r/s stars. As the abundance of Sr could not be estimated for the object HE~0017+0055,     [Sr/Ba]  could not be used to classify this object. 

The object CD$-$27~14351 is found to satisfy the criteria for the CEMP-s stars of all    four classification schemes, hence this object with  [Ba/Fe] = 1.82, [Eu/Fe]~$<$~0.39, [Ba/Eu]~$>$~1.43, [La/Eu]~$>$~1.17, and [Sr/Ba] = $-$0.08  is classified as a  CEMP-s star.
\begin{figure*}
     \begin{center}
\centering
        {%
                 \label{fig:hs_histogram}
            \includegraphics[height=6.5cm,width=7cm]{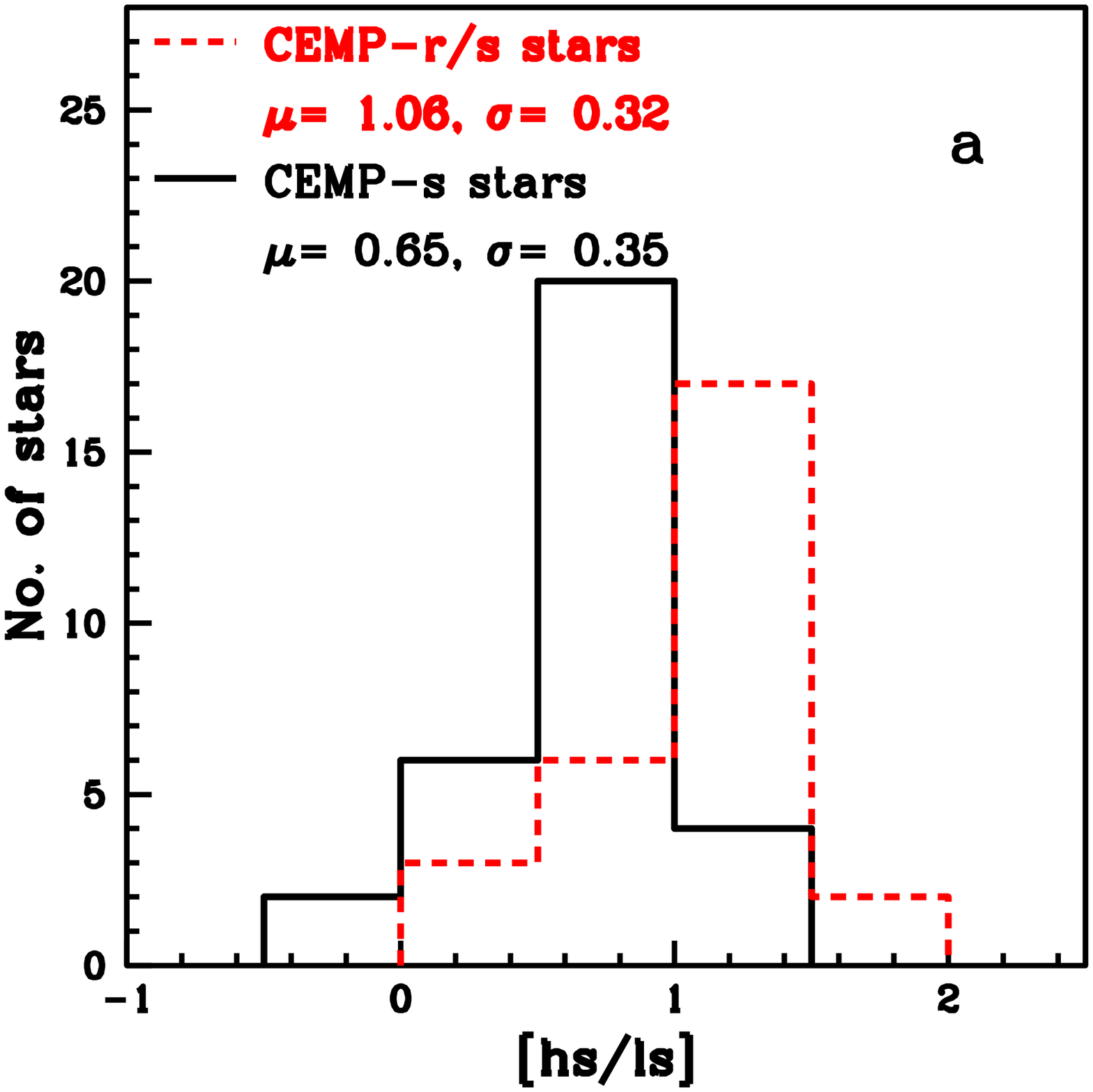}
        }%
        {%
      \label{fig:hs}
            \includegraphics[height=6.5cm,width=7cm]{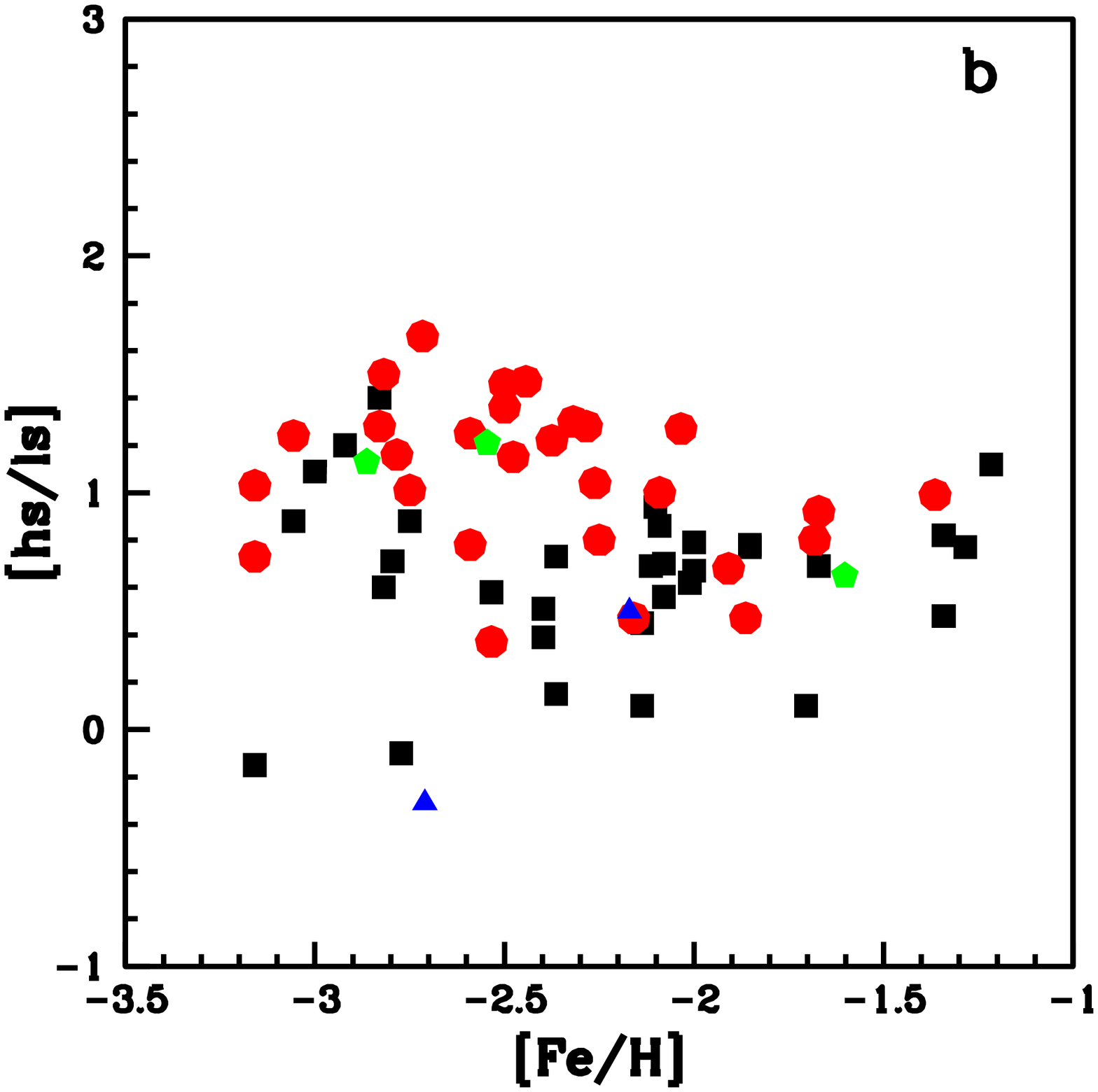}
        }
        
        {%
            \label{fig:hseu}
            \includegraphics[height=6.5cm,width=7cm]{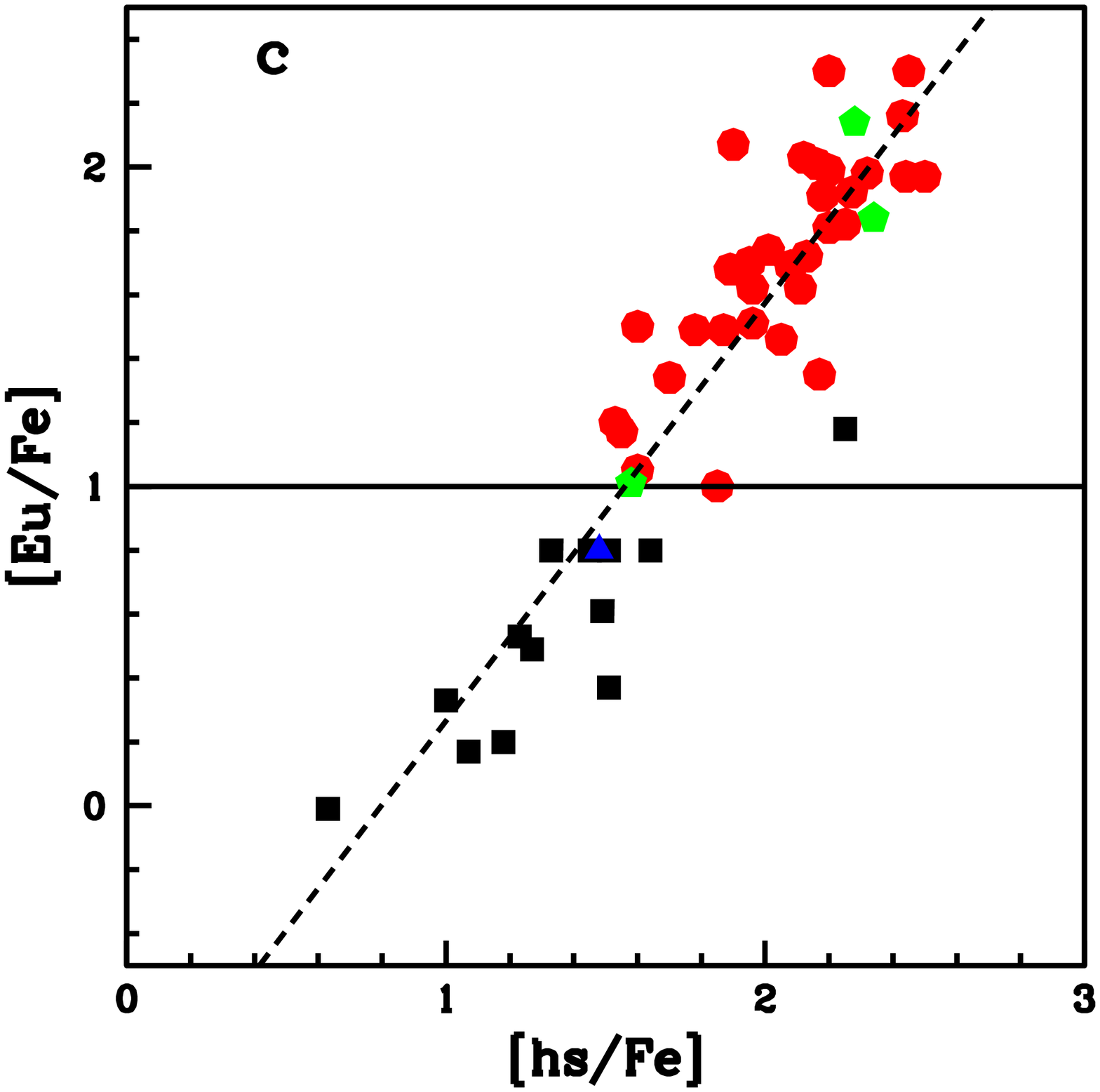}
        }%
        {%
            \label{fig:hsls}
            \includegraphics[height=6.5cm,width=7cm]{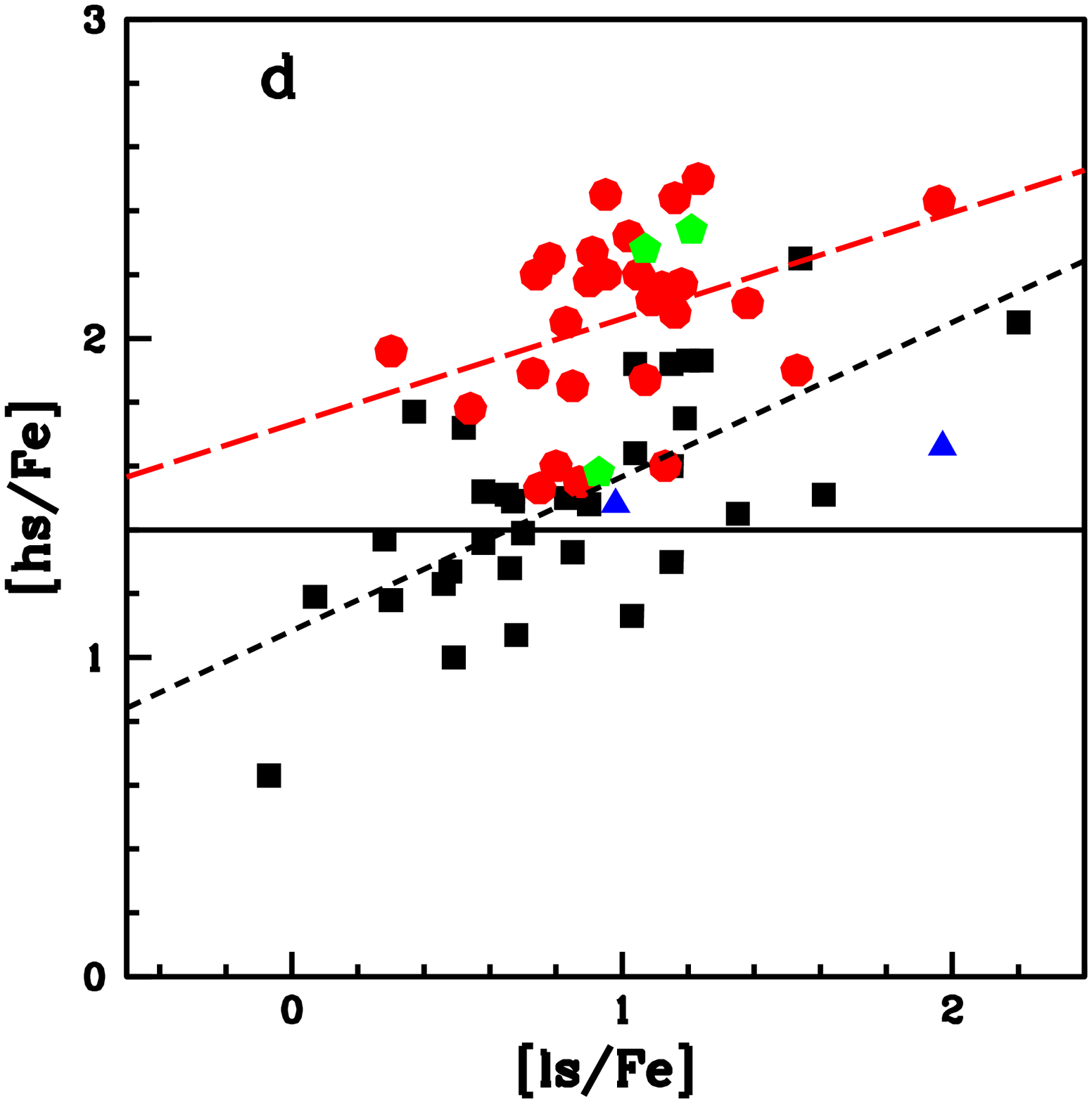}
        }
    \caption{Probing [hs/ls] as a classifier of CEMP-s and CEMP-r/s stars. The filled red circles represent CEMP-r/s stars, filled black squares represent CEMP-s stars, and filled blue triangles and filled green pentagons respectively represent CEMP-s and CEMP-r/s stars in this work. In panel (a) $\mu$ and $\sigma$ represent the mean and standard deviation of [hs/ls] respectively for CEMP-s and CEMP-r/s stars. In panel (c) the short-dashed black line  represents the correlation between [hs/Fe] and [Eu/Fe] for CEMP-s and CEMP-r/s stars. In panel (d) the short-dashed black line and long-dashed red line represent the correlation between [hs/Fe] and [ls/Fe] for CEMP-s and CEMP-r/s stars, respectively.
     }%
   \label{fig:subfigures_hs}
       \end{center}
\end{figure*}

\begin{figure*}
     \begin{center}
\centering
        {%
            \label{fig:srba}
            \includegraphics[height=6.5cm,width=7cm]{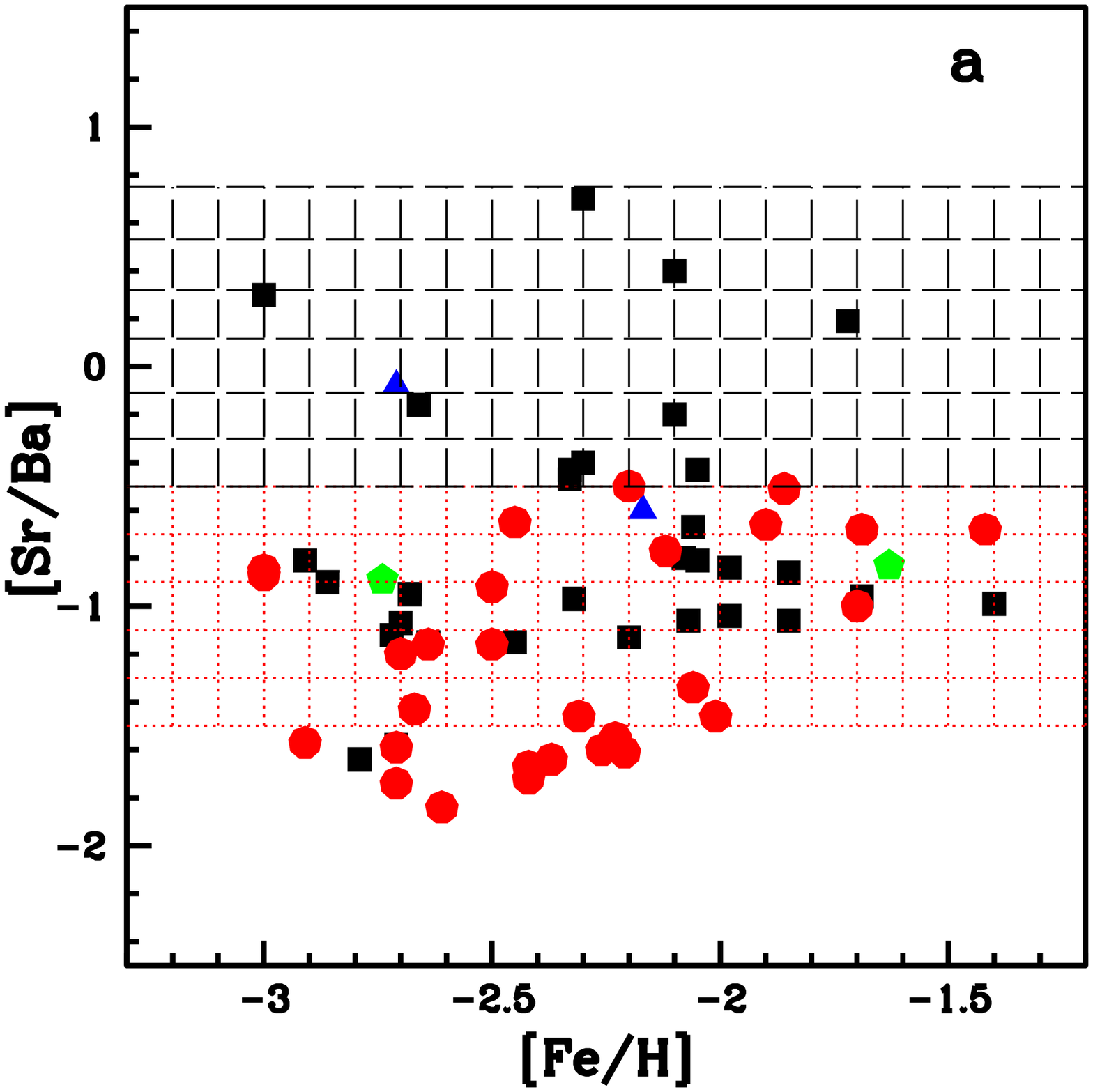}
        }%
        {%
            \label{fig:srba_eu}
            \includegraphics[height=6.5cm,width=7cm]{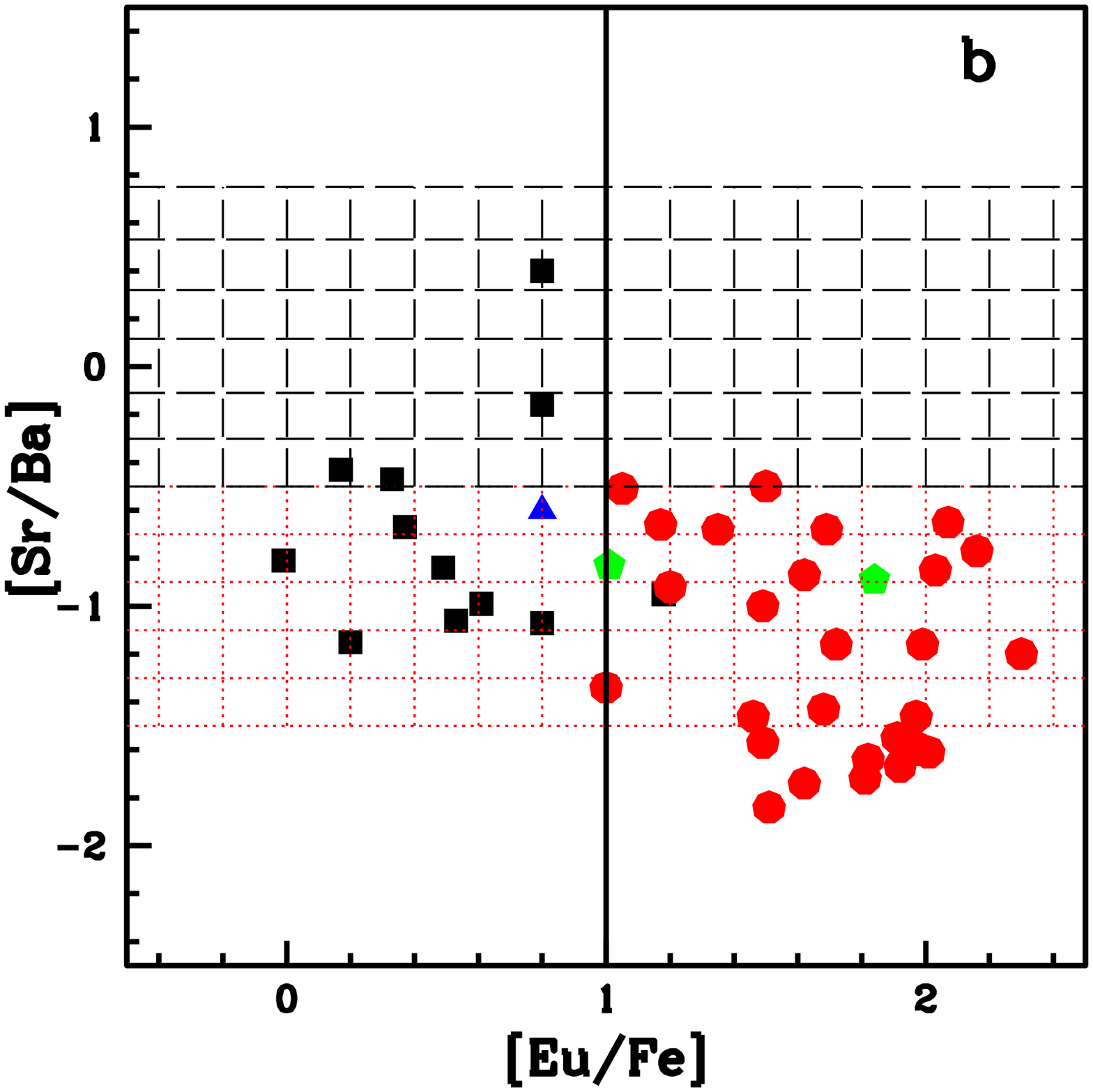}
        }
    \caption{[Sr/Ba] as a classifier of CEMP-s and CEMP-r/s stars. The filled red circles represent CEMP-r/s stars, filled black squares represent CEMP-s stars, and filled blue triangles and filled green pentagons respectively represent CEMP-s and CEMP-r/s stars in this work. The grid formed by the dashed black lines represents the region of CEMP-s stars, and the grid formed by the dotted red lines represents the region of CEMP-r/s stars put forward by \citet{hansen2019abundances}. In panel (b) the solid black line at [Eu/Fe] = 1.0 separates the CEMP-s and CEMP-r/s stars according to the classification criteria adopted by \citet{abate2016cemp-rs}.
     }%
   \label{fig:subfigures_srba}
       \end{center}

\end{figure*}

\begin{figure*}
     \begin{center}
\centering
        {%
                 \label{fig:baeu}
            \includegraphics[height=6.5cm,width=7cm]{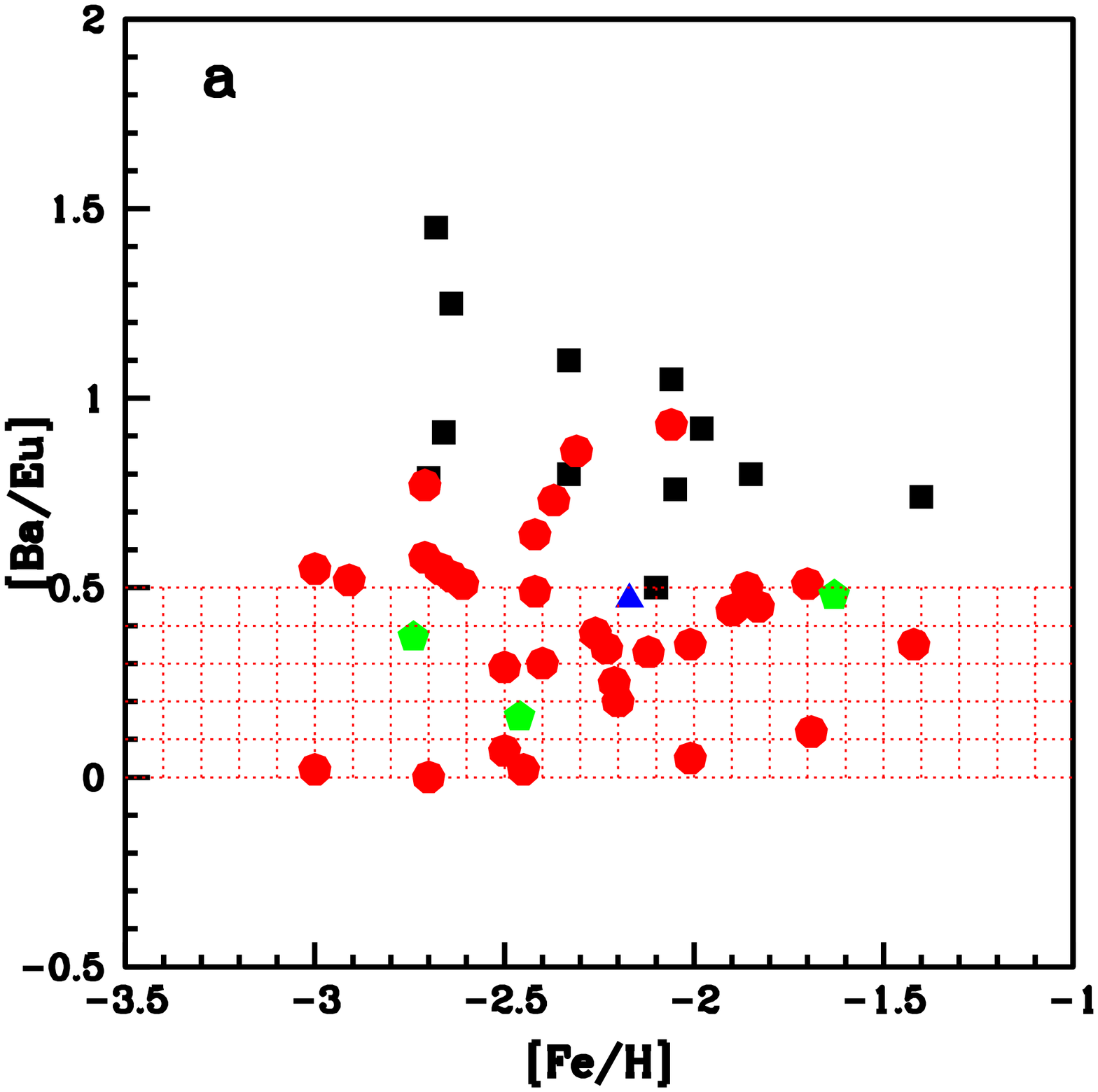}
        }%
        {%
      \label{fig:baeu_eu}
            \includegraphics[height=6.5cm,width=7cm]{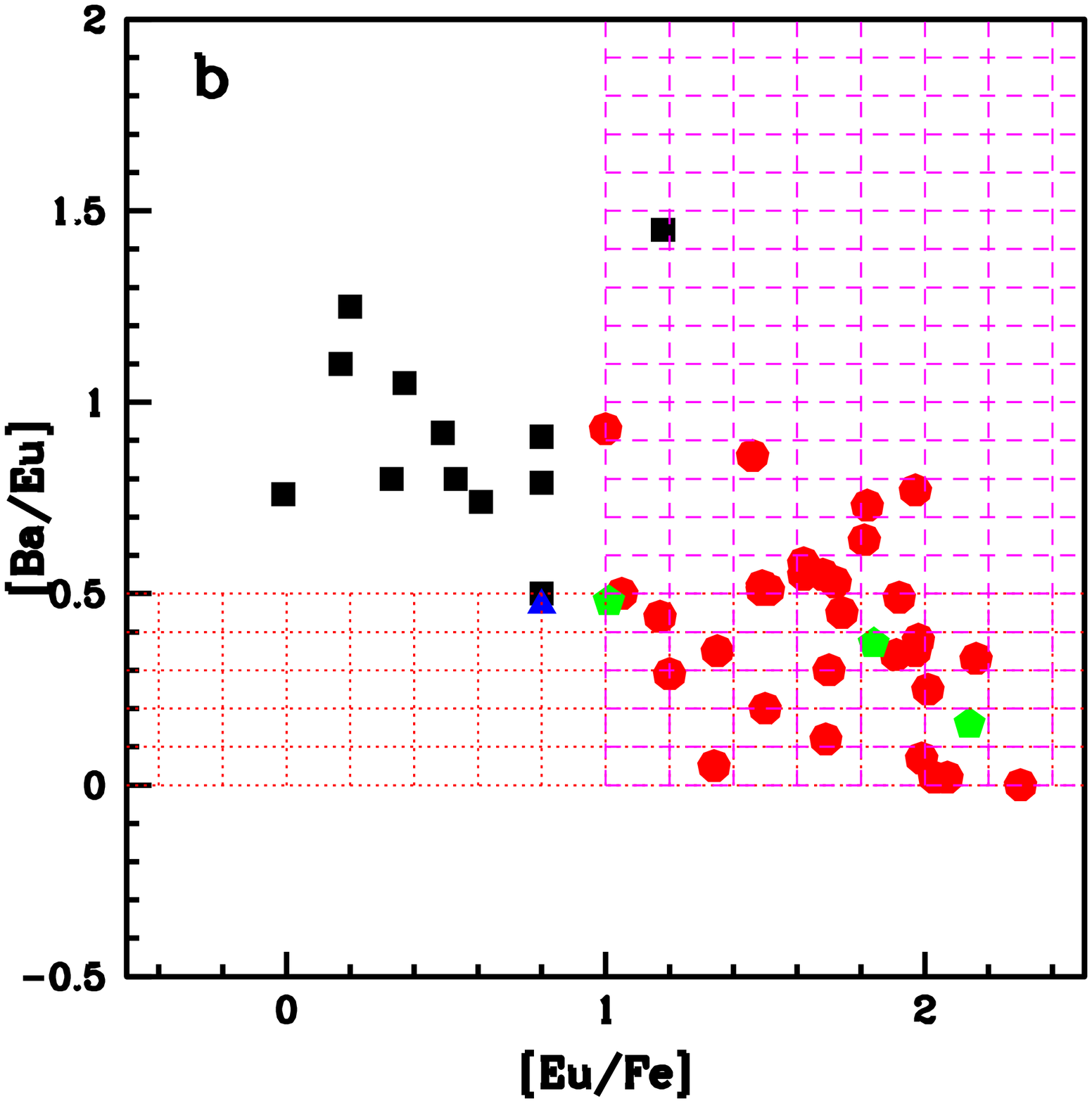}

        }
    \caption{Same as Fig.~\ref{fig:subfigures_srba}, but for [Ba/Eu]. The filled red circles represent CEMP-r/s stars, filled black squares represent CEMP-s stars, and filled blue triangles and filled green pentagons respectively represent CEMP-s and CEMP-r/s stars in this work. The grid formed by the dotted red lines represents the region of CEMP-r/s stars put forward by \citet{beers2005discovery}. The grid formed by the dashed magenta lines represents the region of CEMP-r/s stars put forward by \citet{abate2016cemp-rs}.
     }%
   \label{fig:subfigures_baeu}
       \end{center}
\end{figure*}

\begin{figure*}
     \begin{center}
\centering
        {%
      \label{fig:laeu}
            \includegraphics[height=6.5cm,width=7cm]{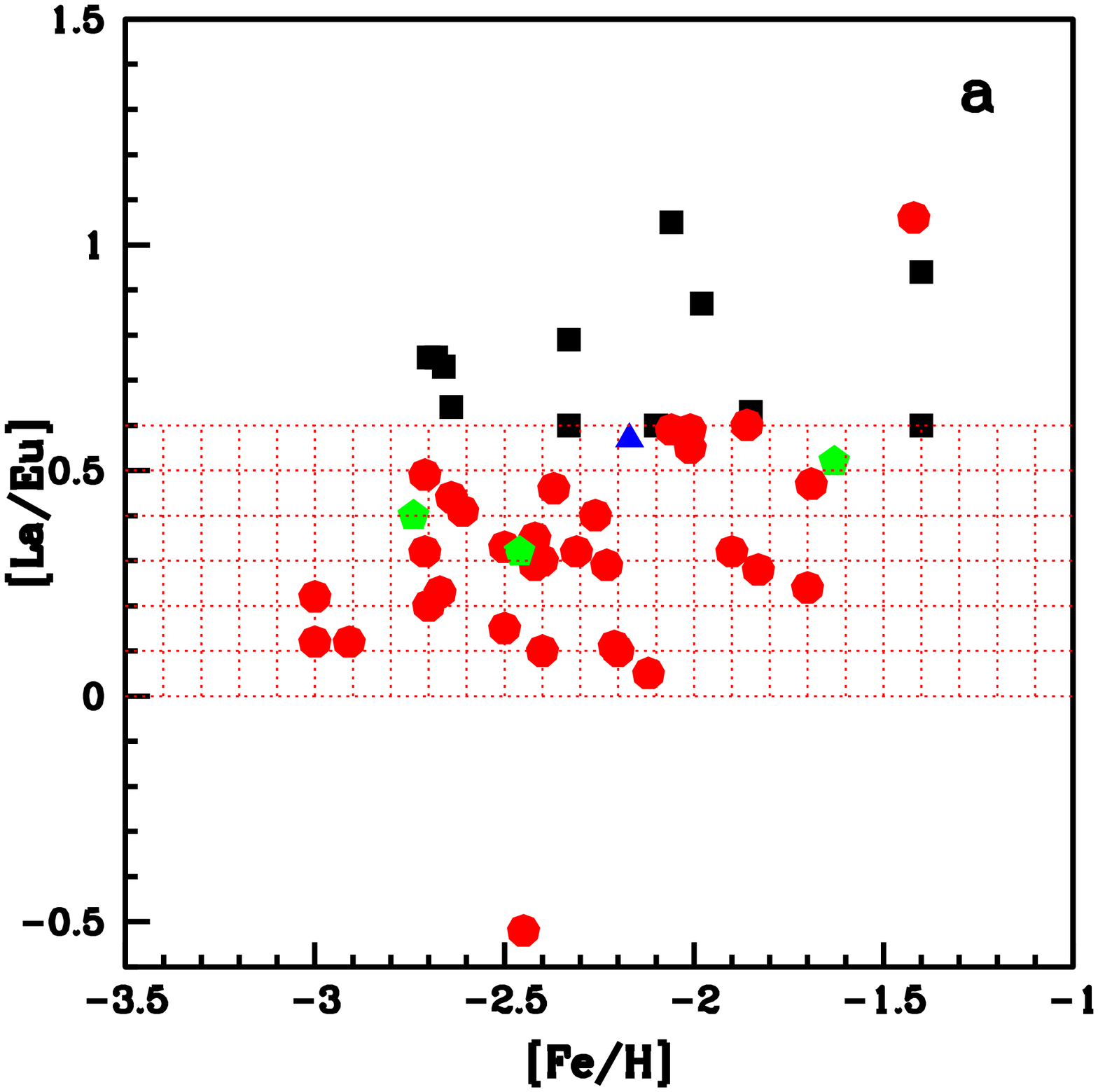}
        }%
        {%
      \label{fig:laeubaeu}
            \includegraphics[height=6.5cm,width=7cm]{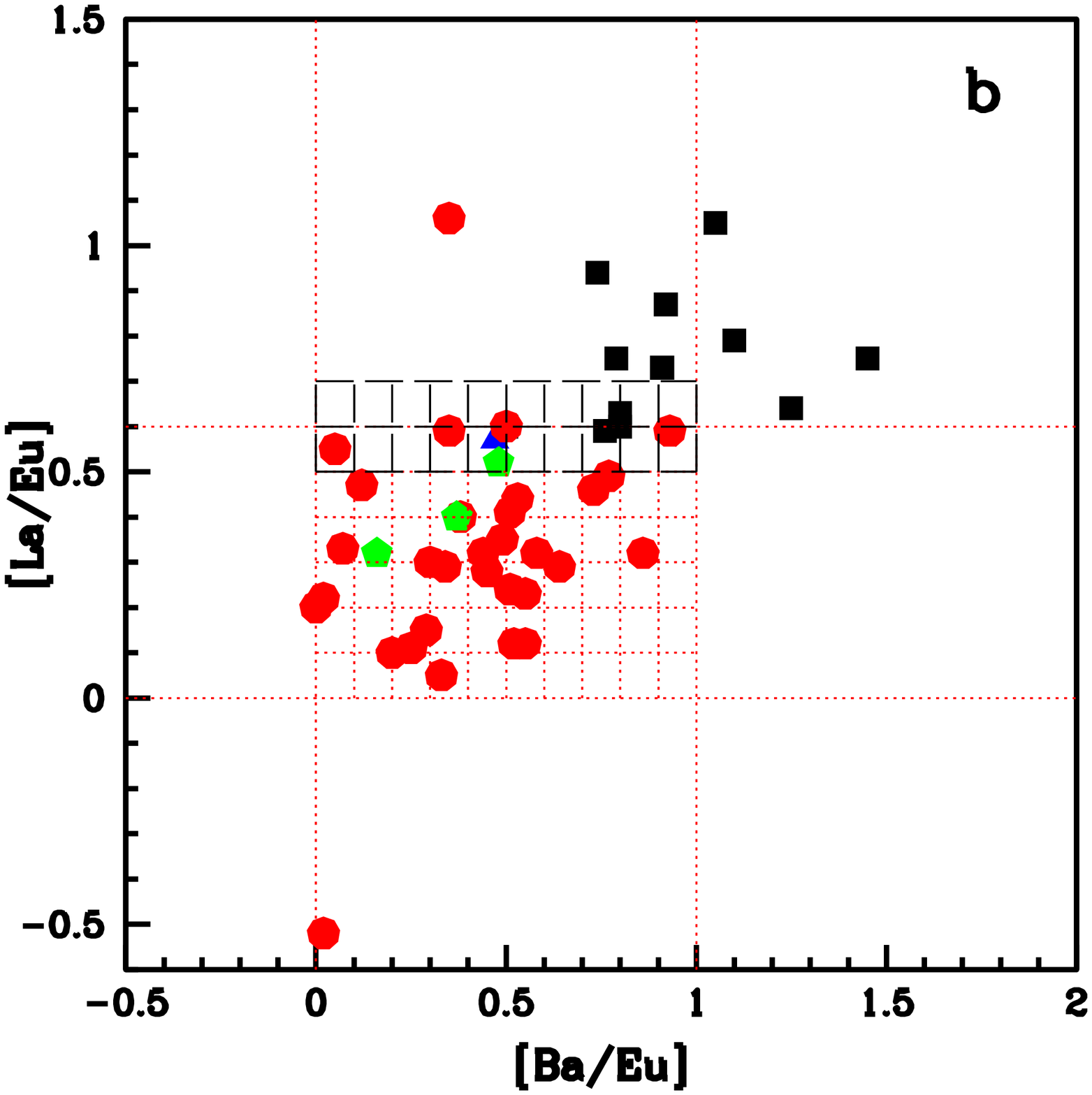}

        }
    \caption{Same as Fig.~\ref{fig:subfigures_srba}, but for [La/Eu]. The filled red circles represent CEMP-r/s stars, filled black squares represent CEMP-s stars, and filled blue triangles and filled green pentagons respectively represent CEMP-s and CEMP-r/s stars in this work. In panel (a) the grid formed by the dotted red lines represents the region of CEMP-r/s stars put forward by \citet{Frebel_review_2018}. In panel (b) we can see that   all the CEMP-r/s stars but two fall inside the grid formed by the dotted red lines bound by 0.0 $<$ [La/Eu] $<$ 0.6 and 0.0 $<$ [Ba/Eu] $<$ 1.0. The grid formed by the black dashed lines bound by 0.5 $<$ [La/Eu] $<$ 0.7 represents the uncertain region in the upper boundary of the classification put forward by \citet{Frebel_review_2018} for CEMP-r/s stars.
     }%
   \label{fig:subfigures_laeu}
       \end{center}
\end{figure*}

\subsubsection{Probing [hs/ls] as a classifier}

It has been noted from various studies in the past
 \citep{bisterzo2011s, bisterzo2012, Abate2015_carbon-enhanced, 
 abate2016cemp-rs, Hampel2016}  that [hs/ls] exhibit higher values for 
 CEMP-r/s stars than that of CEMP-s stars. In this section we  explore whether [hs/ls] can be used to distinguish CEMP-r/s and  CEMP-s stars, and hence if
 this ratio can be used as a classifier.

To accomplish  this we  carried out a literature survey and compiled  
data for 40  CEMP-s and 32 CEMP-r/s stars, and 
calculated [hs/ls], [Sr/Ba], and  [Ba/Eu] for these stars on a 
homogeneous scale. We  adopt the CEMP classifications of the literature 
objects given by the original studies and without any ambiguity. However, for 
some objects different  groups  have provided different atmospheric parameters and elemental 
abundances, and  have assigned different classes to them. For instance, 
\citet{allen2012elemental} considered HE~0336+0113 to be a CEMP-s star, but 
\citet{masseron2010aholistic} considered it   a CEMP-r/s star. HE~0336+0113 exhibits 
[Eu/Fe]~$>$~1.0 and shows the highest value of [Ba/Eu] (= 1.45) among the CEMP stars \citep{cohen2006carbon}. 
\citet{Hampel_2019}, after comparing the object's observed abundances with i-process 
models, reported that the characteristic properties of HE~0336+0113 are more of 
s-process than i-process. This analysis prompted us to classify HE~0336+0113   as a 
CEMP-s star. \citet{bisterzo2011s, bisterzo2012} studied most of the CEMP-s stars 
in Table~\ref{tab:lit1} and explained their abundance peculiarities with the help 
of AGB models with initial masses of 1.3 $-$ 2.0 M$\odot$. Except for SDSS~J1349$-$0229, 
which is classified as a CEMP-r/s star \citep{behara2010three}, all other CEMP-r/s
stars in Table~\ref{tab:lit1} have been examined with i-process model predictions 
and their abundance patterns are well reproduced with i-process model yields 
\citep{Hampel2016, Hampel_2019, Goswami_Goswami_1_2020}.

The elemental abundances 
of the objects reported by different authors differ from each other. As different 
authors use different codes, model atmospheres, and line lists 
for their abundance analyses, it is not surprising to see the differences in their 
results; however,  it is not an  easy task to select one analysis over the 
others. Thus, for   the objects for which the elemental abundance 
results from several different groups are available,   we 
 take the mean value of [Fe/H], and similarly the mean abundance for 
each element. For these  cases, we   use the mean [X/Fe] to 
calculate [hs/ls], [Sr/Ba], and [Ba/Eu]. To calculate [ls/Fe], we   
take the mean of the abundances of light-s elements (Sr, Y, Zr);  
similarly, the abundances of heavy-s elements (Ba, La, Ce, Nd) are averaged 
out to calculate [hs/Fe]. We do not consider the  two second peak s-process 
elements Pr and Sm, as r-process contributes  more than s-process to the 
isotopic abundances of these elements \citep{arlandini1999}. 
\citet{bisterzo2012} noted that light-s elements when plotted against 
each other show large scatter. Heavy-s elements also show 
scatter for some objects, although not as large. In order to reduce the systematic error, we   
calculated [ls/Fe] and [hs/Fe] only for those stars for which abundance 
estimates are available at least for two light s-process and two
heavy s-process elements.  To calculate [ls/Fe] and [hs/Fe], we also  
excluded those elements  for which only the upper limit of abundance 
is mentioned in the literature.  Table~\ref{tab:lit1} presents the list of 
objects and their elemental abundance ratios of C, N, and heavy elements 
used for this analysis. Our estimates of [hs/ls] are also presented in 
this table. Estimates of [hs/ls], when plotted with respect to [Fe/H], show 
a large scatter (Figure~\ref{fig:subfigures_hs}(b)). The use of different 
sets of elemental abundances instead of using a consistent set of elements 
for hs and ls abundances may also contribute to this observed scatter. 
Nevertheless, it is not always possible to have the same set of hs and ls 
elements for all stars, and using a consistent set of elements will 
tremendously affect the number of stars used to carry out the test.
As metallicity decreases, the neutron-to-iron seed ratio increases and hence heavier elements are synthesised to a greater extent (i.e. [hs/ls] increases as metallicity decreases; \citealt{gallino1998evolution}). In Figure~\ref{fig:subfigures_hs}(b) we can clearly see this trend for both CEMP-s (black filled squares) and CEMP-r/s (red filled circles) stars. However, even though they have the same metallicity range, we note that the CEMP-r/s stars exhibit higher values of [hs/ls] than the CEMP-s stars \citep{bisterzo2011s, bisterzo2012}. This implies that neutron density has to be higher for the formation of CEMP-r/s stars than of CEMP-s stars. Similarly, in Figure~\ref{fig:subfigures_hs}(d), we can see that the mean abundances of ls elements have the same range in the cases of   CEMP-s and CEMP-r/s stars, but in CEMP-r/s stars the hs elements are more enhanced than in  CEMP-s stars. This indicates that the heavy elements observed in CEMP-r/s stars are synthesised with a higher neutron density than those in CEMP-s stars. This is because  the higher neutron density can easily overcome the barrier ($\tau$ $\sim$ 0.05 mb$^{-1}$) at the magic neutron number 50 (the ls peak), and effectively produces the hs peak at the magic neutron number 82. As seen from Figure~\ref{fig:subfigures_hs}(c), [Eu/Fe] increases with [hs/Fe], showing a tight correlation between [Eu/Fe] and [hs/Fe]. This correlation between heavy-s elements and Eu observed in the CEMP-r/s stars can only be explained with the help of a formation mechanism where both the s-process and r-process elements are produced in a single stellar site, which provides much  support for i-process as a possible formation mechanism for CEMP-r/s stars. 
 As [hs/ls] does not depend upon the dilution of the AGB processed material on the companion star, using a sample of CEMP-s and CEMP-r/s stars from the literature, we   examined whether this  ratio could be used  to distinguish between CEMP-s and CEMP-r/s stars. We find  that CEMP-s stars peak at around [hs/ls]$\sim$0.65 with a standard deviation of 0.35, and CEMP-r/s stars peak at [hs/ls]$\sim$1.06 with a standard deviation of 0.32. However, even though they peak at different values of [hs/ls], we see an overlap of CEMP-s and CEMP-r/s stars in the range 0.0$<$[hs/ls]$<$1.5 (Figure~\ref{fig:subfigures_hs}(a) and \ref{fig:subfigures_hs}(b)).
 Hence, [hs/ls]   cannot be used  as a useful classifier for CEMP-s and CEMP-r/s stars. In  Figure~\ref{fig:subfigures_hs}(d)     all the CEMP-r/s stars lie above the solid black line at [hs/Fe]$\sim$1.4; however, a limiting value of [hs/Fe]$\geq$1.4 also cannot be used as a classifier as  many CEMP-s stars also exhibit  [hs/Fe] $\geq$1.4.\\\\

\subsubsection{Identification of classifier based on available data}

In order to explain the nature of the emerging CEMP 
stars, \citet{beers2005discovery} introduced for the first time  four subclasses
of CEMP stars using only two neutron-capture  elements, Ba and Eu (due to the 
observational limitations). This scheme of classification is a  guide to  the 
community to use the unifying nomenclature and to facilitate further advances.
However, since the classification of \citet{beers2005discovery}, various authors \citep{jonsell2006, 
masseron2010aholistic, Norris_et_al_2010, bisterzo2011s, Bonifacio_2015, 
maeder_&_meynet_2015, Yoon_et_al_2016,  abate2016cemp-rs, Hansen_2016_III, 
Frebel_review_2018, hansen2019abundances, Skuladottir_et_al.2020} put forward different criteria for identification and classification of the CEMP stars. The community has not had much time to   understand the   physics 
behind the observed abundance patterns. The ever-growing data sets and the
advancement of theoretical predictions require further revisions of the
classification schemes, and thus revisiting these classification schemes 
is extremely important and unavoidable.

We have seen that one of our programme stars, HD~145777, is classified as a 
CEMP-r/s star by three classification schemes 
\citep{beers2005discovery, Frebel_review_2018, hansen2019abundances} and 
as a CEMP-s star by the criteria discussed by \citet{abate2016cemp-rs}. So, 
we tried to figure out which  of the classification scheme discussed above 
fits   the literature data. As the programme stars are CEMP-s 
and CEMP-r/s types, we  focus on  the classification schemes that 
discuss  the criteria to distinguish these two subclasses.

\vskip 0.4cm
\noindent
{\bf{[Sr/Ba] as a classifier}}\\
\vskip 0.01cm
\noindent
We first discuss the most recent classifier, [Sr/Ba]. When we plot [Sr/Ba] 
versus [Fe/H] (Figure~\ref{fig:subfigures_srba}(a)), we can see that there 
is a clear overlap of CEMP-s and CEMP-r/s stars with respect to [Sr/Ba] 
(as seen in the case of [hs/ls]) in the range $-$1.6~$<$~[Sr/Ba]~$<$~$-$0.5. In Figure~\ref{fig:subfigures_srba}(a)~
and~\ref{fig:subfigures_srba}(b) the grid formed by the dashed black 
lines and the grid formed by the dotted red lines represent the region 
of CEMP-s and CEMP-r/s stars, respectively, put forward by the classification 
scheme of \citet{hansen2019abundances}. In 
Figure~\ref{fig:subfigures_srba}(b), it is seen  that a few CEMP 
stars (Table~\ref{tab:lit1}) with an underabundance of 
Eu ([Eu/Fe] $<$ 1.0) also fall in the region defined for CEMP-r/s 
stars. The number difference of CEMP-s stars in 
Figure~\ref{fig:subfigures_srba}(a) and \ref{fig:subfigures_srba}(b) falling 
in the CEMP-r/s region indicates that many CEMP stars with the absence of 
Eu also fall in the category of CEMP-r/s stars by this classification 
criterion. We cannot separate CEMP-s and CEMP-r/s stars with the help 
of the [Sr/Ba] criterion  \citep{hansen2019abundances}, and hence it  cannot be used effectively. Moreover, it is worth mentioning that the abundances of Sr and Ba are highly uncertain due to NLTE effects \citep{bisterzo2012}. However, as mentioned by \citet{hansen2019abundances}, this criterion may be useful to roughly distinguish between the different CEMP subclasses in low-resolution, low S/N large surveys as absorption lines of Eu are much weaker than Sr.

\vskip 0.4cm
\noindent
{\bf{[Ba/Eu] as a classifier}}\\
\vskip 0.01cm
\noindent
In Figure~\ref{fig:subfigures_baeu}(a) and \ref{fig:subfigures_baeu}(b) the grid formed by the dotted red lines bound the region (0.0~$<$~[Ba/Eu]~$<$~0.5) allowed by \citet{beers2005discovery} for CEMP-r/s stars. Although most of the CEMP-r/s stars (filled red circles), including our programme stars (filled green pentagons), reside inside the region, some fall outside the boundary. 
\citet{abate2016cemp-rs} did not put an upper limit to [Ba/Eu] for  CEMP-s and CEMP-r/s stars, so this criterion  could be used more effectively (HE~0336+0113 is the only outlier). 
According to this criterion, CEMP-r/s stars are a subclass of CEMP-s stars and the criterion [Eu/Fe]~$>$~1.0 separates the CEMP-r/s stars from the CEMP-s stars \citep{abate2016cemp-rs}.  \citet{jonsell2006} and \citet{masseron2010aholistic} also adopted the same classification scheme as \citet{abate2016cemp-rs} to distinguish between CEMP-s and CEMP-r/s stars; however, their criteria to classify the other two subclasses (CEMP-r \& CEMP-no) differ. Abundances of many of the CEMP-r/s stars, (e.g. CS~22881$-$036, CS~29497$-$030, HE~0143$-$0441, HE~2258$-$6358, and LP~625$-$44) having [Ba/Eu]~$>$~0.5 are found to fit well with the i-process model \citep{Hampel2016} of higher neutron density (n$\sim$10$^{14}$ cm$^{-3}$) \citep{Hampel_2019}. This is in contrast to  0.0~$<$~[Ba/Eu]~$<$~0.5 of  \citet{beers2005discovery} for CEMP-r/s stars and also [Ba/Eu]~$>$~0.5 of \citet{Frebel_review_2018} for CEMP-s stars.

\vskip 0.4cm
\noindent
{\bf{[La/Eu] as a classifier}}\\
\vskip 0.01cm
\noindent
Unlike other classification schemes, \citet{Frebel_review_2018} used 
different sets of elements to classify CEMP-s and CEMP-r/s stars. While Ba, 
Eu, and Pb are used in the case of CEMP-s stars, La, Eu, Hf, and Ir are used to 
classify CEMP-r/s stars. The abundances of Hf and Ir are usually  derived from  the  absorption lines of Hf and Ir that are normally found  in the NUV spectra. As we could not get the abundances for these two elements in the literature,   the  criterion [Hf/Ir] $\sim$ 1.0 could not be tested with our compiled data. 
Theoretical validation of this criterion can be found in   
Table~4 of \citet{Hampel2016}. In Figure~\ref{fig:subfigures_laeu}(a),
 two CEMP-r/s stars  are found to lie  outside the region 
(the grid formed by dotted red lines) defined by 0.0 $<$ [La/Eu] $<$ 0.6, 
and two CEMP-s stars are located  near the upper boundary 
([La/Eu] $\sim$ 0.6) of  the region. 

\vskip 0.4cm
\noindent
{\bf{ Modifications to the classifiers [Ba/Eu] and [La/Eu]}}\\
\vskip 0.01cm
\noindent
\citet{Frebel_review_2018} kept provisions for future adjustments to the 
classifying criteria. In Figure~\ref{fig:subfigures_laeu}(b), except 
for two stars (CD$-$28~1082 and HD~209621), all the  CEMP-r/s stars fall 
inside the region (the grid formed by the dotted red lines) bound by 
0.0~$<$~[La/Eu]~$<$~0.6 and 0.0~$<$~[Ba/Eu]~$<$~1.0. As a few 
CEMP-(s \& r/s) stars also fall very close to the upper boundary, we can 
put uncertainty on the upper limit of [La/Eu] 
(i.e. [La/Eu]~$<$~0.6 $\pm$ 0.1). We suggest that an object satisfying 
the criteria 0.0~$<$~[Ba/Eu]~$<$~1.0 and 0.0~$<$~[La/Eu]~$<$~0.5 can be 
classified as a CEMP-r/s star; however, if  [La/Eu] = 0.6 $\pm$ 0.1 
the condition  [Eu/Fe]~$>$~1.0 also needs to be satisfied for the star to be a 
CEMP-r/s star. The two outliers exhibit differences in the range
0.5$-$0.7 dex 
between the abundances of La and the other hs elements. In Figure~\ref{fig:subfigures_baeu}(b) 
we can see that a CEMP-s star HE~0336+0113, which is Eu ($>$~1.0) enhanced with very high value for 
[Ba/Eu] (=1.45), falls in the region allowed for CEMP-r/s stars. 
The upper limit of [Ba/Eu] ($<$~1.0) is chosen so as to avoid misclassifying such stars 
as CEMP-r/s stars.

\vskip 0.4cm
\noindent
{\bf{[La/Ce] as a classifier}}\\
\vskip 0.01cm
\noindent
 As models with higher neutron densities are required to reproduce the observed overabundance of heavy elements in the CEMP-r/s stars,  the possibility of $^{22}$Ne($\alpha$, n)$^{25}$Mg being the primary neutron source cannot be discarded as it can produce a higher neutron density than the $^{13}$C($\alpha$, n)$^{16}$O reaction. \citet{masseron2010aholistic} discussed the requirement of the $^{22}$Ne($\alpha$, n)$^{25}$Mg neutron source for the production of heavy elements  in CEMP-r/s stars on the basis of [La/Ce]. In general, [La/Ce]  gives higher values in CEMP-r/s stars, which is possible only when $^{22}$Ne($\alpha$, n)$^{25}$Mg operates. We have found [La/Ce]~$>$~0.0 only in HE~0017+0055 among our programme stars. \citet{jorissen2016HE0017} also  recorded a positive value for [La/Ce] for this object. In the case of the other two CEMP-r/s stars (HE~2144$-$1832 and HE~2339$-$0837), [La/Ce] is found to be negative. In the literature, such negative values are  seen in a few other CEMP-r/s stars as well (Table~\ref{tab:lit1}). Furthermore, from the simulations of \citet{Hampel2016} at higher neutron densities (n $\sim$ 10$^{12-15}$ cm$^{-3}$), it can be seen that [La/Ce] does not yield a positive value for all the neutron densities (see Table~4 of \citealt{Hampel2016}). For n $\sim$ 10$^{12}$ and 10$^{13}$ cm$^{-3}$ [La/Ce] gives positive values, but for n $\sim$ 10$^{14}$ and 10$^{15}$ cm$^{-3}$ [La/Ce] $<$ 0. So we propose that [La/Ce] needs to be reconsidered as  an indicator of neutron source.

\vskip 0.4cm
\noindent
{\bf{[As/Ge] and [Se/Ge] as classifiers}}\\
\vskip 0.01cm
\noindent
\citet{Roederer_2016} suggested that supersolar [As/Ge] and solar 
or subsolar [Se/Ge] could indicate the operation of i-process. 
Although, due to the lack of high-quality NUV spectra, the abundances of 
these elements (As, Se, Ge)  are currently  limited, in the future  the 
neutron-capture elements found in the NUV region are likely to play  
important roles in characterising the i-process. This will also help 
to validate the claim that [As/Ge] and  [Se/Ge] could be used as classifiers 
of i-process.

\vskip 0.2cm
The difficulty in drawing sharp boundaries to distinguish CEMP-s and CEMP-r/s stars is clearly visible in all the classifiers. Moreover, the smooth transition of CEMP-s and CEMP-r/s stars, as seen in   Figure~\ref{fig:subfigures_hs}(c), may indicate that s- and i-process neutron densities are not distinctly different and  have a continuous range from n $\sim$ 10$^{7}$ to 10$^{15}$ cm$^{-3}$.

\section{Conclusions}
\label{sec:conclusion}
We have estimated the atmospheric parameters and abundances of twenty-four elements including C, N, $\alpha$-elements, Fe-peak elements, and neutron-capture elements for HD~145777, CD$-$27~14351, HE~0017+0055, HE~2144$-$1832, and HE~2339$-$0837. The object HE~2144$-$1832 is found to be a metal-poor  star with 
[Fe/H] ${\sim}$ $-$1.63, and the other four  belong to the very metal-poor  class with [Fe/H]$<-2$. 
The kinematic analysis shows that  HD 145777, HE~2144$-$1832, and HE~0017$-$0055 belong to the  thick disc population, and the objects HE~2339$-$0837 and CD$-$27~14351 are members of the  halo population. 

 All the programme stars are found to be bright giants with   surface gravity 0.60$\leq$ log g $\leq$1.40 and  enhanced in carbon with a maximum of [C/Fe] ${\sim}$ 2.98 for CD$-$27~14351, and a minimum of ${\sim}$ 1.80 for HE~2144$-$1832. While nitrogen is mildly enhanced in HD~145777 and HE~2144-1832, it shows overabundance with [N/Fe]$\sim$2.83 and 1.88 in HE~0017+0055 and CD$-$27~14351, respectively. In the stars HD~145777 and HE~2144$-$1832, the $\alpha$-elements Mg and Ca show mild overabundance ([X/Fe] $\sim$ 0.5). 
Neutron-capture elements Sr, Y, Zr, Ba, La, Ce, Nd, Sm, Pr, and Eu show mild to high enhancements in all the programme stars. 

 Adopting the classification of \citet{abate2016cemp-rs}, we have found that HE~2144$-$1832, HE~0017+0055, and HE~2339$-$0837 with [Ba/Fe]~$>$~1, [Eu/Fe]~$>$~1, and [Ba/Eu]~$>$~0 are CEMP-r/s stars and HD~145777 and CD$-$27~14351 with [Ba/Fe]~$>$~1, [Eu/Fe]~$<$~1, and [Ba/Eu]~$>$~0 are CEMP-s stars. We do not confirm the high abundance of Eu reported  in a previous study \citep{Drisya2017} for the object CD$-$27~14351. 
 
 In the context of  the double enhancement (enhanced in both s- and r-process elements) seen in four of the programme stars, we have discussed different formation scenarios of CEMP-r/s and CEMP-s stars. The absence of Tc lines and the low values of $^{12}$C/$^{13}$C   indicate the extrinsic nature of the programme stars. We have seen that i-process models of \citet{Hampel2016} can satisfactorily reproduce the observed overabundance of heavy elements in the CEMP-r/s stars. The i-process model \citep{Hampel2016} with neutron density, n$\sim$10$^{13}$ cm$^{-3}$, n$\sim$10$^{15}$ cm$^{-3}$, and n$\sim$10$^{14}$ cm$^{-3}$ give the best fit to the observed overabundance of heavy elements in the stars HE~2144$-$1832, HE~0017+0055, and HE~2339$-$0837, respectively. Such a high neutron density explains the enhancement of both s- and r-process elements in these stars. For HD~145777, in order to fit the observed abundances, an i-process model with neutron density of 10$^{10}$ cm$^{-3}$ is needed. This is towards the higher limit of the neutron density required for s-process nucleosynthesis. Although the Eu abundance in HD~145777 is not   high enough to make it a CEMP-r/s star, Eu is  still  enhanced, and this  enhancement can be attributed to this high neutron density.

With the help of a sample of 72 CEMP-s and CEMP-r/s stars from  the literature, we have critically analysed the different criteria used by various authors for CEMP-s and CEMP-r/s stars. We have found that the set of criteria adopted by \citet{abate2016cemp-rs} is suitable  and  fits well with the literature data. The criterion 0.0~$<$~[La/Eu]~$<$~0.6 put forward by \citet{Frebel_review_2018} to classify CEMP-r/s stars is also found to be suitable with a slight modification to the upper limit. As [hs/ls] gives higher values for CEMP-r/s stars than CEMP-s stars, we have examined whether  [hs/ls] can be used as a classifier. Even though they peak at different values of [hs/ls], CEMP-s and CEMP-r/s stars show an overlap in the range 0.0~$<$~[hs/ls]~$<$1.5. Thus, this ratio alone cannot be used as  a definitive classifier of 
CEMP-s and CEMP-r/s stars. The best criteria to distinguish the CEMP-s and CEMP-r/s stars are found to be the following:
\begin{itemize}
      \item CEMP: [C/Fe] $\geq$ 0.7
\vskip 0.1cm
      \item CEMP-r/s: [Ba/Fe] $\geq$ 1.0, [Eu/Fe] $\geq$ 1.0
\vskip 0.1cm
      \begin{itemize}
            \item [i)] 0.0~$\leq$~[Ba/Eu]~$\leq$~1.0 and/or 0.0~$\leq$~[La/Eu]~$\leq$~0.7;
      \end{itemize}
\vskip 0.1cm
      \item CEMP-s: [Ba/Fe]~$\geq$~1.0 
\vskip 0.1cm
      \begin{itemize}
            \item [i.)] [Eu/Fe]~$<$~1.0, [Ba/Eu]~$>$~0.0 and/or  [La/Eu]~$>$~0.5;
            \item [ii.)] [Eu/Fe] $\geq$ 1.0, [Ba/Eu] $>$ 1.0 and/or  [La/Eu] $>$ 0.7.
      \end{itemize}
\end{itemize}

\begin{acknowledgements}
We thank the staff at IAO and at the remote control station at CREST, Hosakote, for assisting during the observations. Funding from DST SERB project No. EMR/2016/005283 is gratefully acknowledged. We are thankful to Melanie Hampel for providing us with the  i-process yields in the form of number fractions. We thank the anonymous referee for many constructive suggestions and useful comments, which significantly improved the readability of the paper.
This work made use of the SIMBAD astronomical database, operated at CDS, Strasbourg, France, the NASA ADS, USA and data from the European Space Agency (ESA) mission Gaia (\url{https://www.cosmos.esa.int/gaia}), processed by the Gaia Data Processing and Analysis Consortium (DPAC, \url{https://www.cosmos.esa.int/web/gaia/dpac/consortium}). RSR thank IISER-Pune and IIA-Bangalore for providing the platform to do the research work and DST for INSPIRE scholarship support during the project.
\end{acknowledgements}

\bibliographystyle{aa}
\bibliography{sample}

\begin{appendix}

\onecolumn
\section{}

{\footnotesize
\begin{table*}[ht]
\caption{Differential abundance ($\Delta$log$\epsilon$) of different species due to the variations in stellar atmospheric parameters for HE~2144$-$1832.}
\label{tab:error}
\scalebox{0.75}{
\begin{tabular}{ l c c c c c c c c c c c c }
\hline
\hline
Element & $\Delta$T$_{eff}$ & $\Delta$T$_{eff}$ & $\Delta$log g & $\Delta$log g & $\Delta\zeta$ & $\Delta\zeta$ & $\Delta$[Fe/H] & $\Delta$[Fe/H] & $(\Sigma \sigma_{i}^{2})^{1/2}$ & $(\Sigma \sigma_{i}^{2})^{1/2}$ & $\sigma$[X/Fe] & $\sigma$[X/Fe]  \\
 &(+100 K)&($-$100 K)&(+0.2 dex)&($-$0.2 dex)&(+0.2 kms$^{-1}$)&($-$0.2 kms$^{-1}$)&(+0.2 dex)&($-$0.2 dex)&(+$\Delta$)&($-\Delta$)&(+$\Delta$)&($-$$\Delta$)\\
         &        &        &        &        &        &         &         &         &         &         &        &                \\
\hline
C (C$_{2}$, 5165 $\AA$) &  -0.03 &    0.02&   0.01 &   0.00 &   0.00 &    0.00 &   0.00  &    0.01 &   0.03  &   0.02  &  0.15  &   0.15    \\
C (C$_{2}$, 5635 $\AA$) &  -0.04 &    0.05&   0.03 &  -0.02 &   0.00 &    0.00 &  -0.01  &    0.01 &   0.05  &   0.05  &  0.16  &   0.16    \\
C (CH, 4310 $\AA$) &  -0.04 &    0.04&   0.01 &   0.00 &   0.00 &    0.00 &  -0.04  &   0.03 &   0.06  &   0.05  &  0.16  &   0.16         \\
N        &   0.10 &   -0.10&  -0.15 &   0.13 &   0.00 &    0.00 &   0.00  &    0.00 &   0.18  &   0.16  &  0.23  &   0.22         \\
Na I     &   0.10 &   -0.10&  -0.03 &   0.03 &  -0.04 &    0.04 &  -0.04  &    0.04 &   0.12  &   0.12  &  0.20  &   0.20         \\
Mg I     &   0.09 &   -0.08&  -0.04 &   0.03 &  -0.09 &    0.08 &  -0.03  &    0.03 &   0.14  &   0.12  &  0.22  &   0.21         \\   
Ca I     &   0.16 &   -0.16&  -0.03 &   0.03 &  -0.10 &    0.11 &  -0.05  &    0.05 &   0.20  &   0.20  &  0.25  &   0.26         \\
Ti I     &   0.24 &   -0.22&  -0.01 &   0.02 &  -0.06 &    0.08 &  -0.03  &    0.05 &   0.25  &   0.24  &  0.30  &   0.29         \\
Ti II    &  -0.03 &    0.04&   0.05 &  -0.03 &  -0.09 &    0.12 &   0.03  &   -0.01 &   0.11  &   0.13  &  0.19  &   0.20         \\
V I      &   0.22 &   -0.21&   0.00 &   0.00 &  -0.03 &    0.03 &  -0.04  &    0.04 &   0.23  &   0.22  &  0.27  &   0.26         \\
Cr I     &   0.23 &   -0.20&   0.00 &   0.02 &  -0.14 &    0.18 &  -0.03  &    0.05 &   0.27  &   0.27  &  0.32  &   0.32         \\
Mn I     &   0.15 &   -0.12&  -0.02 &   0.04 &   0.02 &   -0.02 &  -0.04  &    0.05 &   0.16  &   0.14  &  0.22  &   0.20         \\
Fe I     &   0.10 &   -0.10&   0.10 &  -0.08 &  -0.12 &    0.14 &   0.20  &   -0.20 &   0.27  &   0.27  &        &                \\
Fe II    &  -0.13 &    0.18&   0.13 &  -0.09 &  -0.06 &    0.09 &   0.20  &   -0.20 &   0.28  &   0.30  &        &                \\
Co I     &   0.10 &   -0.08&   0.00 &   0.01 &  -0.02 &    0.02 &   0.00  &    0.01 &   0.10  &   0.08  &  0.18  &   0.17         \\
Ni I     &   0.06 &   -0.04&   0.00 &   0.01 &  -0.07 &    0.07 &  -0.01  &    0.01 &   0.09  &   0.08  &  0.19  &   0.19         \\
Zn I     &  -0.09 &    0.10&   0.03 &  -0.01 &  -0.06 &    0.06 &   0.01  &    0.00 &   0.11  &   0.12  &  0.19  &   0.19         \\
Sr I     &   0.20 &   -0.20&   0.00 &   0.00 &  -0.05 &    0.05 &  -0.03  &    0.03 &   0.21  &   0.21  &  0.26  &   0.26         \\
Y II     &  -0.01 &    0.02&   0.06 &  -0.05 &  -0.14 &    0.16 &   0.04  &   -0.03 &   0.16  &   0.17  &  0.22  &   0.23         \\
Zr I     &   0.25 &   -0.24&  -0.01 &   0.01 &  -0.04 &    0.05 &  -0.03  &    0.04 &   0.26  &   0.25  &  0.31  &   0.31         \\
Ba II    &   0.01 &   -0.02&   0.03 &  -0.03 &  -0.16 &    0.16 &   0.02  &   -0.03 &   0.16  &   0.17  &  0.22  &   0.22         \\
La II    &   0.10 &   -0.05&   0.10 &  -0.05 &  -0.15 &    0.25 &   0.00  &    0.00 &   0.21  &   0.26  &  0.25  &   0.30         \\
Ce II    &   0.02 &   -0.02&   0.04 &  -0.03 &  -0.12 &    0.15 &   0.02  &   -0.02 &   0.13  &   0.16  &  0.21  &   0.23         \\
Pr II    &   0.03 &   -0.02&   0.07 &  -0.07 &  -0.09 &    0.11 &   0.05  &   -0.04 &   0.13  &   0.14  &  0.22  &   0.22         \\
Nd II    &   0.02 &   -0.02&   0.06 &  -0.05 &  -0.13 &    0.16 &   0.03  &   -0.03 &   0.15  &   0.17  &  0.22  &   0.24         \\
Sm II    &   0.04 &   -0.04&   0.05 &  -0.05 &  -0.11 &    0.12 &   0.03  &   -0.03 &   0.13  &   0.14  &  0.23  &   0.24         \\
Eu II    &   0.00 &    0.00&   0.10 &  -0.05 &   0.00 &    0.02 &   0.05  &   -0.02 &   0.11  &   0.06  &  0.18  &   0.16         \\
\hline
\end{tabular}}

\end{table*}
}

{\footnotesize
\begin{table}
\caption{CEMP-s and CEMP-r/s stars from the literature: 1. \citet{aoki2001neutron}, 2. \citet{aoki2002december}, 3. \citet{aoki2008carbon}, 4. \citet{barbuy2005new}, 5. \citet{barklem2005}, 6. \citet{behara2010three}, 7. \citet{cohen2003abundance}, 8. \citet{cohen2006carbon}, 9. \citet{drisya2014chemical}, 10. \citet{goswami2006a_high}, 11. \citet{hill2000heavy}, 12. \citet{ishigaki2010}, 13. \citet{ivans2005near}, 14. \citet{johnson&bolte2002}, 15. \citet{johnson&bolte2004}, 16. \citet{jonsell2005chemical}, 17. \citet{jonsell2006}, 18. \citet{lucatello2003stellar}, 19. \citet{meenakshi2019chIII}, 20. \citet{preston2001}, 21. \citet{pereira2009high}, 22. \citet{sivarani2004first}, 23. \citet{van2003more}, 24. \citet{zacs1998}, 25. \citet{zhang2009chemical}, 26. \citet{jorissen2016HE0017}, 27. this work, 28. \citet{goswami2010hd209621}, 29. \citet{liu2012abundances}, 30. \citet{allen2012elemental}, 31. \citet{hansen2019abundances}, 32. mean of the presented data, 33. \citet{burris2000neutron} 34. \citet{vanture1992}, 35. \citet{masseron2010aholistic}, 36. \citet{placco2013metal}.}.
\label{tab:lit1}
\scalebox{0.56}{
\begin{tabular}{ l c c c c c c c c c c c c c c c c c c c c }

\hline
\hline
    &      &      &      &      &      &      &      & CEMP-s stars &      &      &      &      &      &      &      &     &     &      &    & \\

Star Name&[Fe/H]&[C/Fe]&[N/Fe]&[C/N]&[Sr/Fe]&[Y/Fe]&[Zr/Fe]&[Ba/Fe]&[La/Fe]&[Ce/Fe]&[Nd/Fe]&[Sm/Fe]&[Eu/Fe]&[Pb/Fe]&[ls/Fe]&[hs/Fe]&[hs/ls]&[Ba/Eu]&[Sr/Ba]&Ref \\
\hline
BD+04 2466      &-1.99 & 1.17 & 1.10 & 0.07 &  -   & 0.52 & 0.79 & 1.49 & 1.20 & 1.07 & 1.35 &  -   &  -   & 1.92 & 0.66 & 1.28& 0.62&  -   & - &32\\
                &-1.92 & 1.17 & 1.10 & 0.07 &  -   & 0.47 & 0.79 & 1.70 & 1.20 & 1.07 & 1.35 &  -   &  -   & 1.92 &  -   &  -  &  -  &  -   & - &21\\
                &-2.10 &  -   &  -   &  -   &  -   & 0.54 &  -   & 1.36 &  -   &  -   &  -   &  -   &  -   &  -   &  -   &  -  &  -  &  -   & - &12\\
                &-1.92 &  -   &  -   &  -   &  -   & 0.55 &  -   & 1.31 &  -   &  -   &  -   &  -   &  -   &  -   &  -   &  -  &  -  &  -   & - &25\\
                &-2.00 &  -   &  -   &  -   &  -   &  -   &  -   & 1.60 &  -   &  -   &  -   &  -   &  -   &  -   &  -   &  -  &  -  &  -   & - &33\\
BD-01 2582      &-2.25 &  -   &  -   &  -   &  -   &  -   &  -   & 1.50 &  -   &  -   &  -   &  -   &  -   &  -   &  -   &  -  &  -  &  -   & - &33\\
BS 16077-0077$^{d,e}$   &-2.05$^{a}$& 2.25 &  -   &  -   &-0.06 &-0.12 &-0.02 & 0.75 & 0.58 & 0.60 & 0.60 &  -   &-0.01 &  -   &-0.07 & 0.63& 0.70& 0.76 &-0.81 &30\\
CD-38 2151      &-1.72 & 1.50 & 1.40 & 0.10 & 1.16 & 0.44 & 1.50 & 0.97 & 1.09 & 1.30 & 1.17 & 1.07 &  -   &  -   & 1.03 & 1.13& 0.10&  -   &0.19 &32\\
                &-2.03 & 1.50 & 1.40 & 0.10 & 1.16 & 0.57 & 2.00 & 0.97 & 0.88 & 1.24 & 1.34 & 1.53 &  -   &  -   &  -   &  -  &  -  &  -   & - &19\\
                &-1.40 &  -   &  -   &  -   &  -   & 0.30 & 1.00 &  -   & 1.30 & 1.35 & 1.00 & 0.61 &  -   &  -   &      &     &     &      & - &34\\
CS 22880-074$^{d}$    &-1.85 & 1.41 & 0.05 & 1.36 & 0.27 & 0.38 & 0.73 & 1.33 & 1.16 & 1.22 & 1.20 &  -   & 0.53 & 1.90 & 0.46 & 1.23& 0.77& 0.80 &-1.06 &32\\
                &-1.76 & 1.51 & 0.20 & 1.31 & 0.14 & 0.60 & 0.73 & 1.34 & 1.24 &  -   &  -   &  -   & 0.55 &  -   &  -   &  -  &  -  &  -   & - &20\\
                &-1.93 & 1.30 &-0.10 & 1.40 & 0.39 & 0.16 &  -   & 1.31 & 1.07 & 1.22 & 1.20 &  -   & 0.50 & 1.90 &  -   &  -  &  -  &  -   & - &2\\
CS 22942-019    &-2.66 & 2.10 & 0.50 & 1.60 & 1.55 & 1.58 & 1.69 & 1.71 & 1.53 & 1.54 & 1.26 & 1.64 & 0.80 &$\leq$1.60 & 1.61 & 1.51&-0.10& 0.91 &-0.16&32\\
                &-2.67 & 2.20 & 0.70 & 1.50 & 1.40 &  -   &  -   & 1.50 & 1.85 &  -   &  -   &  -   & 0.80 &  -   &  -   &  -  &  -  &  -   & - &20\\
                &-2.64 & 2.00 & 0.30 & 1.70 & 1.70 & 1.58 & 1.69 & 1.92 & 1.20 & 1.54 & 1.26 & 1.64 & 0.79 &$\leq$1.60 &  -   &  -  &  -  &  -   & - &2\\
CS 22949-008$^{d}$    &-2.45$^{a}$& 2.00 &  -   &  -   &-0.17 & 1.25 & 1.61 & 0.98 &  -   & 1.49 & 1.98 & 0.61 &  -   &  -   & 0.90 & 1.48& 0.58&  -  &-1.15 &30\\
CS 29512-073$^{d}$    &-2.06$^{a}$& 1.40 &  -   &  -   &0.75$^{b}$& 0.60 & 0.59 & 1.42 & 1.42 & 1.63 & 1.57 & 1.88 & 0.37 &  - & 0.65 & 1.51& 0.86& 1.05&-0.67&30\\
CS 30301-015$^{d}$    &-2.64 & 1.60 & 0.60 & 1.00 & 0.30 & 0.29 &  -   & 1.45 & 0.84 & 1.16 & 1.25 & 0.85 & 0.20 & 1.70 & 0.30 & 1.18& 0.88& 1.25  &-1.15 &2\\
G 18-24         &-1.62 &  -   &  -   &  -   &  -   & 0.58 &  -   & 1.17 &  -   &  -   &  -   &  -   &  -   &  -   &  -   &  -  &  -  &  -    & - &12\\
G 24-25$^{d}$         &-1.40 & 1.03 &  -   &  -   &0.36$^{b}$& 0.70 &0.95$^{b}$& 1.35 & 1.55 & 1.59 & 1.47 & 1.40 & 0.61 & 1.68 & 0.67 & 1.49&0.82&0.74&-0.99 &29\\
HD 5223         &-2.05 & 1.57 &  -   &  -   & 1.39 & 0.63 & 1.57 & 1.82 & 1.76 & 1.87 & 1.54 & 1.68 &  -   &$<$2.21 & 1.19 & 1.75& 0.56&  -   &-0.43 &10\\
HD 30443$^{d}$        &-1.69 & 1.68 & 0.40 & 1.28 & 0.74 & 1.37 & 1.60 & 1.70 &  -   & 1.62 & 2.47 & 2.24 &  -   &  -   & 1.24 & 1.93& 0.69&  -   &-0.96 &19\\
HD 196944       &-2.33 & 1.31 & 1.30 & 0.01 & 0.84 & 0.57 & 0.63 & 1.27 & 0.96 & 1.20 & 0.83 & 0.69 & 0.17 & 2.00 & 0.68 & 1.07& 0.39& 1.10  &-0.43 &32\\
                &-2.45 & 1.42 &  -   &  -   &  -   & 0.58 &  -   & 1.56 &  -   & 1.49 & 0.94 &  -   &  -   &  -   &  -   &  -  &  -  &  -    & - &24\\
                &-2.25 & 1.20 & 1.30 &-0.10 & 0.84 & 0.56 & 0.66 & 1.10 & 0.91 & 1.01 & 0.86 & 0.78 & 0.17 & 1.90 &  -   &  -  &  -  &  -    & - &2\\
                &-2.23 &  -   &  -   &  -   &  -   &  -   &  -   & 1.14 &  -   &  -   &  -   &  -   &  -   &  -   &  -   &  -  &  -  &  -    & - &16\\
                &-2.40 &  -   &  -   &  -   &  -   &  -   & 0.60 &  -   & 1.00 & 1.10 & 0.70 & 0.60 &  -   & 2.10 &  -   &  -  &  -  &  -    & - &23\\
HD 198269       &-1.40 &  -   &  -   &  -   &  -   & 0.50 & 1.20 &  -   & 1.40 & 1.60 & 1.00 & 0.90 & 0.80 &  -   & 0.85 & 1.33& 0.48&  -    & - &34\\
HD 201626       &-1.35 &  -   &  -   &  -   &  -   & 1.00 & 1.30 & 2.12 & 1.68 & 1.90 & 1.97 & 1.57 &  -   &  -   & 1.15 & 1.92& 0.77&  -    & - &32\\
                &-1.39 &  -   &  -   &  -   &  -   &  -   &  -   & 2.12 & 1.76 & 1.89 & 2.24 & 1.63 &  -   &  -   &  -   &  -  &  -  &  -    & - &9\\
                &-1.30 &  -   &  -   &  -   &  -   & 1.00 & 1.30 &  -   & 1.60 & 1.90 & 1.70 & 1.50 &  -   &  -   &  -   &  -  &  -  &  -    & - &34\\
\hline
\end{tabular}}

{\textit{a}} $-$ [Fe/H] is from Fe I lines; {\textit{b}} $-$ Average abundance from neutral and ionised species; {\textit{c}} $-$ Outliers of the classifier of \citet{beers2005discovery}; {\textit{d}} $-$ Outliers of the classifier of \citet{hansen2019abundances}; {\textit{e}} $-$  Outliers of the classifier of \citet{Frebel_review_2018}; {\textit{f}} $-$ Outlier of the classifier of \citet{abate2016cemp-rs}.\\
\end{table}
}

{\footnotesize
\begin{table*}
\label{tab:continued}
{\bf{Table A.2 }}\textit{-- continued}\\
\scalebox{0.56}{
\begin{tabular}{ l c c c c c c c c c c c c c c c c c c c c }
\hline
\hline
    &      &      &      &      &      &      &      & CEMP-s stars &      &      &      &      &      &      &      &     &     &      &    & \\

Star Name&[Fe/H]&[C/Fe]&[N/Fe]&[C/N]&[Sr/Fe]&[Y/Fe]&[Zr/Fe]&[Ba/Fe]&[La/Fe]&[Ce/Fe]&[Nd/Fe]&[Sm/Fe]&[Eu/Fe]&[Pb/Fe]&[ls/Fe]&[hs/Fe]&[hs/ls]&[Ba/Eu]&[Sr/Ba]&Ref \\
\hline
HE 0012-1441    &-2.52$^a$ & 1.59 & 0.64 & 0.95 &  -   &  -   &  -   & 1.15 &  -   &  -   &  -   &  -   &  -   &$<$1.92 &  -   &  -  &  -  &  -  & - &8\\
HE 0024-2523$^{d}$    &-2.72$^a$& 2.60 & 2.10 & 0.50 & 0.34 &$<$0.91 &$<$1.22 & 1.46 & 1.80 &  -   &  -   &  -   &$<$1.10 & 3.30 &  -   & 1.63&- &-&-1.12 &18\\
HE 0202-2204$^{d}$    &-1.98 & 1.16 &  -   &  -   & 0.57 & 0.41 & 0.47 & 1.41 & 1.36 & 1.30 & 1.02 & 1.03 & 0.49 &  -   & 0.48 & 1.27& 0.79& 0.92  &-0.84 &5\\
HE 0231-4016$^{d}$    &-2.08 & 1.36 &  -   &  -   & 0.67 & 0.72 &  -   & 1.47 & 1.22 & 1.53 & 1.35 &  -   &  -   &  -   & 0.70 & 1.39& 0.69&  -    &-0.80 &5\\
HE 0253-6024    &-2.10 & 1.30 & 0.20 & 1.10 & 1.50 & 0.80 &  -   & 1.70 & 1.50 & 1.20 & 2.00 &  -   &$<$1.00 &  -   & 1.15 & 1.60& 0.45&  -  &-0.20 &31\\
HE 0317-4705    &-2.30 & 1.40 & 0.40 & 1.00 & 1.70 & 0.60 &  -   & 1.00 & 1.40 & 1.50 & 1.30 &  -   &$<$1.00 &  -   & 1.15 & 1.30& 0.15&  -  & 0.70&31\\
HE 0336+0113$^{f}$ &-2.68$^a$ & 2.25 & 1.60 & 0.65 & 1.68 & 1.40 &  -   & 2.63 & 1.93 & 2.30 & 2.12 &  -   & 1.18 &$<$2.28 & 1.54 & 2.25& 0.71& 1.45 &-0.95&8\\
HE 0430-4404$^{d}$    &-2.07 & 1.44 &  -   &  -   & 0.56 & 0.60 &  -   & 1.62 & 1.41 &  -   &  -   &  -   &  -   &  -   & 0.58 & 1.52& 0.94&  -    &-1.06&5\\
HE 1031-0020$^{d}$    &-2.86$^a$ & 1.63 & 2.48 &-0.85 & 0.31 & 0.25 &  -   & 1.21 & 1.16 & 1.40 & 1.72 &  -   &$<$0.87 & 2.66 & 0.28 & 1.37& 1.09& -   &-0.90&8\\
HE 1135+0139    &-2.33 & 1.19 &  -   &  -   & 0.66 & 0.36 & 0.46 & 1.13 & 0.93 & 1.17 & 0.77 &  -   & 0.33 &  -   & 0.49 & 1.00& 0.51& 0.80  &-0.47&5\\
HE 1152-0355    &-1.29 & 0.58 &  -   &  -   &  -   & 0.14 & 0.00 & 1.58 & 1.57 &  -   & 0.43 & 0.87 &  -   &  -   & 0.07 & 1.19& 1.12&  -    & - &10\\
HE 1430-1123$^{d}$    &-2.71 & 1.84 &  -   &  -   & 0.24 & 0.50 &  -   & 1.82 &  -   &  -   & 1.72 &  -   &  -   &  -   & 0.37 & 1.77& 1.40&  -    &-1.58&5\\
HE 1434-1442    &-2.39$^a$ & 1.95 & 1.40 & 0.55 &  -   & 0.37 &  -   & 1.23 &  -   &  -   & 1.70 &  -   &  -   & 2.18 &  -   & 1.47&  -  &  -    & - &8\\
HE 1443+0113    &-2.07$^a$ & 1.84 &  -   &  -   &  -   &  -   &  -   & 1.40 &  -   &  -   &  -   &  -   &  -   &  -   &  -   &  -  &  -  &  -    & - &8\\
HE 1509-0806$^{d}$    &-2.91$^a$ & 1.98 & 2.23 &-0.25 & 1.12 & 0.95 &  -   & 1.93 & 1.67 & 1.89 & 2.18 &  -   &$<$0.93 & 2.61 & 1.04 & 1.92& 0.88& -   &-0.81&8\\
HE 2138-3336$^{d}$    &-2.79 & 2.43 & 1.66 & 0.77 & 0.27 & 0.48 & 0.81 & 1.91 & 1.60 & 1.81 & 1.57 & 1.53 &$<$1.09 & 3.84 & 0.52 & 1.72& 1.20& -   &-1.64&36\\
HE 2150-0825$^{d}$    &-1.98 & 1.35 &  -   &  -   & 0.66 & 0.85 & 0.97 & 1.70 & 1.41 & 1.48 & 1.42 &  -   &  -   &  -   & 0.83 & 1.50& 0.67&  -    &-1.04&5\\
HE 2158-0348$^{d}$    &-2.70$^a$ & 1.87 & 1.52 & 0.35 & 0.52 & 0.87 & 1.74 & 1.59 & 1.55 & 1.89 & 1.51 &$<$2.40 & 0.80 & 2.60 & 1.04 & 1.64& 0.60& 0.79 &-1.07&8\\
HE 2158-5134    &-3.00 & 2.60 & 0.80 & 1.80 & 2.60 & 1.80 &  -   & 2.30 &$<$2.00 & $<$2.20 & 1.80 &  -   &  -   &  -   & 2.20 & 2.05&-0.15&  -  & 0.30&31\\
HE 2227-4044$^{d}$    &-2.32 & 1.67 &  -   &  -   & 0.41 &  -   &  -   & 1.38 & 1.28 &  -   &  -   &  -   &  -   &  -   &  -   & 1.33&  -  &  -    &-0.97&5\\
HE 2232-0603$^{d}$    &-1.85$^a$ & 1.22 & 0.47 & 0.75 & 0.55 & 0.60 &  -   & 1.41 & 1.23 & 1.45 &  -   &  -   &  -   & 1.55 & 0.58 & 1.36& 0.78&  -    &-0.86&8\\
HE 2240-0412$^{d}$    &-2.20 & 1.35 &  -   &  -   & 0.24 &  -   &  -   & 1.37 &  -   &  -   &  -   &  -   &  -   &  -   &  -   &  -  &  -  &  -    &-1.13&5\\
HE 2258-4427    &-2.10 & 1.40 &-0.10 & 1.50 & 1.70 & 1.00 &  -   & 1.30 & 1.40 & 1.60 & 1.50 &  -   & 0.80 &  -   & 1.35 & 1.45& 0.10& 0.50  &0.40&31\\
HE 2339-4240    &-2.30 & 1.70 & 0.60 & 1.10 & 1.60 & 0.80 &  -   & 2.00 & 2.00 & 1.70 & 2.00 &  -   &  -   &  -   & 1.20 & 1.93& 0.73&  -    &-0.40&31\\
\hline
\hline
  &      &      &      &      &      &      &      & CEMP-r/s stars &      &      &      &      &      &      &      &     &     &     &   & \\

Star Name&[Fe/H]&[C/Fe]&[N/Fe]&[C/N]&[Sr/Fe]&[Y/Fe]&[Zr/Fe]&[Ba/Fe]&[La/Fe]&[Ce/Fe]&[Nd/Fe]&[Sm/Fe]&[Eu/Fe]&[Pb/Fe]&[ls/Fe]&[hs/Fe]&[hs/ls]&[Ba/Eu]&[Sr/Ba]&Ref \\
\hline
BS 16080-175$^{e}$    &-1.86 & 1.75 &$<$0.80&  -  & 1.04 & 1.07 & 1.29 & 1.55 & 1.65 &  -   &  -   &  -   & 1.05 & 2.60 & 1.13 & 1.60& 0.47& 0.50  &-0.51 &30\\
BS 17436-058    &-1.90 & 1.50 & 1.25 & 0.25 & 0.95 & 0.73 & 0.93 & 1.61 & 1.49 &  -   &  -   &  -   & 1.17 & 2.26 & 0.87 & 1.55& 0.68& 0.44  &-0.66 &30\\
CD-28 1082$^{e}$      &-2.45 & 2.19 & 2.73 &-0.54 & 1.44 & 1.61 &  -   & 2.09 & 1.55 & 1.97 & 1.99 & 2.29 & 2.07 &  -   & 1.53 & 1.90& 0.37& 0.02  &-0.65 &19\\
CS 22183-015$^{c,d}$    &-2.91 & 2.13 & 1.78 & 0.35 & 0.44 & 0.50 & 0.68 & 2.01 & 1.61 & 1.72 & 1.78 &  -   & 1.49 & 2.99 & 0.54 & 1.78& 1.24& 0.52  &-1.57 &32\\
                &-2.75$^a$ & 1.92 & 1.77 & 0.15 & 0.34 & 0.52 &  -   & 2.04 & 1.70 & 1.88 & 1.91 &  -   & 1.70 & 2.79 &  -   & -   &  -  &  -    & - &8\\
                &-3.12 &  -   &  -   &  -   &  -   & 0.45 & 0.62 & 2.09 & 1.59 & 1.55 & 1.65 &  -   & 1.39 & 3.17 &  -   & -   &  -  &  -    & - &14\\
                &-2.85 & 2.34 & 1.79 & 0.55 & 0.54 & 0.54 & 0.74 & 1.89 & 1.53 &  -   &  -   &  -   & 1.37 & 3.00 &  -   & -   &  -  &  -    & - &30\\
CS 22881-036$^{c}$    &-2.06 & 1.96 & 1.00 & 0.96 & 0.59 & 1.01 & 0.95 & 1.93 & 1.59 &  -   & 2.04 &  -   & 1.00 &  -   & 0.85 & 1.85& 1.00& 0.93  &-1.34&20\\
CS 22887-048$^{c}$    &-1.70 & 1.84 &$<$1.29&  -  & 1.00 & 0.99 & 1.23 & 2.00 & 1.73 &  -   &  -   &  -   & 1.49 & 3.40 & 1.07 & 1.87& 0.80& 0.51  &-1.00&30\\
CS 22898-027$^{d}$    &-2.23 & 2.16 & 1.11 & 1.05 & 0.70 & 0.76 & 1.23 & 2.25 & 2.20 & 2.13 & 2.12 & 2.08 & 1.91 & 2.87 & 0.90 & 2.18& 1.28& 0.34  &-1.55&32\\
                &-2.15 & 1.95 & 1.20 & 0.75 & 0.60 & 0.95 & 1.39 & 2.27 & 2.28 &  -   & 2.00 &  -   & 1.94 &  -   &  -   & -   &  -  &  -    & - &20\\
                &-2.30 & 2.34 & 1.24 & 1.10 & 0.59 & 0.60 & 1.28 & 2.26 & 2.19 &  -   &  -   &  -   & 1.91 & 2.89 &  -   & -   &  -  &  -    & - &30\\
                &-2.25 & 2.20 & 0.90 & 1.30 & 0.92 & 0.73 & 1.01 & 2.23 & 2.13 & 2.13 & 2.23 & 2.08 & 1.88 & 2.84 &  -   & -   &  -  &  -    & - &2\\
CS 22948-027    &-2.50 & 2.19 & 1.55 & 0.64 & 0.90 & 1.00 &  -   & 2.06 & 2.32 & 2.20 & 2.22 &  -   & 1.99 & 2.72 & 0.95 & 2.20& 1.25& 0.07  &-1.16&32\\
                &-2.47 & 2.43 & 1.75 & 0.68 & 0.90 & 1.00 &  -   & 2.26 & 2.32 & 2.20 & 2.22 &  -   & 1.88 & 2.72 &  -   &  -  &  -  &  -    & - &4\\
                &-2.57 & 2.10 & 1.10 & 1.00 &  -   &  -   &  -   & 1.67 &  -   &  -   &  -   &  -   &  -   &  -   &  -   &  -  &  -  &  -    & - &20\\
                &-2.47$^{a}$& 2.05 & 1.80 & 0.25 & 0.90 & 1.00 &  -   & 2.26 & 2.32 & 2.20 & 2.22 &  -   & 2.10 &  -   &  -  &  - &  -  &  - & - &11\\
CS 29497-034    &-3.00 & 2.61 & 2.16 & 0.45 & 1.20 & 0.97 &  -   & 2.05 & 2.25 & 2.00 & 2.16 &  -   & 2.03 & 2.95 & 1.09 & 2.12& 1.03& 0.02  &-0.85 &32\\
                &-2.90 & 2.63 & 2.38 & 0.25 & 1.00 & 1.10 &  -   & 2.03 & 2.12 & 1.95 & 2.09 &  -   & 1.80 & 2.95 &  -   &  -  &  -  &   -   & - &4\\
                &-2.90$^{a}$& 2.50 & 2.30 & 0.20 & 1.00 & 1.10 &  -   & 2.03 & 2.12 & 1.95 & 2.09 &  -   & 2.25 &  -   &  -   &  -  &  - &  -& -  &11\\
                &-3.20 & 2.70 & 1.80 & 0.90 & 1.60 & 0.70 &  -   & 2.10 & 2.50 & 2.10 & 2.30 &  -   &  -   &  -   & -    & -  &  -  &  -    & - &31\\
CS 29497-030$^{c}$    &-2.64 & 2.43 & 2.00 & 0.43 & 1.09 & 0.84 & 1.42 & 2.25 & 2.16 & 2.12 & 2.00 &  -   & 1.72 & 3.60 & 1.12 & 2.13& 1.01& 0.53 &-1.16 &32\\
                &-2.57 & 2.47 & 2.12 & 0.35 & 1.34 & 0.97 & 1.40 & 2.32 & 2.22 & 2.10 & 2.14 &  -   & 1.99 & 3.65 &  -   &  -  &  -  &  -     & - &13\\
                &-2.70 & 2.38 & 1.88 & 0.50 & 0.84 & 0.71 & 1.43 & 2.17 & 2.10 & 2.14 & 1.85 &  -   & 1.44 & 3.55 &  -   &  -  &  -  &  -    & - &22\\
CS 29503-010    &-1.69$^{a}$& 1.65 &  -   &  -   &1.13$^{b}$& 1.09 & 1.26 & 1.81 & 2.16 & 2.05 & 2.31 & 2.34 & 1.69 &  -   & 1.16 &2.08& 0.92&0.12&-0.68&30\\
CS 29526-110$^{d}$    &-2.21 & 2.22 & 1.40 & 0.82 & 0.65 & 1.32 & 1.39 & 2.26 & 2.12 & 2.15 & 2.10 &  -   & 2.01 & 3.23 & 1.12 & 2.16& 1.04& 0.25  &-1.61&32\\
                &-2.07 & 2.07 &  -   &  -   & 0.77 & 1.34 & 1.26 & 2.39 & 2.09 &  -   &  -   &  -   &  -   & 3.16 &  -   &  -  &  -  &  -    & - &3\\
                &-2.38 & 2.20 & 1.40 & 0.80 & 0.88 &  -   & 1.11 & 2.11 & 1.69 & 2.01 & 2.01 &  -   & 1.73 & 3.30 &  -   &  -  &  -  &  -    & - &2\\
                &-2.19$^{a}$& 2.38 &  -   &  -   & 0.29 & 1.29 & 1.79 & 2.29 & 2.57 & 2.29 & 2.19 &  -   & 2.28 &  -   &  -  &  -  &  -  &  -  & - &30\\
CS 29528-028    &-2.12$^{a}$& 2.76 &  -   &  -   &1.72$^{b}$& 1.99 & 2.17 & 2.49 & 2.21 & 2.47 & 2.54 &  -   & 2.16 &  -   & 1.96 &2.43&0.47&0.33 &-0.77&30\\
CS 31062-050$^{c,d}$    &-2.37 & 1.91 & 1.20 & 0.71 & 0.91 & 0.48 & 0.94 & 2.55 & 2.28 & 2.06 & 2.12 & 2.06 & 1.82 & 2.86 & 0.78 & 2.25& 1.47& 0.73  &-1.64&32\\
                &-2.41 & 1.82 &  -   &  -   &  -   & 0.48 & 0.85 & 2.80 & 2.12 & 2.02 & 1.99 & 1.96 & 1.79 & 2.81 &  -   &  -  &  -  &  -    & - &15\\
                &-2.32 & 2.00 & 1.20 & 0.80 & 0.91 &  -   & 1.02 & 2.30 & 2.44 & 2.10 & 2.24 & 2.15 & 1.84 & 2.90 &  -   &  -  &  -  &  -    & - &2\\
HD 187861       &-2.01 & 2.02 & 2.18 &-0.16 &  -   & 0.00 &  -   & 1.39 & 1.89 & 1.69 & 1.80 & 0.70 & 1.34 & 2.86 &  -   & 1.70&  -  & 0.05  & - &32\\
                &-1.65 &  -   &  -   &  -   &  -   & 0.00 &  -   &  -   & 2.05 & 2.00 & 1.80 & 0.70 &   -  &  -   &  -   &  -  &  -  &  -    & - &34\\
                &-2.36 & 2.02 & 2.18 &-0.16 &  -   &  -   &  -   & 1.39 & 1.73 & 1.37 &   -  &   -  & 1.34 & 2.86 &  -   &  -  &  -  &  -    & - &35\\
HD 209621$^{e}$       &-1.42 & 1.25 &  -   &  -   & 1.02 & 0.73 & 1.80 & 1.70 & 2.41 & 2.42 & 2.14 & 1.46 & 1.35 & 1.88 & 1.18 & 2.17& 0.99& 0.35  &-0.68&32\\
                &-1.93 & 1.25 &  -   &  -   & 1.02 & 0.36 & 1.80 & 1.70 & 2.41 & 2.04 & 1.87 & 1.46 & 1.35 & 1.88 &  -   &  -  &  -  &  -    & - &28\\
                &-0.90 &  -   &  -   &  -   &  -   & 1.10 &  -   &  -   &  -   & 2.80 & 2.40 &  -   &  -   &  -   &  -   &  -  &  -  &  -    & - &34\\
HD 224959       &-1.83 & 1.77 & 1.88 &-0.11 &  -   & 0.80 &  -   & 2.19 & 2.02 & 2.01 & 1.80 & 1.40 & 1.74 & 3.06 &  -   & 2.01&  -  & 0.45  & - &32\\
                &-1.60 &  -   &  -   &  -   &  -   & 0.80 &  -   &  -   & 2.00 & 2.10 & 1.80 & 1.40 &  -   &  -   &  -   &  -  &  -  &  -    & - &34\\
                &-2.06 & 1.77 & 1.88 &-0.11 &  -   &  -   &  -   & 2.19 & 2.03 & 1.91 &  -   &  -   & 1.74 & 3.06 &  -   &  -  &  -  &  -    & - &35\\
HE 0002-1037    &-2.40 & 1.90 & 1.00 & 0.90 &$<$1.00 & 0.40 &  -   & 2.00 & 2.00 & 1.70 & 2.10 &  -   & 1.70 &  -   &  - & 1.95&  -  & 0.30  & - &31\\
HE 0017+0055    &-2.40 & 2.17 & 2.47 &-0.30 &  -   & 0.50 & 1.60 &$>$1.90 & 2.40 & 2.00 & 2.20 & 1.90 & 2.30 &  -   & 1.05 & 2.20& 1.15&  -    & - &26\\
HE 0059-6540    &-2.20 & 1.30 & 1.20 & 0.10 & 1.20 & 0.40 &  -   & 1.70 & 1.60 & 1.40 & 1.70 &  -   & 1.50 &  -   & 0.80 & 1.60& 0.80& 0.20  &-0.50&31\\
HE 0131-3953$^{c,d}$    &-2.71 & 2.45 &  -   &  -   & 0.46 &  -   &  -   & 2.20 & 1.94 & 1.93 & 1.76 &  -   & 1.62 &  -   &  -   & 1.96&  -  & 0.58  &-1.74&5\\
HE 0143-0441$^{c}$    &-2.31$^a$ & 1.98 & 1.73 & 0.25 & 0.86 & 0.59 & 1.05 & 2.32 & 1.78 & 1.93 & 2.17 &  -   & 1.46 & 3.11 & 0.83 & 2.05& 1.22& 0.86  &-1.46&8\\
HE 0151-6007    &-2.70 & 1.40 & 0.20 & 1.20 & 1.10 & 0.80 &  -   & 2.30 & 2.50 & 2.40 & 2.60 &  -   & 2.30 &  -   & 0.95 & 2.45& 1.50& 0.00  &-1.20&31\\
HE 0338-3945$^{d}$    &-2.42 & 2.10 & 1.55 & 0.55 & 0.74 & 0.78 & 1.20 & 2.41 & 2.27 & 2.19 & 2.20 & 2.25 & 1.92 & 3.10 & 0.91 & 2.27& 1.36& 0.49  &-1.67&32\\
                &-2.41 & 2.07 &  -   &  -   & 0.73 & 0.73 &  -   & 2.41 & 2.26 & 2.21 & 2.09 &  -   & 1.89 &  -   &  -   &  -  &  -  &  -    & - &5\\
                &-2.42 & 2.13 & 1.55 & 0.58 & 0.74 & 0.83 & 1.20 & 2.41 & 2.28 & 2.16 & 2.30 & 2.25 & 1.94 & 3.10 &  -   &  -  &  -  &  -    & - &17\\
HE 1105+0027$^{c,d}$    &-2.42 & 2.00 &  -   &  -   & 0.73 & 0.75 &  -   & 2.45 & 2.10 &  -   & 2.06 &  -   & 1.81 &  -   & 0.74 & 2.20& 1.46& 0.64  &-1.72&5\\
HE 1305+0007    &-2.01 & 1.84 &  -   &  -   & 0.86 & 0.73 & 2.09 & 2.32 & 2.56 & 2.53 & 2.59 & 2.60 & 1.97 & 2.37 & 1.23 & 2.50& 1.27& 0.35  &-1.46&10\\
HE 2148-1247$^{d}$    &-2.26 & 1.91 & 1.65 & 0.26 & 0.76 & 0.83 & 1.47 & 2.36 & 2.38 & 2.28 & 2.27 & 1.99 & 1.98 & 3.12 & 1.02 & 2.32& 1.30& 0.38  &-1.60&7\\
HE 2258-6358$^{c}$    &-2.67 & 2.42 & 1.44 & 0.98 & 0.80 & 0.70 & 0.69 & 2.23 & 1.91 & 1.66 & 1.76 & 1.81 & 1.68 & 3.32 & 0.73 & 1.89& 1.16& 0.55  &-1.43&36\\
LP 625-44$^{c,d}$       &-2.71 & 2.10 & 1.00 & 1.10 & 1.15 & 0.99 & 1.34 & 2.74 & 2.46 & 2.27 & 2.30 & 2.21 & 1.97 & 2.55 & 1.16 & 2.44& 1.28& 0.77  &-1.59&1\\
LP 706-7$^{c,d}$        &-2.61 & 2.13 & 1.50 & 0.63 & 0.18 & 0.42 &$<$1.16 & 2.02 & 1.92 & 1.99 & 1.90 &$<$2.21 & 1.51 & 2.40 & 0.30 & 1.96& 1.66& 0.51 & -1.84&32\\
                &-2.53 & 2.14 &  -   &  -   & 0.08 &  -   &  -     & 2.08 & 1.92 &  -   &  -   &  -     &  -   & 2.53 &  -   &  -  &  -  &  -   & - &3\\
                &-2.55 & 2.10 & 1.20 & 0.90 & 0.30 & 0.59 &  -   & 1.98 & 2.02 & 2.12 & 1.79 &  -   & 1.62 & 2.40 &  -   &  -  &  -  &  -    & - &2\\
                &-2.74 & 2.15 & 1.80 & 0.35 & 0.15 & 0.25 &$<$1.16 & 2.01 & 1.81 & 1.86 & 2.01 &$<$2.21 & 1.40 & 2.28 &  -   &  -  &  -  &  -   & - &1\\
SDSS J0912+0216 &-2.50 & 2.17 & 1.75 & 0.42 & 0.57 & 0.61 & 1.08 & 1.49 & 1.35 & 2.17 & 1.12 & 2.60 & 1.20 & 2.33 & 0.75 & 1.53& 0.78& 0.29 &-0.92&6\\
SDSS J1349-0229$^{c}$ &-3.00 & 2.82 & 1.60 & 1.22 & 1.30 & 1.29 & 1.56 & 2.17 & 1.74 & 2.63 & 1.91 & 2.35 & 1.62 & 3.09 & 1.38 & 2.11& 0.73& 0.55  &-0.87&6\\
\hline
\hline
  &      &      &      &      &      &      &      & Our Work &      &      &      &      &      &      &      &     &     &   &   &   \\
Star Name&[Fe/H]&[C/Fe]&[N/Fe]&[C/N]&[Sr/Fe]&[Y/Fe]&[Zr/Fe]&[Ba/Fe]&[La/Fe]&[Ce/Fe]&[Nd/Fe]&[Sm/Fe]&[Eu/Fe]&[Pb/Fe]&[ls/Fe]&[hs/Fe]&[hs/ls]&[Ba/Eu]&[Sr/Ba]&Ref \\
\hline
HD 145777$^{c,d,e}$       &-2.17 & 2.43 & 0.67 & 1.76 & 0.67 & 1.22 & 1.05 & 1.27 & 1.37 & 1.79 & 1.48 & 1.63 & 0.80 &  -   & 0.98 & 1.48& 0.50  & 0.47 &-0.60&27\\
CD-27 14351     &-2.71 & 2.98 & 1.88 & 1.10 & 1.74 & 1.97 & 2.21 & 1.82 & 1.56 & 1.89 & 1.37 &  -  &$<$0.39&  -   & 1.97 &1.66&$-$0.31&$>$1.43&-0.08&27\\
HE 0017+0055    &-2.46 & 2.73 & 2.83 &-0.10 &   -  & 0.58 & 1.55 & 2.30 & 2.46 & 2.11 & 2.25 & 1.98 & 2.14 &  -   & 1.07 & 2.28& 1.21  & 0.16  & - &27\\
HE 2144-1832    &-1.63 & 1.85 & 0.50 & 1.35 & 0.66 & 1.16 & 0.97 & 1.49 & 1.53 & 1.70 & 1.61 & 1.78 & 1.01 &  -   & 0.93 & 1.58& 0.65  & 0.48 &-0.83&27\\
HE 2339-0837    &-2.74 & 3.04 &  -   &  -   & 1.32 & 0.67 & 1.64 & 2.21 & 2.24 & 2.37 & 2.55 & 2.29 & 1.84 &  -   & 1.21 & 2.34& 1.13  & 0.37 &-0.89&27\\
\hline
\hline
\end{tabular}}

{\textit{a}} $-$ [Fe/H] is from Fe I lines; {\textit{b}} $-$ Average abundance from neutral and ionised species; {\textit{c}} $-$ Outliers of the classifier of \citet{beers2005discovery}; {\textit{d}} $-$ Outliers of the classifier of \citet{hansen2019abundances}; {\textit{e}} $-$  Outliers of the classifier of \citet{Frebel_review_2018}; {\textit{f}} $-$ Outlier of the classifier of \citet{abate2016cemp-rs}.\\
\end{table*}
}
\end{appendix}

{\footnotesize
\begin{table*}[ht]
{\bf{Table 4. Equivalent widths (in m\r{A}) of Fe lines used for deriving atmospheric parameters.}}\\\\
\scalebox{0.87}{
\begin{tabular}{cccccccccc}
\hline
Wavelength   &Element    &E$_{low}$ &   log gf  &  HD~145777     & CD$-$27~14351 &HE~0017$+$0055 & HE~2144$-$1832 & HE~2339$-$0837   & References  \\
(\r{A})      &           & (eV)     &           &                &               &               &               &                &        \\
\hline 
4466.573     &  Fe I     & 0.11     & $-$4.464  &      -         &       -       &  101.9 (5.08) &       -       &        -        &     1    \\
4476.019     &           & 2.85     & $-$0.570  &      -         &  112.1 (4.71) &        -      &       -       &        -        &     2   \\
4871.318     &           & 2.87     & $-$0.410  &  157.0 (5.63)  &       -       &        -      &  152.9 (5.67) &        -        &     1   \\
4882.143     &           & 3.42     & $-$1.640  &   49.3 (5.65)  &       -       &        -      &   54.6 (5.73) &        -        &     1   \\
4890.755     &           & 2.88     & $-$0.430  &      -         &  131.8 (4.77) &        -      &       -       &        -        &     1   \\
4903.310     &           & 2.88     & $-$1.080  &   93.9 (5.07)  &   67.9 (4.62) &   72.3 (4.85) &  133.9 (5.93) &  25.7 (4.69)    &     1   \\
4924.770     &           & 2.28     & $-$2.220  &      -         &   80.7 (5.10) &   66.1 (5.10) &       -       &        -        &     1   \\
4982.499     &           & 4.10     &    0.164  &      -         &   55.8 (4.79) &        -      &       -       &        -        &     1   \\
4994.130     &           & 0.92     & $-$3.080  &  155.7 (5.43)  &  134.0 (4.72) &  119.1 (4.95) &  169.2 (5.85) &  44.3 (4.72)    &     1   \\
5006.119     &           & 2.83     & $-$0.615  &      -         &       -       &        -      &       -       &  53.6 (4.68)    &   1     \\
5079.223     &           & 2.20     & $-$2.067  &  108.6 (5.32)  &   66.6 (4.66) &   83.7 (5.08) &  133.9 (5.88) &  20.7 (4.75)    &     1   \\
5166.282     &           & 0.00     & $-$4.195  &  171.9 (5.44)  &  167.7 (4.91) &  145.8 (5.19) &       -       &        -        &     1   \\
5171.596     &           & 1.49     & $-$1.793  &  180.9 (5.38)  &  179.0 (4.84) &  162.5 (5.26) &       -       &  84.4 (4.77)    &     1   \\
5192.344     &           & 3.00     & $-$0.421  &      -         &       -       &        -      &       -       &  53.9 (4.66)    &   1     \\
5194.941     &           & 1.56     & $-$2.090  &  156.3 (5.28)  &  131.4 (4.54) &        -      &       -       &  63.0 (4.77)    &     1   \\
5198.711     &           & 2.22     & $-$2.135  &   98.7 (5.23)  &   58.1 (4.65) &        -      &  137.6 (6.01) &        -        &     1   \\
5202.336     &           & 2.18     & $-$1.838  &      -         &  122.6 (5.03) &  100.2 (5.06) &       -       &  35.4 (4.80)    &     1   \\
5216.274     &           & 1.61     & $-$2.150  &  132.8 (4.95)  &  118.4 (4.50) &        -      &       -       &  48.2 (4.65)    &     1   \\
5217.389     &           & 3.21     & $-$1.162  &      -         &   73.1 (5.09) &        -      &       -       &        -        &   1     \\
5226.862     &           & 3.04     & $-$0.555  &  115.4 (5.07)  &  106.9 (4.71) &  114.5 (5.19) &       -       &  45.4 (4.69)    &   1     \\
5227.189     &           & 1.56     & $-$1.228  &      -         &       -       &  168.7 (4.89) &       -       &        -        &   1     \\
5232.940     &           & 2.94     & $-$0.190  &      -         &       -       &        -      &       -       &  78.2 (4.79)    &     1   \\
5242.491     &           & 3.63     & $-$0.840  &   45.0 (5.05)  &       -       &   50.0 (5.21) &   91.6 (5.79) &        -        &     1   \\
5247.050     &           & 0.09     & $-$4.946  &  126.1 (5.40)  &   96.6 (4.89) &   83.5 (5.00) &       -       &        -        &     1   \\
5253.462     &           & 3.28     & $-$1.670  &      -         &       -       &        -      &   85.7 (6.03) &        -        &     1   \\
5266.555     &           & 3.00     & $-$0.490  &  132.8 (5.26)  &   93.4 (4.43) &  104.4 (4.88) &  150.2 (5.70) &  61.4 (4.85)    &     1   \\
5281.790     &           & 3.04     & $-$1.020  &  107.2 (5.38)  &   65.8 (4.69) &   73.6 (4.97) &  144.6 (6.17) &  25.9 (4.78)    &     1   \\
5283.621     &           & 3.24     & $-$0.630  &  122.3 (5.54)  &  100.1 (4.97) &        -      &       -       &        -        &     1   \\
5307.361     &           & 1.61     & $-$2.987  &   91.2 (5.08)  &   70.5 (4.79) &        -      &  130.4 (5.80) &        -        &     1   \\
5324.179     &           & 3.21     & $-$0.240  &  127.1 (5.20)  &  119.5 (4.78) &  105.8 (4.94) &  149.8 (5.72) &  54.0 (4.71)    &     1   \\
5328.531     &           & 1.56     & $-$1.850  &  180.2 (5.46)  &  165.3 (4.72) &  142.9 (4.94) &       -       &        -        &   1     \\
5339.929     &           & 3.27     & $-$0.680  &  103.5 (5.29)  &   94.5 (4.97) &        -      &       -       &        -        &     1   \\
5341.024     &           & 1.61     & $-$2.060  &  152.0 (5.18)  &  147.2 (4.75) &        -      &       -       &        -        &     1   \\
5367.466     &           & 4.42     &    0.350  &   74.0 (5.32)  &   42.9 (4.80) &   57.9 (5.13) &  106.7 (5.91) &        -        &     1   \\
5369.961     &           & 4.37     &    0.350  &   67.8 (5.17)  &       -       &        -      &   91.0 (5.56) &        -        &     1   \\
5415.199     &           & 4.39     &    0.500  &      -         &   60.6 (4.84) &        -      &       -       &        -        &     1   \\
5466.396     &           & 4.37     & $-$0.630  &      -         &       -       &        -      &   48.3 (5.83) &        -        &     1   \\
5586.756     &           & 3.37     & $-$0.210  &  126.8 (5.31)  &   87.4 (4.53) &  108.5 (5.11) &  153.5 (5.90) &  48.2 (4.75)    &     1   \\
5638.262     &           & 4.22     & $-$0.870  &      -         &       -       &        -      &   56.6 (5.99) &        -        &     1   \\
5753.122     &           & 4.26     & $-$0.760  &   35.6 (5.60)  &       -       &        -      &   69.5 (6.13) &        -        &     1   \\
6027.051     &           & 4.08     & $-$1.210  &      -         &       -       &        -      &   53.5 (6.07) &        -        &     1   \\
6056.005     &           & 4.73     & $-$0.460  &      -         &       -       &        -      &   30.1 (5.77) &        -        &     1   \\
6136.615     &           & 2.45     & $-$1.400  &  157.8 (5.62)  &  119.7 (4.78) &        -      &       -       &  55.2 (4.93)    &     1   \\
6137.691     &           & 2.59     & $-$1.403  &  128.1 (5.33)  &  100.7 (4.76) &        -      &  167.5 (6.06) &  43.6 (4.91)    &     1   \\
6151.617     &           & 2.18     & $-$3.299  &   30.4 (5.24)  &       -       &        -      &   84.0 (5.98) &        -        &     1   \\
6180.203     &           & 2.73     & $-$2.780  &   28.2 (5.43)  &       -       &        -      &   50.6 (5.76) &        -        &     1   \\
6230.722     &           & 2.56     & $-$1.281  &  132.0 (5.21)  &  121.2 (4.81) &  111.8 (5.03) &       -       &  43.7 (4.75)    &     1   \\
6252.555     &           & 2.40     & $-$1.687  &  138.9 (5.49)  &   91.9 (4.69) &   91.5 (4.92) &  164.5 (5.98) &  32.4 (4.77)    &     1   \\
6318.018     &           & 2.45     & $-$2.228  &      -         &       -       &        -      &  111.9 (5.81) &        -        &     1   \\
6335.330     &           & 2.20     & $-$2.230  &  118.6 (5.42)  &   74.0 (4.77) &   75.5 (4.96) &  137.8 (5.74) &        -        &     1   \\
6421.350     &           & 2.28     & $-$2.027  &      -         &  111.8 (5.05) &        -      &       -       &        -        &     1   \\
6593.870     &           & 2.43     & $-$2.422  &   79.0 (5.37)  &       -       &        -      &  112.5 (5.83) &        -        &     1   \\
4233.172     &  Fe II    & 2.58     & $-$2.000  &   75.3 (5.39)  &       -       &        -      &       -       &        -        &     1   \\
4515.339     &           & 2.84     & $-$2.480  &   39.2 (5.40)  &       -       &        -      &       -       &        -        &     1   \\
4520.224     &           & 2.81     & $-$2.600  &   20.4 (4.98)  &       -       &        -      &       -       &        -        &     1   \\
4583.837     &           & 2.81     & $-$2.020  &      -         &    54.7 (4.74)&   58.4 (5.08) &       -       &  49.6 (4.87)    &     1   \\
4923.927     &           & 2.89     & $-$1.320  &   98.1 (5.56)  &    96.9 (4.81)&        -      &       -       &  71.7 (4.73)    &     1   \\
5197.577     &           & 3.23     & $-$2.100  &      -         &       -       &   29.5 (5.03) &   62.0 (5.87) &        -        &     1   \\
5234.625     &           & 3.22     & $-$2.050  &      -         &       -       &        -      &   65.2 (5.88) &  24.4 (4.76)    &     1   \\
5276.002     &           & 3.20     & $-$1.940  &      -         &    40.5 (4.87)&   37.6 (5.03) &       -       &  26.4 (4.68)    &     1   \\
\hline
\end{tabular}}

The numbers in the parenthesis in columns 5 to 9 give the derived abundances from the respective line.\\
References: 1. \citet{FMW1988}, 2. \citet{BK1974}. \\

\end{table*}}

{\footnotesize
\begin{table*}[ht]
{\bf{Table 6. Equivalent widths (in m\r{A}) of lines used for calculation of elemental abundances.}}\\\\
\scalebox{0.87}{
\begin{tabular}{cccccccccc}
\hline
Wavelength   &Element    &E$_{low}$ &   log gf  &  HD~145777     & CD$-$27~14351 &HE~0017$+$0055 &HE~2144$-$1832 & HE~2339$-$0837 & References  \\
(\r{A})      &           & (eV)     &           &                &               &               &               &                &             \\
\hline 
5682.633     &  Na I     & 2.10     & $-$0.700  &  40.2 (4.61)   &  52.9 (5.05)  &      -        &  69.8 (5.05)  &  36.3 (5.06)   &     1     \\
5688.205     &           & 2.10     & $-$0.450  &  45.1 (4.44)   &  51.0 (4.77)  &      -        &  93.7 (5.16)  &  47.2 (5.02)   &     1     \\
6160.747     &           & 2.10     & $-$1.260  &     -          &       -       &      -        &  45.1 (5.20)  &       -        &     1     \\
4571.096     &  Mg I     & 0.00     & $-$5.691  &     -          &       -       &      -        &      -        &       -        &     2     \\
5528.405     &           & 4.35     & $-$0.620  & 149.7 (6.42)   & 170.9 (6.70)  &      -        & 176.5 (6.71)  & 142.2 (7.03)   &     3     \\
5711.088     &           & 4.35     & $-$1.833  &  53.9 (6.09)   &       -       &      -        &  94.4 (6.56)  &       -        &     3     \\
4318.652     &  Ca I     & 1.90     & $-$0.208  & 106.5 (4.45)   &  79.3 (4.36)  &      -        & 138.1 (5.08)  &  53.4 (4.10)   &     4     \\
5349.465     &           & 2.71     & $-$1.178  &  36.5 (4.91)   &       -       &      -        &      -        &       -        &     5     \\
5512.980     &           & 2.93     & $-$0.290  &      -         &       -       &      -        &  67.3 (4.82)  &       -        &     4     \\
5588.749     &           & 2.53     &    0.210  &      -         &       -       &      -        &      -        &  36.6 (3.93)   &     4     \\
5590.114     &           & 2.52     & $-$0.710  &  57.6 (4.50)   &       -       &      -        &  86.3 (4.97)  &       -        &     4     \\
5594.462     &           & 2.52     & $-$0.050  &      -         &       -       &      -        &      -        &  40.8 (4.27)   &     4     \\
5857.451     &           & 2.93     & 0.230     & 101.4 (4.80)   &       -       &      -        &      -        &       -        &     4     \\
6102.723     &           & 1.88     & $-$0.890  &  97.8 (4.31)   &  79.6 (4.34)  &      -        & 151.4 (5.15)  &       -        &     4     \\
6166.439     &           & 2.52     & $-$0.900  &  61.7 (4.68)   &  42.9 (4.67)  &      -        & 100.1 (5.28)  &       -        &     4     \\
6169.042     &           & 2.52     & $-$0.550  &  94.8 (4.84)   &  76.5 (4.79)  &      -        & 113.6 (5.16)  &       -        &     4     \\
6169.563     &           & 2.53     & $-$0.270  &  93.4 (4.54)   &       -       &      -        & 125.1 (5.07)  &       -        &     4     \\
6439.075     &           & 2.53     &  0.470    & 169.9 (4.87)   & 135.4 (4.56)  &      -        &      -        &  58.0 (4.06)   &     4     \\
6572.779     &           & 0.00     & $-$4.290  &      -         &       -       &      -        & 147.5 (5.34)  &       -        &     4     \\
4840.874     &  Ti I     & 0.90     & $-$0.509  &      -         &  82.1 (3.34)  &      -        & 123.3 (3.54)  &       -        &     6     \\
4926.148     &           & 0.82     & $-$2.170  &      -         &       -       &      -        &  38.8 (3.70)  &       -        &     6     \\
4999.503     &           & 0.83     &    0.250  & 165.8 (3.32)   & 123.1 (2.97)  &      -        &      -        &       -        &     6     \\
5024.844     &           & 0.82     & $-$0.602  &      -         &  83.2 (3.27)  &      -        & 131.1 (3.57)  &       -        &     6     \\
5210.385     &           & 0.05     & $-$0.884  &      -         &       -       & 118.3 (3.13)  &      -        &       -        &     6     \\
5460.499     &           & 0.05     & $-$2.880  &      -         &       -       &      -        &  75.5 (3.60)  &       -        &     7     \\
5918.535     &           & 1.07     & $-$1.460  &      -         &       -       &      -        &  47.3 (3.33)  &       -        &     6     \\
5922.109     &           & 1.05     & $-$1.466  &      -         &       -       &      -        &  58.8 (3.46)  &       -        &     6     \\
5978.541     &           & 1.87     & $-$0.496  &  49.5 (3.56)   &       -       &      -        &      -        &       -        &     6     \\
4443.794     &  Ti II    & 1.08     & $-$0.700  &      -         &       -       &      -        &      -        & 101.3 (2.65)   &     6     \\
4563.761     &           & 1.22     & $-$0.960  &      -         &       -       &      -        &      -        &  93.6 (2.82)   &     6     \\
4764.526     &           & 1.24     & $-$2.770  &  60.3 (3.43)   &       -       &      -        &      -        &       -        &     5     \\
4779.985     &           & 2.05     & $-$1.370  &  72.8 (3.39)   &  82.0 (3.27)  &  66.1 (3.19)  &      -        &       -        &     6     \\
4798.521     &           & 1.08     & $-$2.430  &  86.2 (3.35)   &  92.9 (3.18)  &      -        &      -        &       -        &     6     \\
4805.085     &           & 2.06     & $-$1.100  &  93.4 (3.56)   &  82.2 (3.01)  &  75.6 (3.12)  &  90.3 (3.54)  &       -        &     6     \\
4865.612     &           & 1.12     & $-$2.610  &      -         &       -       &      -        &  78.2 (3.46)  &       -        &     6     \\
5185.913     &           & 1.89     & $-$1.350  &  94.4 (3.44)   &       -       &      -        &      -        &       -        &     6     \\
5226.543     &           & 1.57     & $-$1.300  &      -         &       -       & 104.3 (3.10)  &      -        &  56.5 (2.53)   &     6     \\
5336.771     &           & 1.58     & $-$1.700  &      -         & 109.7 (3.18)  &      -        & 102.6 (3.53)  &  33.5 (2.49)   &     6     \\
5381.015     &           & 1.57     & $-$2.080  &      -         &  99.7 (3.37)  &      -        &      -        &       -        &     6     \\
5206.037     &  Cr I     & 0.94     &    0.019  &      -         &       -       &      -        &      -        &  58.9 (2.56)   &     6     \\
5296.691     &           & 0.98     & $-$1.400  & 104.2 (3.31)   &       -       &      -        & 145.6 (4.04)  &       -        &     6     \\
5298.272     &           & 0.98     & $-$1.150  & 122.1 (3.34)   &  65.4 (2.95)  &      -        &      -        &       -        &     6     \\
5300.745     &           & 0.98     & $-$2.120  &  49.0 (3.22)   &       -       &      -        & 101.7 (4.00)  &       -        &     6     \\
5345.796     &           & 1.00     & $-$0.980  & 136.3 (3.41)   &  86.1 (3.07)  &      -        &      -        &       -        &     6     \\
5348.315     &           & 1.00     & $-$1.290  &  91.8 (3.03)   &       -       &  33.4 (2.68)  & 129.9 (3.65)  &       -        &     6     \\
5409.784     &           & 1.03     & $-$0.720  & 142.5 (3.26)   & 121.6 (3.30)  &      -        & 172.3 (3.89)  &       -        &     6     \\
4813.467     &  Co I     & 3.22     &    0.050  &      -         &       -       &      -        &  44.9 (3.46)  &       -        &     5     \\
5342.695     &           & 4.02     &    0.690  &      -         &       -       &      -        &  24.9 (3.41)  &       -        &     5     \\
5483.344     &           & 1.71     & $-$1.490  &  51.4 (2.85)   &       -       &      -        &      -        &       -        &     8     \\
6116.996     &           & 1.79     & $-$2.490  &      -         &       -       &      -        &  22.4 (3.39)  &       -        &     8     \\
4756.510     &  Ni I     & 3.48     & $-$0.270  &      -         &       -       &      -        &  80.4 (4.85)  &       -        &     9     \\
5146.480     &           & 3.71     &    0.120  &      -         &       -       &      -        &  94.4 (4.87)  &       -        &     10    \\
6175.360     &           & 4.09     & $-$0.530  &      -         &       -       &      -        &  57.3 (5.16)  &       -        &     8     \\
6176.807     &           & 4.09     & $-$0.530  &      -         &       -       &      -        &  58.7 (5.19)  &       -        &     9     \\
6177.236     &           & 1.83     & $-$3.500  &      -         &       -       &      -        &  51.5 (4.91)  &       -        &     8     \\
4810.528     &  Zn I     & 4.08     & $-$0.137  &  58.0 (3.02)   &       -       &  45.0 (2.62)  &  49.6 (2.85)  &       -        &     11    \\
4854.863     &  Y II     & 0.99     & $-$0.380  &      -         &       -       &  74.1 (0.42)  &      -        &  41.4 (0.18)   &     12    \\
5200.406     &           & 0.99     & $-$0.570  &      -         &       -       &  63.0 (0.27)  & 130.8 (1.71)  &  35.8 (0.21)   &     12    \\
5205.724     &           & 1.03     & $-$0.340  &      -         &       -       &  73.1 (0.30)  &      -        &  35.8 (0.03)   &     12    \\
5289.815     &           & 1.03     & $-$1.850  &      -         &  63.2 (1.36)  &      -        &      -        &       -        &     12    \\
5402.774     &           & 1.84     & $-$0.510  &  56.4 (1.25)   &  90.9 (1.50)  &      -        &  83.5 (1.83)  &       -        &     13    \\
5662.925     &           & 1.94     &    0.160  &  89.7 (1.28)   & 132.6 (1.56)  &      -        & 105.3 (1.68)  &       -        &     13    \\
4739.480     &  Zr I     & 0.65     &    0.230  &  57.0 (1.55)   &       -       &      -        &      -        &       -        &     14    \\
4772.323     &           & 0.62     &    0.040  &      -         &       -       &      -        &  79.1 (2.03)  &       -        &     14    \\

\hline
\end{tabular}}

The numbers in the parenthesis in columns 5 to 9 give the derived abundances from the respective line.\\
References: 1. \citet{KP1975}, 2. \citet{LV1974}, 3. \citet{LZ1971}, 4. NBS in kurucz database, 5. K88 in kurucz database, 6. MFW in kurucz database, 7. \citet{SK1978}, 8. \citet{FMW1988}, 9. \citet{LW1970}, 10. \citet{HEISE1974}, 11. \citet{WAR1968}, 12. \citet{HL1982}, 13. \citet{CC1983}, 14. \citet{BG1981}. \\

\end{table*}}

{\footnotesize
\begin{table*}[ht]
{\bf{Table 6}} --{\textit{continued}}\\\\
\scalebox{0.87}{
\begin{tabular}{cccccccccc}
\hline
Wavelength   &Element    &E$_{low}$ &   log gf  &  HD~145777     & CD$-$27~14351 &HE~0017$+$0055 &HE~2144$-$1832 & HE~2339$-$0837 & References  \\
(\r{A})      &           & (eV)     &           &                &               &               &                &               &             \\
\hline 
4805.889     &           & 0.69     & $-$0.420  &      -         &       -       &      -        &  27.0 (1.72)  &       -        &     14    \\
6134.585     &           & 0.00     & $-$1.280  &  29.4 (1.36)   &  34.9 (2.07)  &      -        &  79.7 (2.02)  &       -        &     14    \\
4048.680     &  Zr II    & 0.80     & $-$0.345  &      -         &       -       &      -        &      -        &  75.3 (1.48)   &     13    \\
4457.418     &           & 1.37     & $-$0.610  &      -         &       -       &  51.0 (1.64)  &      -        &      -         &     15    \\
4962.310     &           & 0.97     & $-$2.000  &      -         &       -       &  27.1 (1.49)  &      -        &      -         &     15    \\
5112.297     &           & 1.67     & $-$0.590  &      -         & 102.1 (2.09)  &  73.2 (1.89)  &      -        &      -         &     16    \\
4257.119     &  Ce II    & 0.46     & $-$1.116  &      -         &       -       &      -        &      -        &  30.2 (1.18)   &     17    \\
4336.244     &           & 0.70     & $-$0.564  &      -         &       -       &      -        &  74.2 (1.45 ) &      -         &     17    \\
4427.916     &           & 0.54     & $-$0.460  &  89.1 (1.29)   &       -       &      -        &      -        &      -         &     17    \\
4460.207     &           & 0.17     &    0.478  &      -         & 160.4 (0.79)  &      -        &      -        &      -         &     17    \\
4483.893     &           & 0.86     &    0.010  &  88.0 (1.26)   &  83.3 (0.91)  &      -        & 103.1 (1.70)  &      -         &     17    \\
4486.909     &           & 0.30     & $-$0.474  &      -         &       -       &      -        & 107.2 (1.43)  &      -         &     17    \\
4562.359     &           & 0.48     &    0.081  & 119.2 (1.23)   &       -       &      -        &      -        &      -         &     17    \\
4725.069     &           & 0.52     & $-$1.204  &      -         &       -       &      -        &  77.2 (1.71)  &  34.0 (1.35)   &     17    \\
4873.999     &           & 1.11     & $-$0.892  &      -         &       -       &      -        &  58.2 (1.80)  &      -         &     6     \\
5187.458     &           & 1.21     & $-$0.104  &  66.1 (1.16)   &  45.6 (0.59)  &  66.8 (1.19)  &  93.9 (1.80)  &      -         &     17    \\
5191.633     &           & 0.87     & $-$0.689  &      -         &       -       &  72.0 (1.28)  &      -        &      -         &     17    \\
5330.556     &           & 0.87     & $-$0.760  &  53.0 (1.07)   &       -       &      -        &      -        &  27.5 (1.10)   &     17    \\
5259.728     &  Pr II    & 0.63     &    0.080  &      -         &       -       &      -        &  64.4 (0.79)  &  37.8 (0.09)   &     9     \\
5322.772     &           & 0.48     & $-$0.315  &  70.5 (0.22)   &       -       &  90.6 (0.68)  & 106.1 (0.99)  &      -         &     17    \\
6165.891     &           & 0.92     & $-$0.205  &  46.8 (0.22)   &  44.3 (-0.03) &      -        &  68.1 (0.64)  &  26.9 (0.39)   &     17    \\
4451.980     &  Nd II    & 0.00     & $-$1.340  &      -         &  30.3 (-0.09) &      -        &      -        &      -         &     17    \\
4594.447     &           & 0.20     & $-$1.550  &      -         &       -       &  68.3 (1.07)  &      -        &      -         &     17    \\
4645.760     &           & 0.56     & $-$0.750  &      -         &       -       &      -        &  90.2 (1.40)  &      -         &     17    \\
4706.543     &           & 0.00     & $-$0.880  &      -         &       -       &      -        &      -        &  77.6 (1.31)   &     17    \\
4797.153     &           & 0.56     & $-$0.950  &  47.3 (0.61)   &       -       &  81.0 (1.35)  &  75.1 (1.22)  &      -         &     17    \\
4811.342     &           & 0.06     & $-$1.140  &  91.9 (0.91)   &  57.1 (0.12)  &      -        &      -        &  67.0 (1.31)   &     17    \\
4825.478     &           & 0.18     & $-$0.860  &      -         &  78.0 (0.32)  &      -        & 126.2 (1.65)  &  64.8 (1.10)   &     17    \\
5212.361     &           & 0.20     & $-$0.870  &      -         &  55.1 (-0.09) &      -        &      -        &      -         &     17    \\
5255.505     &           & 0.20     & $-$0.820  & 104.7 (0.87)   &  68.1 (0.03)  &      -        &      -        &  74.8 (1.28)   &     17    \\
5293.163     &           & 0.82     & $-$0.060  &  90.4 (0.76)   &  73.5 (0.18)  & 108.5 (1.20)  & 120.7 (1.48)  &  70.7 (1.15)   &     18    \\
5356.967     &           & 1.26     & $-$0.250  &      -         &       -       &      -        &  71.3 (1.30)  &      -         &     18    \\
5442.264     &           & 0.68     & $-$0.910  &      -         &       -       &      -        &  94.1 (1.53)  &      -         &     18    \\
5688.518     &           & 0.99     & $-$0.250  &  56.7 (0.50)   &       -       &  96.2 (1.23)  &  91.9 (1.21)  &      -         &     18    \\
4318.927     &  Sm II    & 0.28     & $-$0.613  &      -         &       -       &      -        &      -        &  47.3 (0.37)   &     17    \\
4424.337     &           & 0.48     & $-$0.260  &      -         &       -       &      -        &      -        &  62.5 (0.65)   &     17    \\
4458.509     &           & 0.10     & $-$1.110  &  69.5 (0.44)   &       -       &  65.0 (0.45)  & 106.5 (1.34)  &  41.9 (0.51)   &     17    \\
4704.400     &           & 0.00     & $-$1.562  &      -         &       -       &      -        &  84.7 (1.03)  &      -         &     17    \\
4815.805     &           & 0.18     & $-$1.420  &      -         &       -       &  54.3 (0.50)  &      -        &      -         &     17    \\
4854.368     &           & 0.38     & $-$1.873  &  17.3 (0.40)   &       -       &      -        &  39.2 (0.95)  &      -         &     17    \\

\hline
\end{tabular}}

The numbers in the parenthesis in columns 5 to 9 give the derived abundances from the respective line.\\
References: , 6. MFW in kurucz database, 9. \citet{LW1970}, 13. \citet{CC1983}, 14. \citet{BG1981}, 15. CB in kurucz database, 16. \citet{SW1979}, 17. MC in kurucz database, 18. \citet{WVA1985}.\\

\end{table*}}

\label{lastpage}

\end{document}

%% file: 1.pdf_tex
\begingroup%
  \makeatletter%
  \providecommand\color[2][]{%
    \errmessage{(Inkscape) Color is used for the text in Inkscape, but the package 'color.sty' is not loaded}%
    \renewcommand\color[2][]{}%
  }%
  \providecommand\transparent[1]{%
    \errmessage{(Inkscape) Transparency is used (non-zero) for the text in Inkscape, but the package 'transparent.sty' is not loaded}%
    \renewcommand\transparent[1]{}%
  }%
  \providecommand\rotatebox[2]{#2}%
  \ifx\svgwidth\undefined%
    \setlength{\unitlength}{720bp}%
    \ifx\svgscale\undefined%
      \relax%
    \else%
      \setlength{\unitlength}{\unitlength * \real{\svgscale}}%
    \fi%
  \else%
    \setlength{\unitlength}{\svgwidth}%
  \fi%
  \global\let\svgwidth\undefined%
  \global\let\svgscale\undefined%
  \makeatother%
  \begin{picture}(1,0.5625)%
    \put(0,0){\includegraphics[width=\unitlength,page=1]{1.pdf}}%
  \end{picture}%
\endgroup%

%% file: 2.pdf_tex
\begingroup%
  \makeatletter%
  \providecommand\color[2][]{%
    \errmessage{(Inkscape) Color is used for the text in Inkscape, but the package 'color.sty' is not loaded}%
    \renewcommand\color[2][]{}%
  }%
  \providecommand\transparent[1]{%
    \errmessage{(Inkscape) Transparency is used (non-zero) for the text in Inkscape, but the package 'transparent.sty' is not loaded}%
    \renewcommand\transparent[1]{}%
  }%
  \providecommand\rotatebox[2]{#2}%
  \ifx\svgwidth\undefined%
    \setlength{\unitlength}{720bp}%
    \ifx\svgscale\undefined%
      \relax%
    \else%
      \setlength{\unitlength}{\unitlength * \real{\svgscale}}%
    \fi%
  \else%
    \setlength{\unitlength}{\svgwidth}%
  \fi%
  \global\let\svgwidth\undefined%
  \global\let\svgscale\undefined%
  \makeatother%
  \begin{picture}(1,0.5625)%
    \put(0,0){\includegraphics[width=\unitlength,page=1]{2.pdf}}%
  \end{picture}%
\endgroup%

%% file: 3.pdf_tex
\begingroup%
  \makeatletter%
  \providecommand\color[2][]{%
    \errmessage{(Inkscape) Color is used for the text in Inkscape, but the package 'color.sty' is not loaded}%
    \renewcommand\color[2][]{}%
  }%
  \providecommand\transparent[1]{%
    \errmessage{(Inkscape) Transparency is used (non-zero) for the text in Inkscape, but the package 'transparent.sty' is not loaded}%
    \renewcommand\transparent[1]{}%
  }%
  \providecommand\rotatebox[2]{#2}%
  \ifx\svgwidth\undefined%
    \setlength{\unitlength}{720bp}%
    \ifx\svgscale\undefined%
      \relax%
    \else%
      \setlength{\unitlength}{\unitlength * \real{\svgscale}}%
    \fi%
  \else%
    \setlength{\unitlength}{\svgwidth}%
  \fi%
  \global\let\svgwidth\undefined%
  \global\let\svgscale\undefined%
  \makeatother%
  \begin{picture}(1,0.5625)%
    \put(0,0){\includegraphics[width=\unitlength,page=1]{3.pdf}}%
  \end{picture}%
\endgroup%

%% file: 4.pdf_tex
\begingroup%
  \makeatletter%
  \providecommand\color[2][]{%
    \errmessage{(Inkscape) Color is used for the text in Inkscape, but the package 'color.sty' is not loaded}%
    \renewcommand\color[2][]{}%
  }%
  \providecommand\transparent[1]{%
    \errmessage{(Inkscape) Transparency is used (non-zero) for the text in Inkscape, but the package 'transparent.sty' is not loaded}%
    \renewcommand\transparent[1]{}%
  }%
  \providecommand\rotatebox[2]{#2}%
  \ifx\svgwidth\undefined%
    \setlength{\unitlength}{720bp}%
    \ifx\svgscale\undefined%
      \relax%
    \else%
      \setlength{\unitlength}{\unitlength * \real{\svgscale}}%
    \fi%
  \else%
    \setlength{\unitlength}{\svgwidth}%
  \fi%
  \global\let\svgwidth\undefined%
  \global\let\svgscale\undefined%
  \makeatother%
  \begin{picture}(1,0.5625)%
    \put(0,0){\includegraphics[width=\unitlength,page=1]{4.pdf}}%
  \end{picture}%
\endgroup%

%% file: 5.pdf_tex
\begingroup%
  \makeatletter%
  \providecommand\color[2][]{%
    \errmessage{(Inkscape) Color is used for the text in Inkscape, but the package 'color.sty' is not loaded}%
    \renewcommand\color[2][]{}%
  }%
  \providecommand\transparent[1]{%
    \errmessage{(Inkscape) Transparency is used (non-zero) for the text in Inkscape, but the package 'transparent.sty' is not loaded}%
    \renewcommand\transparent[1]{}%
  }%
  \providecommand\rotatebox[2]{#2}%
  \ifx\svgwidth\undefined%
    \setlength{\unitlength}{720bp}%
    \ifx\svgscale\undefined%
      \relax%
    \else%
      \setlength{\unitlength}{\unitlength * \real{\svgscale}}%
    \fi%
  \else%
    \setlength{\unitlength}{\svgwidth}%
  \fi%
  \global\let\svgwidth\undefined%
  \global\let\svgscale\undefined%
  \makeatother%
  \begin{picture}(1,0.5625)%
    \put(0,0){\includegraphics[width=\unitlength,page=1]{5.pdf}}%
  \end{picture}%
\endgroup%

%% file: 6.pdf_tex
\begingroup%
  \makeatletter%
  \providecommand\color[2][]{%
    \errmessage{(Inkscape) Color is used for the text in Inkscape, but the package 'color.sty' is not loaded}%
    \renewcommand\color[2][]{}%
  }%
  \providecommand\transparent[1]{%
    \errmessage{(Inkscape) Transparency is used (non-zero) for the text in Inkscape, but the package 'transparent.sty' is not loaded}%
    \renewcommand\transparent[1]{}%
  }%
  \providecommand\rotatebox[2]{#2}%
  \ifx\svgwidth\undefined%
    \setlength{\unitlength}{720bp}%
    \ifx\svgscale\undefined%
      \relax%
    \else%
      \setlength{\unitlength}{\unitlength * \real{\svgscale}}%
    \fi%
  \else%
    \setlength{\unitlength}{\svgwidth}%
  \fi%
  \global\let\svgwidth\undefined%
  \global\let\svgscale\undefined%
  \makeatother%
  \begin{picture}(1,0.5625)%
    \put(0,0){\includegraphics[width=\unitlength,page=1]{6.pdf}}%
  \end{picture}%
\endgroup%